\documentclass[13pt]{article}
\usepackage{amssymb,amsmath,amstext,amsgen,amsopn,amsxtra,indentfirst,natbib}
\numberwithin{equation}{section}
\input psfig
\newcommand{\beq}{\begin{equation}}
\newcommand{\eeq}{\end{equation}}

\newcommand{\f}{\frac}

\newcommand{\po}{\left(}
\newcommand{\pc}{\right)}

\newcommand{\bs}{\thickspace}
\newcommand{\bbs}{\bs\bs}

\newcommand{\prt}{\partial}

\newcommand{\itR}{\mathcal{R}}
\newcommand{\itC}{\mathcal{C}}
\newcommand{\itG}{\mathcal{G}}

\newcommand{\itH}{\mathcal{H}}

\newcommand{\itD}{\mathcal{D}}
\newcommand{\itM}{\mathcal{M}}
\newcommand{\absl}{\left|}
\newcommand{\absr}{\right|}

\newcommand{\jfir}{j_{1}}
\newcommand{\jsec}{j_{2}}
\newcommand{\jthi}{j_{1,\varpi}}
\newcommand{\jfou}{j_{2,\varpi}}
\newcommand{\jfif}{j_{1,\Omega}}
\newcommand{\jsix}{j_{2,\Omega}}

\newcommand{\jifir}{j_{1}^{(i)}}
\newcommand{\jisec}{j_{2}^{(i)}}
\newcommand{\jithi}{j_{1,\varpi}^{(i)}}
\newcommand{\jifou}{j_{2,\varpi}^{(i)}}
\newcommand{\jifif}{j_{1,\Omega}^{(i)}}
\newcommand{\jisix}{j_{2,\Omega}^{(i)}}
\newcommand{\jik}{j_{k}^{(i)}}
\newcommand{\jiklilo}{j_{k,\varpi}^{(i)}}
\newcommand{\jikbigo}{j_{k,\Omega}^{(i)}}

\newcommand{\xin}{X^{(i,p)}}

\newcommand{\libarg}{e_{1}^{|\jithi|} 
e_{2}^{|\jifou|} s_{1}^{|\jifif|}  s_{2}^{|\jisix|}}

\newcommand{\aone}{a_{1}}
\newcommand{\atwod}{\aone^{-1}}
\newcommand{\atwodh}{a_{k}^{-\f{1}{2}}}

\newcommand{\Akfir}{A_{k,1}}
\newcommand{\Aksec}{A_{k,2}}
\newcommand{\Akthi}{A_{k,3}}
\newcommand{\Akfou}{A_{k,4}}
\newcommand{\Akfif}{A_{k,5}}
\newcommand{\Aksix}{A_{k,6}}

\newcommand{\Bkfir}{B_{k,1}^{(i,p)}}
\newcommand{\Bksec}{B_{k,2}^{(i,p)}}
\newcommand{\Bkthi}{B_{k,3}^{(i,p)}}
\newcommand{\Bkfou}{B_{k,4}^{(i,p)}}
\newcommand{\Bkfif}{B_{k,5}^{(i,p)}}
\newcommand{\Bksix}{B_{k,6}^{(i,p)}}

\begin{document}

{}
\bigskip
\bigskip
\bigskip

\centerline{\bf \Huge A resonant-term-based model including}

\smallskip
\smallskip

\centerline{\bf \Huge a nascent disk, precession, and oblateness:}

\smallskip
\smallskip

\centerline{\bf \Huge application to GJ 876}

\bigskip
\bigskip

\centerline{\Large by}

\bigskip

\centerline{\Large Dimitri Veras}

\bigskip
\bigskip
\bigskip
\bigskip
\bigskip
\bigskip
\bigskip
\bigskip
\bigskip
\bigskip
\bigskip
\bigskip

{ \Large Running head:  Term-based Resonant Model}

\bigskip
\bigskip
\bigskip
\bigskip
\bigskip
\bigskip
\bigskip
\bigskip
\bigskip
\bigskip
\bigskip
\bigskip
\bigskip
\bigskip
\bigskip
\bigskip
\bigskip
\bigskip
\bigskip

CORRESPONDENCE FOR AUTHOR:

\bigskip

All the work for this paper was completed at:

JILA, University of Colorado, 440 UCB, Boulder, CO, 80309-0440, USA

Department of Astrophysical and Planetary Sciences, University of Colorado, 
Boulder, CO, 80309-0391, USA

\bigskip

{\it current address}: Department of Astronomy, University of Florida, 211 Bryant Space Science Center, Gainesville, 

FL, 32611-2055, USA

{\it current email}: veras@astro.ufl.edu

\newpage

\baselineskip=16pt

\begin{abstract}

\Large
\baselineskip=24pt

Investigations of two resonant planets orbiting a star or two resonant satellites 
orbiting a planet often rely on a few resonant and secular terms in order to 
obtain a representative quantitative description of the system's dynamical evolution.
We present a semianalytic model which traces the orbital evolution of any two 
resonant bodies in a first- through fourth-order eccentricity or inclination-based 
resonance dominated by the resonant and secular arguments of the 
user's choosing.  By considering the variation of libration width with different
orbital parameters, we identify regions of phase space which give rise
to different resonant ``depths,'' and propose methods to model libration profiles.  
We apply the model 
to the GJ 876 extrasolar planetary system, quantify the relative importance of the 
relevant resonant and secular contributions, and thereby assess the goodness of the 
common approximation of representing the system by just the presumably dominant terms. 
We highlight the danger in using ``order'' as the metric for accuracy
in the orbital solution by revealing the unnatural libration centers produced
by the second-order, but not first-order, solution, and by demonstrating that the
true orbital solution lies somewhere ``in-between'' the third- and fourth-order 
solutions.  We also present formulas used to
incorporate perturbations from central-body 
oblateness and precession, and a protoplanetary or protosatellite thin disk 
with gaps, into a resonant system.  We quantify the contributions of these
perturbations into the GJ 876 system, and thereby highlight the conditions
which must exist for multi-planet exosystems to be significantly influenced
by such factors.  We find that massive enough 
disks may convert resonant libration into circulation; such disk-induced 
signatures may provide constraints for future studies of exoplanet systems.

\end{abstract}

\bigskip

\noindent{}Keywords: RESONANCES, ORBITS, PLANETARY DYNAMICS, EXTRASOLAR PLANETS, PROTOPLANETARY DISKS

\newpage

\large
\baselineskip=24pt

\section{Introduction}

Resonances have played an increasingly important role in the dynamical 
analysis of planetary and satellite systems.
The ongoing discoveries of extrasolar planets, Solar System satellites, 
Kuiper Belt Objects,
and asteroids have sparked a resurgence of interest in resonant systems.
The variety of resonances observed or thought to exist in nature 
showcases the utility of a versatile model which can help determine
what approximations are sufficient or inadequate for future detailed studies.

The term ``resonance'' typically refers to any system which features a 
commensurability of frequencies.  These frequencies may refer to 
an object's electromagnetic forcing, 
the so-called ``Lorentz'' resonance
\citep{buretal1985,schbur1992,hambur1993,hamilton1994,buretal2004}, 
an object's spin \citep{golpea1966,murdock1978,
celletti1993,biachi2002,flysah2005}, or an object's orbital elements 
\citep{goldreich1965b,peale1976,peale1986,greenberg1977,
malhotra1994,morbidelli2002a}.  Commensurabilities between 
orbital frequencies of any number of bodies may exist.
Most known cases of orbital resonance occur between two objects revolving
around a massive central object; our model evolves such two-body 
``orbit-orbit'' resonances.  Two-body orbit-orbit resonances have been 
subdivided into a variety of designations, but naturally are split into
``mean-motion'' and ``secular'' resonances.  Secular resonances 
occasionally occur inside mean-motion resonances, but not vice-versa.
``Secondary'' resonances, which arise when the libration frequency of 
the primary resonance is commensurate with the circulation frequencies 
of the secular mode of motion, also lie just inside mean-motion resonances.
Some secular resonances include ``eccentricity-type'' resonances, 
which involve commensurabilities among longitudes of
pericenters, ``inclination-type'' resonances, which involve commensurabilities among
longitudes of ascending nodes, and ``mixed'' resonances, which include relations among
both types of angles 
\citep[][p. 359]{murder1999}\footnote{\cite{vakhidov2001} 
classifies three-body resonances among Jupiter, Saturn
and an asteroid as ``mixed.''}.

No orbit-orbit resonances exist among the 8 planets of the Solar System, 
in contrast to the several resonances present in recently discovered 
multiple-planet extrasolar systems.  
Jupiter and Saturn, however, are in a $5$:$2$ near-resonance known 
as ``The Great Inequality'', which has been known since the time of Laplace
and recently studied in \cite{micfer2001} and \cite{frasop2003}.  Relative 
to known exosystem resonances, Solar System resonances involve comparatively 
small eccentricities, are better identified due to their proximity to Earth, 
and have undergone scrutiny for a longer period of time.  
The ongoing discovery of extrasolar planets has led to at least $20$ confirmed 
multiple-planet exosystems\footnote{From the on-line Extrasolar Planets Encyclopedia,
at http://vo.obspm.fr/exoplanetes/encyclo/catalog.php} 
almost half of which (see Figs. 1 and 2 of \citealt{verarm2007})
reside close to a 1st-4th order mean motion resonance.
GJ 876 b and c are two of the most well-studied resonant
exoplanets, and are widely thought to reside in the 
strong $2$:$1$ resonance.

\cite{maretal2001a} reported the presence of two planets in the GJ 876 system
with orbital parameters suggestive of the presence of a mean motion resonance.
The system has since become a catalyst for more careful studies of
the $2$:$1$ resonance, which represents
a crucial dynamical marker for stability and possible prior evolution 
\citep{beaetal2006a}.  \cite{leepea2002} explore the geometry of the 
$2$:$1$ resonance as applied to GJ 876, and 
\cite{feretal2003}, \cite{lee2004}, \cite{psyhad2005} and \cite{beaetal2006a} 
sample the resonant phase 
space further in order to demonstrate the diversity of the resonant configurations.
Orbital fits suggest that the dominant resonant and secular angles
all librate about $0^{\circ}$ with well-defined amplitudes, and allow for both
resonant planets to harbor a modest nonzero mutual inclination 
\citep{lauetal2005}.
\cite{jietal2002} perform $1$ Myr simulations of the GJ 876 system, assuming
different initial relative inclinations.  

\cite{jietal2003b} suggest that 
the planets in GJ 876 are likely to undergo apsidal alignment 
and derive a criterion for determining if a particular secular 
argument in a two-planet system is 
in libration or circulation.  
\cite{sneetal2001} and 
\cite{kleetal2005}
use hydrodynamic simulations to explore the possibility that
the currently observed eccentricities of the GJ 876 planets are the 
result of differential
migration from the nascent protoplanetary disk.  \cite{ward1981} and 
\cite{kleetal2005} give formulas for the apsidally 
induced precession of a planet due to a disk.
Although \cite{rivetal2005} suggest the presence of a third planet in 
the GJ 876 system
orbiting at $\sim 0.02$ AU, the planet's small ($0.23 m_{Jup}$) mass and 
circular orbit are 
unlikely to affect significantly the resonance between the other planets.

A salient feature of orbit-orbit resonances is the domination of just one or a few 
dynamical terms, often dubbed ``resonant terms'' or as ``secular terms''
(depending on the nature of the terms), in the system 
evolution.  Resonant models for particular systems
are often built around the terms chosen.  Our model can help investigators determine
which terms dominate a particular system by allowing the user to include up to 
twenty resonant and/or secular arguments for each simulation.  Sometimes, 
however, additional
effects, such central-body oblateness 
\citep[][p. 264-270]{murder1999} 
or central-body precession \citep{rubincam2000}, can play a crucial role
in the system evolution.  Further, in the formation stages of a planetary or 
satellite system, a disk may be present.  The mass and breadth of the
disk then can strongly influence the resonant dynamics.  Our code thus includes 
the ability to toggle the effects of oblateness, precession and a nascent disk.
Our approach to evolving the resonant bodies involves retaining a 
classical set of orbital elements
while simultaneously tracking each resonant angle and its time derivative.
The method allows for additional effects to be incorporated through the system's
potential.  By treating the dynamical equations in as general
manner as possible, our code is effective for a wide range of initial
orbital parameters and masses, but cannot accurately model crossing orbits, high
eccentricities, nor high inclinations.

In Section \ref{core}, we derive our core model, connect its 
analytical and numerical aspects,
and relate our formulations to action-angle variables and system constants.  
Section \ref{single} begins the resonant analysis by describing
asteroidal motion in the restricted three-body problem, 
and Section \ref{multi}
illustrates how libration width varies with a variety of parameters
for two massive planets.   We then apply the model to a real exoplanetary
system, GJ 876, in Section \ref{gj876sec}, with a term-based
perturbative treatment.  In Sections \ref{oblsec}, \ref{presec} 
and \ref{disksec}, we present both averaged
and unaveraged formulas expressing the gravitational effects of 
central-body oblateness, precession, and a thin disk, and detail their 
contributions to GJ 876, thereby illustrating the necessary conditions for
their consideration in other exosystems.  We conclude in Section \ref{conc}.

\section{The Core Resonant Model}\label{core}

\subsection{Definitions}

The orbital motions of both resonant bodies in an isolated system
are described completely by ``Lagrange's planetary equations'', which
\cite[][p. 273-284]{brocle1961} derive without approximation.
Despite their name, these equations are not restricted to describing planetary
motion,  and are dependent on two ``disturbing functions,'' which
represent potentials that arise from applying Newton's gravitational force laws
to the central and resonant bodies. 
\cite{ellmur2000} summarize
the history of the development of the disturbing function, and help describe
its modern usage.  The disturbing functions are
infinite linear combinations of cosine terms with arguments of the form
\citep{kaula1961,kaula1962},

\beq
\phi(t) = j_{1}  \lambda_1(t) + j_2 \lambda_2(t) + j_{1,\varpi} \varpi_1(t) + j_{2,\varpi} \varpi_2(t) + j_{1,\Omega} \Omega_1(t) + j_{2,\Omega} \Omega_2(t)
,
\label{phi}
\eeq

\noindent{where} $\lambda$ represents mean longitude, $\varpi$ represents
longitude of pericenter, $\Omega$ represents longitude of ascending node,
the ``$j$'' values represent integer constants, and the subscript 
``$1$'' refers to the outer planet
while the subscript ``$2$'' refers to the inner planet.  Secular arguments
are those for which $j_1 = j_2 = 0$, and arguments which are multiples
of one another are distinguished because they have different coefficients
whose magnitudes may vary drastically (by orders of magnitude).
The set of elements which define each resonant argument are 
subject to constraints known as the
d'Alembert relations, which
require that $j_1 + j_2 + j_{1\varpi} + j_{2\varpi} + 
j_{1,\Omega} + j_{2,\Omega} = 0$ 
and $j_{1,\Omega} + j_{2,\Omega}$ is even 
\citep[e.g.][]{greenberg1977,hamilton1994}.

Typically, orbital elements in celestial mechanics are, by default, assumed to be
osculating, i.e., to satisfy the Lagrange gauge (or Lagrange constraint), which
demands that the elements parameterize instantaneous conics tangent to the
perturbed orbit.  Under this constraint, the dependence of the perturbed velocity
upon the elements has the same functional form as the dependence of the 
unperturbed velocity upon the elements.  Derivation of the standard planetary
equations in the forms of Lagrange or Delaunay is based on this constraint.
This derivation, however, also exploits the assumption that the perturbations
depend solely upon positions, not upon velocities.  In the case of 
velocity-dependent disturbances, the planetary equations for osculating elements
assume a more complicated form.  Specifically, they acquire new terms that are
not parts of the disturbing function.  For the reason, osculating elements become
very mathematically inconvenient when considering velocity-dependent perturbations.
Perturbations of this type arise in problems with relativistic corrections or
with atmospheric drag.  They also emerge when we switch from an inertial reference
frame to a precessing one - below we shall encounter exactly this situation.

Fortunately, even under velocity-dependent perturbations, we can restore the
standard form of the planetary equations.  This, however, can be achieved only
by sacrificing the osculation.  As demonstrated by 
\cite{efrgol2003,efrgol2004} and
\cite{efroimsky2005a,efroimsky2005b}, there exists a non-Lagrange gauge
which returns to the planetary equations their customary form.  The advantage
of this approach is that even the velocity-dependent perturbations appear
in these equations simply as parts of the disturbing function.  The consequence 
is that the elements rendered by these equations are no longer osculating.
Such elements model the orbit with a sequence of conics that are not tangent to this
orbit.  Thereby, these elements return the right position of the satellite but not
its correct velocity.  They are called ``contact elements'' (term offered by
Victor Brumberg).  Though these elements were rigorously defined and comprehensively
studied only very recently, they had appeared in the thitherto 
literature whenever
someone incorporated a velocity-dependent perturbation into the disturbing function
and then substituted this function into the standard planetary equations.  By performing
this sequence of operations, one tacitly postulated a certain non-Lagrange gauge,
i.e., accepted a certain ``amount of nonosculation.''  For the first time, 
this situation
occurred in 
\cite{goldreich1965a} and later in \cite{bruetal1970} and
\cite{kinoshita1993}.  Goldreich and Brumberg noticed 
that the elements furnished by these
written equations were nonosculating.

Though the contact elements differ from the osculating ones already in the first order
(over the perturbation caused by the transition to a precessing reference frame), the
secular parts of contact elements differ from those of their osculating counterparts
only in the second order.  This result, proven by \cite{efroimsky2005b}, is valid only in
the case of uniform precession and only for a solitary satellite, in the absence of any
other disturbances (like the gravitational pull of the Sun or of another satellite).
However, numerical simulation has shown that even in realistic situations of variable
precession, the deviations between the secular parts of the contact and the corresponding
osculating elements accumulate very slowly, for a solitary satellite 
\citep{guretal2006}.  
Thus, in practical cases, one may safely assume that for 
a solitary satellite, the secular parts of the contact elements make a very
good approximation to those of the appropriate osculating ones.  
For our model, the velocity correction between both sets of elements
is on the order of the relativistic correction, an effect orders of
magnitude smaller than any we consider.

We define the ``order'' of a
resonant argument as $|j_1 + j_2|$, and denote each
resonant argument by the set 
$\lbrace j_1, j_2, j_{1,\varpi}, j_{2,\varpi}, j_{1,\Omega}, j_{2,\Omega} \rbrace$.
The mean longitude is directly proportional to mean longitude at
epoch, denoted by $\epsilon_k$, such that 
$\lambda_k \equiv \varpi_k + M_k = \pi_k + \Omega_k + M_{k_0} + 
\int_{t_0}^{t} n_k(t') dt' = \epsilon_k + 
\int_{t_0}^{t} \mu_{k}^{(1/2)} a_k(t')^{(-3/2)} dt'$, where $M_k$ denotes
mean anomaly, $\pi_k$ denotes argument of pericenter,
and $n_k$ denotes mean motion, with $k = 1,2$.
Henceforth, the letter ``$k$'' will represent
the dummy variable which denotes the outer and inner resonant bodies.
The mean motion is related to its semimajor axis and mass through
Kepler's third law by $n_{k}^2 a_{k}^3 = \mu_k$, where 
$\mu_k = \itG \po m_{0} + m_k \pc$, with $m_k$ representing the planet's
mass, $m_{0}$ the central body's mass, and $\itG$ the universal 
gravitational constant.  Using the above result in conjunction with the time 
derivative of Equation (\ref{phi}) (all time derivatives
will henceforth be denoted as overdots) yields:

\beq
\dot{\phi}(t) = \jfir \mu_{1}^{\f{1}{2}} a_{1}^{-\f{3}{2}}(t) + \jfir \dot{\epsilon}_1(t) + \jsec \mu_{2}^{\f{1}{2}} a_{2}^{-\f{3}{2}}(t) + \jsec \dot{\epsilon}_2(t) + \jthi \dot{\varpi}_1(t) + \jfou \dot{\varpi}_2(t) + \jfif \dot{\Omega}_1(t) + \jsix \dot{\Omega}_2(t)
.
\label{phidot}
\eeq

\noindent{}Equation (\ref{phidot}) provides a crucial component of our model;
we will later show (Eq. \ref{phibarag2})  that the right-hand-side may be 
expressed in terms of $\phi$, $a_k$, the eccentricity $e_k$ and the
inclination $I_k$ only.



\subsection{Equations of Motions}

The complete set of Lagrange's planetary equations, with the disturbing
functions denoted by $\itR_k$, can be manipulated to read
\citep[][p. 284-286]{brocle1961},

\begin{subequations}\label{lagagmod}
\begin{align}
\f{d a_k}{d t} &= a_{k}^{\f{1}{2}} \Akfir \f{\prt \itR_k}{\prt \lambda_k}
,
\label{lagagmod1} \\
\f{d e_k}{d t} &= -\atwodh \Aksec \Akthi \f{\prt \itR_k}{\prt \lambda_k} - \atwodh \Aksec \f{\prt \itR_k}{\prt \varpi_k}
,
\label{lagagmod2} \\
\f{d I_k}{dt} &= -\atwodh \Akfou \Akfif \f{\prt \itR_k}{\prt \lambda_k} - \atwodh \Akfou \Akfif \f{\prt \itR_k}{\prt \varpi_k}  -\atwodh \Akfou \Aksix  \f{\prt \itR_k}{\prt \Omega_k}
,
\label{lagagmod3} \\
\f{d \epsilon_{k}}{dt} &= -a_{k}^{\f{1}{2}} \Akfir \f{\prt \itR_k}{\prt a_k} + \atwodh \Aksec \Akthi \f{\prt \itR_k}{\prt e_k} + 
\atwodh \Akfou \Akfif \f{\prt \itR_k}{\prt I_k}
,
\label{lagagmod4} \\
\f{d \varpi_{k}}{dt} &= \atwodh \Aksec \f{\prt \itR_k}{\prt e_k} +
\atwodh \Akfou \Akfif \f{\prt \itR_k}{\prt I_k}
,
\label{lagagmod5} \\ 
\f{d \Omega_k}{dt} &= \atwodh \Akfou \Aksix \f{\prt \itR_k}{\prt I_k}
,
\label{lagagmod5b} \\
\lambda_k &= \int n_k dt + \epsilon_{k} 
,
\end{align}
\label{lagagmod6}
\end{subequations}

\noindent{}where,

\begin{equation}
\begin{split}
\Akfir &= 2\mu_{k}^{-\f{1}{2}},  \\
\Aksec &= \mu_{k}^{-\f{1}{2}} e_{k}^{-1} {\po 1 - e_{k}^2 \pc}^{\f{1}{2}},  \\
\Akthi &= 1 - {\po 1 - e_{k}^2 \pc}^{\f{1}{2}}, \\
\Akfou &= \mu_{k}^{-\f{1}{2}} {\po 1 - e_{k}^2 \pc}^{-\f{1}{2}}, \\
\Akfif &= \tan{\po\f{1}{2} I_k \pc}, \\
\Aksix &= \csc{\po I_k \pc}. 
\end{split}
\label{defA}
\end{equation}

A primary feature of our model is the ability to include whichever, 
and as many, terms in the disturbing function that one wishes depending 
on the resonant situation.  Our heliocentric disturbing function, 
$\itR_{k}^{(H)}$, may be 
written as \citep[][p. 329]{murder1999}:

\beq
\itR_{k}^{(H)} = \atwod \sum_{i=1}^{\infty} 
                 \left[ \sum_{p=1}^{U_i} \itC_{k}^{(i,p)} X^{(i,p)} \right]
                 \cos{\phi^{(i)}}
,
\label{defR}
\eeq

\noindent{such} that $i$ labels each argument $\phi$, and $U_i = 1, 3$ or $11$ 
because the expansion is taken to fourth-order.  We represent the 
$\lbrace 0,0,0,0,0,0 \rbrace$ secular argument, for which $U_i = 11$, as
$\aleph$, and any other secular or resonant argument as 
$\varkappa$.  The expansion of the disturbing function is that 
due to \cite{ellmur2000}.

The quantity $\xin$ represents the infinite linear combination of powers of 
eccentricities and semi-inclinations that accompany each term, and the 
quantities $\itC_{k}^{(i,p)}$ are functions of the semimajor axes alone.
The summation in the brackets indicates the possible presence of more
than one coefficient associated with a single resonant or secular argument.

The form of the disturbing function 
in Eq. (\ref{defR}) fails to model 
accurately resonant bodies in crossing orbits due to the singularity at 
orbit intersection, and the resulting series convergence depends on the 
orbital proximity to this crossing point  
\citep[][p. 250]{murder1999}.  
Therefore, this model cannot reproduce the evolution of resonantly locked 
orbit-crossing pairs such as Neptune and Pluto.  Further, the convergence 
domain of the expansion on which Eq. (\ref{defR}) relies precludes
realistic solutions for high eccentricities. 
Although high eccentricity 
expansions of the disturbing function exist 
\citep{roietal1998,beamic2003}, we investigate the utility of 
\cite{ellmur2000} traditional 
expansion about zero eccentricities and inclinations.  The expansion
converges only for $e \lesssim 0.66$; however, 
the Sundman criterion, applied in
Section \ref{gj876sec}, restricts the magnitude of the eccentricities even more.
We assume,

\beq
X^{(i,1)} = \libarg
,
\label{new6}
\eeq

\noindent{where} for $p > 1$,

\beq
X^{(i,p)} = 
\begin{cases}
 \left\lfloor \f{12 - p}{6}\right\rfloor
  e_{1}^{g_1} e_{2}^{g_2} + 
 \left\lfloor \f{p}{7}\right\rfloor
  s_{1}^{g_1} s_{2}^{g_2}
,
& \aleph \\
& \\
e_{p-1}^{2 \bs 
\left| {\rm sgn} 
\po |j_{1,\varpi}^{(i)}| +  |j_{2,\varpi}^{(i)}| \pc
\right|}
s_{p-1}^{2 \bs 
\left| {\rm sgn} 
\po |j_{1,\Omega}^{(i)}| +  |j_{2,\Omega}^{(i)}| \pc
\right|} X^{(i,1)}
,
& \varkappa
\end{cases}
\label{case2}
\eeq

\beq
\begin{split}
g_1 &= 2 
\left\lfloor \f{\po p+3 \pc \bmod 5}{2} \right\rfloor,
\\
g_2 &= 2 
\left\lfloor \f{\po 11-p \pc \bmod 5}{2} \right\rfloor, 
\label{gs}
\end{split}
\eeq

\noindent{where} the ``semi-inclinations'' are
$s_1 \equiv \sin{(I_1/2)}$ and $s_2 \equiv \sin{(I_2/2)}$.
The quantities $\itC_{k}^{(i,p)}$ are functions of the masses and semimajor axes 
alone:

\begin{subequations}\label{defC2}
\begin{align}
\itC_{1}^{(i,p)} &= \itG m_2 \left[ \alpha^{-2} f_{int}^{(i,p)} + f_{d}^{(i,p)} \right]
,
\label{defC21} \\
\itC_{2}^{(i,p)} &= \itG m_1 \left[ \alpha f_{ext}^{(i,p)} + f_{d}^{(i,p)} \right]
,
\label{defC22} 
\end{align}
\label{defC23}
\end{subequations}

\noindent{where} $\alpha = a_2/a_1$, $f_{int}^{(i,p)}$ and $f_{ext}^{(i,p)}$ are constant 
``indirect'' ``internal'' and ``external'' contributions, and $f_{d}^{(i,p)}$ is the 
``direct'' contribution, which
is a function of $j_1$ and $\alpha$.  The indirect contributions are zero for the
majority of orbital resonances.  In many resonant studies, $f_{d}^{(i,p)}$ is 
treated as constant by setting $\alpha$ as a constant computed from the initial 
semimajor axes; \cite{ferrazmello1988} demonstrated the danger in doing so, and
hence we do not make that assumption here.  We use

\beq
f_{d}^{(i,p)} = f_{d}^{(i,p)}(\alpha(t)) = \sum_{l = 1}^{\infty} \kappa_{l}^{(i,p)} 
\alpha(t)^l
\label{fd}
\eeq

\noindent{where} $\kappa_{l}^{(i,p)}$ are constants {\it specific to each term $(i,p)$},
and are what give each resonance its unique character.
The Appendix describes our method for obtaining $\kappa_{l}^{(i,p)}$ values.
Our code incorporates $\kappa_{l}^{(i,p)}$ 
coefficients for the $\itC_{1}^{(i,p)}$ and $\itC_{2}^{(i,p)}$  terms corresponding to 
all unmixed first thru fourth-order resonances, and for (zeroth-order) 
secular terms up to degree 4 in eccentricities and inclinations.
The disturbing function in Eq. (\ref{defR}) assumes heliocentric 
coordinates, such that 
$\itR_{1}^{(H)}/\itR_{2}^{(H)} = m_2/m_1$.  In Jacobi coordinates 
\citep[][p. 589]{brocle1961},

\beq
\f{\itR_{1}^{(J)}}{\itR_{2}^{(J)}} = \f{m_0 m_2 \po m_0 + m_1 + m_2 \pc}{m_1 \po m_0 + m_2\pc^2}
.
\label{jacratio}
\eeq


By using Eq. (\ref{defR}) and $c_k \equiv \cos{(I_k/2)}$, we can take 
partial derivatives of the disturbing functions and insert 
them into Lagrange's planetary equations, which yields:

\begin{subequations}\label{ntotdervn}
\begin{align}
\f{d a_k}{d t} &= a_{k}^{\f{1}{2}} \atwod 
\sum_{i=1}^{\infty}
\left[ 
\sum_{p=1}^{U_i}
\itC_{k}^{(i,p)} \Bkfir
\right]
\sin{\phi^{(i)}}
,
\label{ntotdervn1} \\
\f{d e_k}{d t} &= a_{k}^{-\f{1}{2}} \atwod 
\sum_{i=1}^{\infty}
\left[ 
\sum_{p=1}^{U_i}
\itC_{k}^{(i,p)} \Bksec 
\right]
\sin{\phi^{(i)}}
,
\label{ntotdervn2} \\
\f{d I_k}{dt} &= a_{k}^{-\f{1}{2}} \atwod 
\sum_{i=1}^{\infty} 
\left[
\sum_{p=1}^{U_i}
\itC_{k}^{(i,p)} \Bkthi 
\right]
\sin{\phi^{(i)}}
,
\label{ntotdervn3} \\
\f{d \epsilon_{1}}{dt} &= 
a_{1}^{\f{1}{2}} \atwod 
\sum_{i=1}^{\infty} 
\left[
\sum_{p=1}^{U_i}
\po \f{\prt \itC_{1}^{(i,p)}}{\prt a_1} - a_{1}^{-1} \itC_{1}^{(i,p)} \pc 
\f{B_{1,1}^{(i,p)}}{\jifir} 
\right]
\cos{\phi^{(i)}}
\label{ntotdervn4}
\\
&+
a_{1}^{-\f{1}{2}} \atwod
\sum_{i=1}^{\infty} 
\left[
\sum_{p=1}^{U_i}
\itC_{1}^{(i,p)} B_{1,4}^{(i,p)} 
\right]
\cos{\phi^{(i)}}
,
\\
\f{d \epsilon_{2}}{dt} &= 
a_{2}^{\f{1}{2}} \atwod
\sum_{i=1}^{\infty} 
\left[
\sum_{p=1}^{U_i}
\f{\prt \itC_{2}^{(i,p)}}{\prt a_2} \f{B_{2,1}^{(i,p)}}{\jisec} 
\right]
\cos{\phi^{(i)}}
+
a_{2}^{-\f{1}{2}} \atwod 
\sum_{i=1}^{\infty} 
\left[
\sum_{p=1}^{U_i}
\itC_{2}^{(i,p)} B_{2,4}^{(i,p)} 
\right]
\cos{\phi^{(i)}}
,
\label{ntotdervn5} \\
\f{d \varpi_{k}}{dt} &= a_{k}^{-\f{1}{2}} \atwod 
\sum_{i=1}^{\infty} 
\left[
\sum_{p=1}^{U_i}
\itC_{k}^{(i,p)} \Bkfif 
\right]
\cos{\phi^{(i)}}
,
\label{ntotdervn6} \\
\f{d \Omega_k}{dt} &= a_{k}^{-\f{1}{2}} \atwod 
\sum_{i=1}^{\infty} 
\left[
\sum_{p=1}^{U_i}
\itC_{k}^{(i,p)} \Bksix
\right]
\cos{\phi^{(i)}}
,
\label{ntotdervn7}
\end{align}
\label{ntotdervn8}
\end{subequations}

\noindent{where}

\begin{subequations}\label{Bval}
\begin{align}
\Bkfir &= -\Akfir \jik \xin
,
\label{Bval1} \\
\Bksec &= \Aksec \Akthi \jik \xin + \Aksec \jiklilo \xin
,
\label{Bval2} \\
\Bkthi &= \Akfou \Akfif \jik \xin +  \jiklilo \Akfou \Akfif \xin + 
          \jikbigo \Akfou \Aksix \xin
,
\label{Bval3} \\
\Bkfou &= h_{k,e}^{(i,p)} \Aksec \Akthi e_{k}^{-1} \xin + 
          \f{1}{2} h_{k,I}^{(i,p)} \Akfou \Akfif s_{k}^{-1} c_{k} \xin
,
\label{Bval4} \\
\Bkfif &= h_{k,e}^{(i,p)} \Aksec e_{k}^{-1} \xin + 
          \f{1}{2} h_{k,I}^{(i,p)} \Akfou \Akfif  s_{k}^{-1} c_{k} \xin
,
\label{Bval5} \\
\Bksix &= 
\f{1}{2} h_{k,I}^{(i,p)} \Akfou \Aksix s_{k}^{-1} c_{k} \xin
,
\end{align}
\label{Bval6}
\end{subequations}

with

\beq
h_{k,e}^{(i,p)} = 
\begin{cases}
\absl \jiklilo \absr
,
&\varkappa \bs {\rm and} \bs p = 1 \\
& \\
\left[ \left| j_{k,\varpi}^{(i)}  \right| +    
2 
\po 1 - \left| k - p + 1 \right|   \pc
\left| {\rm sgn} 
\po |j_{1,\varpi}^{(i)}| +  |j_{2,\varpi}^{(i)}| \pc
\right| \right]
,
&\varkappa \bs {\rm and} \bs p > 1 \\
& \\
g_{k}(p)
,
&\aleph \bs {\rm and} \bs p > 1 \\
& \\
0
,
&\aleph \bs {\rm and} \bs p = 1
,
\end{cases}
\eeq

and

\beq
h_{k,I}^{(i,p)} = 
\begin{cases}
\absl \jikbigo \absr
,
&\varkappa \bs {\rm and} \bs p = 1 \\
& \\
\left[ \left| j_{k,\Omega}^{(i)}  \right| +    
2 
\po 1 - \left| k - p + 1 \right|   \pc
\left| {\rm sgn} 
\po |j_{1,\Omega}^{(i)}| +  |j_{2,\Omega}^{(i)}| \pc
\right| \right]
,
&\varkappa \bs {\rm and} \bs p > 1 \\
& \\
g_{k}(p)
,
&\aleph \bs {\rm and} \bs p > 1 \\
& \\
0
,
&\aleph \bs {\rm and} \bs p = 1
.
\end{cases}
\eeq

Both the $A$ and $B$ auxiliary variables are functions of only the
eccentricities and inclinations.  Now combining Eqs. (\ref{phidot}) and 
(2.11d)-(\ref{ntotdervn7}) provides the following
compact resonant equation:  for $u = 1,... Z$, where $Z$ denotes the number 
of cosine arguments retained in the disturbing function,

\beq
\dot{\phi}^{(u)} = \sum_{i=1}^{Z}  \po D^{(i,u)} \cos{\phi^{(i)}} \pc + E^{(u)} 
,
\label{phibarag2}
\eeq

\noindent{where}

\beq
\begin{split}
D^{(i,u)} &= 
\sum_{p=1}^{U_i}
\Bigg[
j_{1}^{(u)} A_{1,1} X^{(i,p)} a_{1}^{-\f{3}{2}} \itC_{1}^{(i,p)} +
\\
&a_{1}^{-1} \sum_{k=1}^2 \Big\lbrace 
\atwodh \itC_{k}^{(i,p)}  
\po j_{k}^{(u)} \Bkfou + j_{k,\varpi}^{(u)} \Bkfif + j_{k,\Omega}^{(u)} \Bksix  \pc 
- j_{k}^{(u)} A_{k,1} X^{(i,p)} a_{k}^{\f{1}{2}} \f{\prt \itC_{k}^{(i,p)}}{\prt a_k}
\Big\rbrace 
\Bigg]
,
\end{split}
\label{Ddef}
\eeq

\noindent{and},

\beq
 E^{(u)} = \sum_{k = 1}^2 j_{k}^{(u)} \mu_{k}^{\f{1}{2}} a_{k}^{-\f{3}{2}}.
\label{Edef}
\eeq

Equation (\ref{phibarag2}) illustrates that for a disturbing function with $Z$
cosine terms, the time derivative of each argument is a linear combination of
the cosines of all $Z$ arguments, with coefficients that are functions
of $m_0, m_1, m_2, a_1, a_2, e_1, e_2, I_1$ and $I_2$ only.  The arguments may be
resonant or secular.  Equations (\ref{ntotdervn}) and (\ref{phibarag2}) represent a 
self-consistent set of $12 + Z$ first order coupled differential equations, which our model
integrates directly.  We use an adaptive-timestep fourth-order Runge-Kutta 
integrator with user-defined accuracy parameters.

\subsection{System Constants}

Although some quantities, such as energy and angular momentum, 
represent constants of
any isolated physical system, they strictly no longer represent constants when
a finite number of terms in the disturbing function are used to model the evolution.
Further, in the restricted problem, energy and angular momentum no longer represent
constants of the motion.
Regardless, the variation of these quantities may be negligible depending on the
system and the terms chosen.  Further, if only one or a few terms are considered,
then {\it additional} constants may arise, as we detail in this section.
These additional constants prove
useful in studies of, for example, two resonant bodies in which one is much less massive
than the other, such as the case with a main belt asteroid and 
Jupiter, Titan and Hyperion,
or a terrestrial extrasolar planet and a giant extrasolar planet.
The system angular momentum, $c$, may be expressed as:

\beq
|c| = \sqrt{
      c_{1}^2 + c_{2}^2 + 2 c_1 c_2 \left[  
      \sin{I_1} \sin{I_2} \cos{\po \Omega_1 - \Omega_2 \pc}
      + \cos{I_1} \cos{I_2}  \right]}
\label{angmomgen}
\eeq

\noindent{} where,

\beq
c_k = m_k \sqrt{\itG \po m_0 + m_k \pc a_k \po 1 - e_{k}^2 \pc}
,
\label{constants4t}
\eeq

\noindent{} and in Jacobi coordinates,

\beq
\begin{split}
c_1 &= \f{\po m_0 + m_2 \pc m_1}{\sqrt{m_0 + m_1 + m_2}} \sqrt{\itG a_1 \po 1 - e_{1}^2 \pc}, \\
c_2 &= \f{m_0 m_2}{\sqrt{m_0 + m_2}} \sqrt{\itG a_2 \po 1 - e_{2}^2 \pc}
.
\end{split}
\label{angmomgenj}
\eeq


A time-independent Hamiltonian provides the energy constant for the system, and allows
one to construct canonical sets of variables which can reveal additional constants
and provide further insights into the considered system.  In terms of orbital
elements, the Heliocentric Hamiltonian may be approximated as:

\beq
\itH_H \approx -\f{\itG m_0 m_1 }{2a_1} - 
       \f{\itG m_0 m_2}{2a_2} -
       \f{m_1}{a_1} 
       \sum_{i=1}^{\infty} \sum_{p=1}^{U_i} 
         \left[ \itC_{1}^{(i,p)} \xin \right]
        \cos{\phi^{(i)}}
.
\label{hh}
\eeq

One may attempt to avoid the
approximate nature of the Heliocentric Hamiltonian by expressing the 
Hamiltonian in Jacobi coordinates ($\itH_J$).  The Hamiltonian 
for a three-body system in Jacobi coordinates, a form of which has been used by 
several authors 
\citep{harrington1968,harrington1969,sidlichovsky1983, 
konetal2000,foretal2000,leepea2003} and explained from first principles by others
\citep{malder1990,ferosa1994} may be expressed as:

\beq
\itH_{J} = -\f{\itG m_0 m_2}{2a_2} - \f{\itG \po m_0 + m_2\pc m_1}{2a_1} - \itR
,
\label{HamJ1}
\eeq
\noindent{where}
\beq
\itR = \f{\itG m_0 m_1 m_2}{r_1} \sum_{l=2}^{\infty} \po \f{r_2}{r_1} \pc^l  
\mathfrak{M}_l P_l \po \cos{\psi} \pc
,
\label{HamJ2}
\eeq
\noindent{with}
\beq
\mathfrak{M}_l = \f{{m_0}^{l-1} - {\po -m_2 \pc}^{l-1}}{\po m_0 + m_2 \pc^l}
< 1
,
\label{HamJ3}
\eeq
\noindent{where} $\psi$ is the angle subtending $\mathbf{r_1}$ and $\mathbf{r_2}$.  
Note that no indirect terms appear in Eq. (\ref{HamJ1}), thereby eliminating the 
need for ``internal'' and ``external'' nomenclature. We seek to express 
$\itH_{J}$ in terms of $m_k, a_k, e_k, I_k$ and $\phi^{(i)}$ only. 

As partially demonstrated 
by \cite{ling1991}, the use of Jacobi coordinates does not alter the 
derivation of the disturbing function given by
Eq. (6.36) of \citet[][p. 232]{murder1999}, as the coordinate system 
transformation does not alter the angle between the radius vectors nor the 
expansions performed on the orbital elements of the individual orbits.  
Therefore, Eq. (\ref{HamJ2}) may be expressed as 
Eq. (6.36) of \citet[][p. 232]{murder1999}, with $\alpha$ 
replaced by $\alpha_M$, where

\beq
\alpha_{M}^l \equiv \alpha^l \mathfrak{M}_l, \bbs\bbs\bbs l \ge 2 
.
\label{new3a}
\eeq

\noindent{All} other variables in that equation remain unaffected.  
The Jacobi Hamiltonian thus reads:

\beq
\itH_{J} = -\f{\itG m_0 m_2}{2a_2} - \f{\itG \po m_0 + m_2\pc m_1}{2a_1} 
- \f{\itG m_0 m_1 m_2}{a_1} 
\sum_{i=1}^{N}  
\sum_{p=1}^{U_i}
\left[
\po
\sum_{l=2}^{\infty}  
\kappa_{l}^{(i,p)} \alpha_{M}^l 
\pc 
X^{(i,p)} 
\right]
\cos{\phi^{(i)}} 
.
\label{hj}
\eeq

We can establish a set of canonical action angle variables from the Hamiltonian; we
choose the following set:

\begin{align}
\lambda_k &= M_k + \pi_k + \Omega_k      & \Lambda_k &= \itM_k \sqrt{a_k} \\
\gamma_k &= -\pi_k - \Omega_k            & \Gamma_k &= \Lambda_k \po 1 - \sqrt{1 - e_{k}^2}  \pc \notag \\
z_k &= -\Omega_k                         & Z_k &= \Lambda_k \sqrt{(1 - e_{k}^2)} \po 1 - \cos{I_k} \pc, \notag
\end{align}

\noindent{where} $\itM_k$ is a function of the masses.  In the literature, when 
$\itM_k = \sqrt{\itG} {m_{0} m_k}/{\sqrt{m_{0} + m_k}}$, the action-angle 
variables have been classified
as ``Poincar\'{e} variables'' 
\citep[][p. 60]{murder1999}
``mass-weighed Poincar\'{e} elliptic variables'' \citep{micfer2001}, 
and ``modified Delaunay variables'' \citep[][p. 35]{morbidelli2002a}.
In some cases, $\itM_k = {\sqrt{\itG \po m_{0} + m_k \pc}}$ 
\citep[][p. 33]{peale1976,morbidelli2002a},
$\itM_k = {\sqrt{\itG m_{0}}}$ 
\citep{morbidelli2001,euimar2003},
or $\itM_k = m_k \sqrt{\itG m_0}$ \citep{varetal1999}.
The form of $\itM_k$ is chosen based on the Hamiltonian
of the system.  
When the Hamiltonian is written in Jacobi coordinates, 
$\itM_2 = \sqrt{\itG} {m_{0} m_2}/{\sqrt{m_{0} + m_2}}$ while
$\itM_1 = \sqrt{\itG} \po m_0 + m_2 \pc m_1/\sqrt{m_0 + m_1 + m_2}$ 
\citep{harrington1968,sidlichovsky1983}. 
With $\pi_k + \Omega_k = \varpi_k$, the angles correspond to those seen in the 
resonant arguments.  A canonical transformation to the variables 
$(\theta_l, \Theta_l)$, $l = 1...6$, can now be applied to the Hamiltonian such that:

\begin{align} \label{lilthetas}
\theta_1 &=  
            \f{\left[ \jifir \lambda_1 + \jisec \lambda_2 - j_{3-k_{\varpi},\varpi}^{(i)} \gamma_1 + 
             \po \jifir + \jisec +  j_{3-k_{\varpi},\varpi}^{(i)} \pc \gamma_2 \right]}
            {\jifir+\jisec}, \\
\theta_2 &= 
            \f{\left[ \jifir \lambda_1 + \jisec \lambda_2 + \po  \jifir + \jisec \pc \gamma_{k_\varpi} \right]}
            {\jifir+\jisec}, \\
\theta_3 &= 
            \f{\left[ \jifir \lambda_1 + \jisec \lambda_2 - j_{3-k_{\Omega},\Omega}^{(i)} z_1 + 
             \po \jifir + \jisec +  j_{3-k_{\Omega},\Omega}^{(i)} \pc z_2 \right]}
            {\jifir+\jisec}, \\
\theta_4 &= 
            \f{\left[ \jifir \lambda_1 + \jisec \lambda_2 + \po \jifir + \jisec \pc z_{k_\Omega} \right]}
            {\jifir+\jisec}, \\
\theta_5 &= \f{\lambda_1}{\jifir + \jisec}, 
                            \\
\theta_6 &= \f{\lambda_2}{\jifir + \jisec},
\end{align}

\begin{align} \label{bigthetas}
\Theta_1 &= -\f{ \po \jifir + \jisec \pc \Gamma_{3-k_{\varpi}}}{j_{3-k_{\varpi},\varpi}^{(i)}}, 
             \\
\Theta_2 &= \f{\jifir + \jisec + j_{3-k_{\varpi},\varpi}^{(i)}}{j_{3-k_{\varpi},\varpi}^{(i)}}
            \Gamma_{3-k_{\varpi}} + \Gamma_{k_{\varpi}},
             \\
\Theta_3 &= -\f{ \po \jifir + \jisec \pc  Z_{3-k_{\Omega}}}{ j_{3-k_{\Omega},\Omega}^{(i)}}, 
             \\
\Theta_4 &= \f{\jifir + \jisec + j_{3-k_{\Omega},\Omega}^{(i)}}{j_{3-k_{\Omega},\Omega}^{(i)}}
            Z_{3-k_{\Omega}} + Z_{k_{\Omega}},
             \\
\Theta_5 &= \po \jifir + \jisec \pc \Lambda_1 - \jifir 
            \po \Gamma_{k_{\varpi}} + \Gamma_{3-k_{\varpi}} + Z_{k_{\Omega}} + Z_{3-k_{\Omega}} \pc,
             \\
\Theta_6 &= \po \jifir + \jisec \pc \Lambda_2 - \jisec 
            \po \Gamma_{k_{\varpi}} + \Gamma_{3-k_{\varpi}} + Z_{k_{\Omega}} + Z_{3-k_{\Omega}} \pc,
\end{align}

\noindent{with}

\begin{equation}
k_{\varpi} = \begin{cases}
1 , & j_{1,\varpi}^{(i)} = 0 \\
2 , & \bs {\rm otherwise} \bs
\end{cases}
,
\bbs\bbs\bbs\bbs\bbs
k_{\Omega} = \begin{cases}
1 , & j_{1,\Omega}^{(i)} = 0 \\ 
2 , & \bs {\rm otherwise} \bs
\end{cases}
.
\label{kvarom}
\end{equation}

\noindent{}For mixed resonances, $k_{\varpi}$ and $k_{\Omega}$ 
may be set to $2$ when $j_{1,\varpi}^{(i)} = 0$ and $j_{1,\Omega}^{(i)} = 0$;
the choice is arbitrary.

I chose this transformation both for application to 
each Hamiltonian and in order to take into account 
several types of resonances, including unmixed and 
mixed eccentricity and inclination resonances to 
arbitrary order. The transformation is not well suited for the rare coupled 
eccentricity and inclination resonances, which, 
according to the d'Alembert 
rules \citep[][p. 35-36]{morbidelli2002a}, 
must be at least of order 3.  For disturbing functions 
with a single resonant argument, that argument should equal a 
multiple of $\theta_1$ or $\theta_3$.  For disturbing 
functions with several resonant arguments, each argument 
may be equated with an appropriate linear combination of the 
$\theta_l$ values\footnote{See, for example, Michtchenko and 
Ferraz-Mello (2001), who include four resonant arguments 
in their Hamiltonian.}. The number of possible resonant 
arguments associated with given values of $\jifir$ and 
$\jisec$ is $\po \jifir + \jisec + 1 \pc$.  Typically, 
when multiple resonant arguments are present in a disturbing
function, they all have the same values of $\jifir$ 
and $\jisec$. This situation allows one to explore 
the effects of ``resonance splitting'', when 
$j_{1}^{(i)}$ and  $j_{2}^{(i)}$ remain fixed
as $i$ changes but the other $j$ values change as $i$ does.
If $\jifir$ and $\jisec$ are fixed for several $i$, then the 
canonical transformation chosen above allows for 
all such arguments to be represented as a linear 
combination of only $\theta_1$ and $\theta_2$ 
or only $\theta_3$ and $\theta_4$. 

This transformation allows us to obtain constants of the 
motion, as all quantities in the Hamiltonian other than the 
resonant arguments may be expressed in terms of the actions only.  
These constants may be determined immediately from the initial 
conditions, and then used in the subsequent analysis.  
When the disturbing function contains a single resonant argument, 
the Hamiltonian does as well, and several constants of 
the motion exist regardless of the form of $\itM_k$.  For any 
eccentricity-type of resonance, two of 
these constants, $N_1$ and $N_2$, may be expressed in terms 
of $a_k(t)$ and $e_k(t)$ only: 

\beq
\begin{split}
N_k &\equiv 
    \sqrt{a_{k_0}}
    \left[
           j_{k}^{(i)} \po \sqrt{1 - e_{k_0}^2} - 1 \pc - j_{k,\varpi}^{(i)}
    \right]
      \\ 
      &=
    \sqrt{a_k}
    \left[
           j_{k}^{(i)} \po \sqrt{1 - e_{k}^2} - 1 \pc - j_{k,\varpi}^{(i)}
    \right]
,
\end{split}
\label{constants12}
\eeq

\noindent{where} $a_{k_0}$ and $e_{k_0}$ represent initial, known values.
For any inclination type of resonance, the two constants may be 
expressed in terms of $a_k(t)$ and $s_k(t)$ only:  for $k = 1,2$,

\beq
\begin{split}
N_{k} &\equiv 
    \po \f{j_{k,\Omega}^{(i)}}{2 j_{k}^{(i)}} \pc   
    \sqrt{a_{k_0}}
    +
    \sqrt{a_{k_0}} s_{k_0}^2 
    -
    s_{k_0}^2
    \left[
          \sqrt{a_{k_0}}  \po 1 - \sqrt{1 - e_{k_0}^2} \pc 
    \right]
   \\ 
  &= \po \f{j_{k,\Omega}^{(i)}}{2 j_{k}^{(i)}} \pc   
    \sqrt{a_k}
    +
    \sqrt{a_k} s_{k}^2 
    -
    s_{k}^2
    \left[
          \sqrt{a_{k_0}}  \po 1 - \sqrt{1 - e_{k_0}^2} \pc 
    \right]
,
\end{split}
\label{constants12alt}
\eeq

\noindent{where} $s_{k_0}$ similarly represents an initial known value.
The constants in Eqs. (\ref{constants12}) and 
(\ref{constants12alt}) may be derived from 
{\it either} a Hamiltonian or a non-Hamiltonian
approach.  For the former, all $\Theta_l$ except 
either $\Theta_1$ or $\Theta_3$ must be constants: 
for eccentricity resonances, both $Z_k$ actions 
are constants, while for inclination resonances, 
both $\Gamma_k$ actions are constants.  The resulting 5 constants 
may be manipulated to yield 
Eqs. (\ref{constants12})-(\ref{constants12alt}).
Because $\Gamma_k = \Gamma_k(a_k,e_k)$ whereas 
$Z_k = Z_k(a_k,e_k,s_k)$, no inclination terms, constant 
or otherwise, appear in Eq. (\ref{constants12alt}),
whereas constant eccentricity terms do appear 
in Eq. (\ref{constants12alt}).  In the non-Hamiltonian 
derivation, dividing Eq. (\ref{ntotdervn1}) by 
Eq. (\ref{ntotdervn2}), separating variables,
and integrating will yield Eq. (\ref{constants12alt}).  
One follows the same procedure with
Eq. (\ref{ntotdervn1}) and Eq. (\ref{ntotdervn3}) in order 
to yield Eq. (\ref{constants12alt}), with one difference: 
the eccentricity in Eq. (\ref{ntotdervn3}) must first be 
written in terms of $a_k$ using Eq. (\ref{constants12}). 


The quantities $N_1$ and $N_2$ are constants in both the 
heliocentric and Jacobi coordinate
systems, and are independent of any masses.
Each provides a relation among the orbital elements of a 
single body, but does not correlate the parameters of both resonant 
bodies to one another.  
A constant which achieves such a correlation would differ, albeit slightly, 
depending on the coordinate system used, and the form 
of $\itM_k$.  If we take 
$\itM_k = \sqrt{\itG} m_0 m_k/\sqrt{m_0 + m_k}$ 
\citep[][p. 35]{morbidelli2002a}, then 
using only the constants $\Theta_5$ and $\Theta_6$ immediately gives the 
following relation:

\beq
\begin{split}
N_{3} &\equiv 
    \sqrt{a_{2_0}} \f{\jifir m_2}{\jisec m_1}
    -
    \sqrt{a_{1_0}}
      \\ 
      &=
    \sqrt{a_2} \f{\jifir m_2 }{\jisec m_1}
    -
    \sqrt{a_1}
.
\end{split}
\label{constants3h}
\eeq

Whereas for the Jacobi Hamiltonian, 
\beq
\begin{split}
N_{3} &\equiv 
    \sqrt{a_{2_0}} \po \f{\jifir m_2}{\jisec m_1} \pc 
    \f{m_0 \sqrt{\po m_0 + m_1 + m_2 \pc}}{\po m_0 + m_2 \pc^{\f{3}{2}}}
    -
    \sqrt{a_{1_0}}
      \\ 
      &=
    \sqrt{a_2} \po \f{\jifir m_2}{\jisec m_1} \pc 
    \f{m_0 \sqrt{\po m_0 + m_1 + m_2 \pc}}{\po m_0 + m_2 \pc^{\f{3}{2}}}
    -
    \sqrt{a_1}
.
\end{split}
\label{constants3j}
\eeq

\noindent{}Note that in the limit $m_0 \gg m_1, m_2$, both
expressions for $N_3$ tend toward equivalence, as expected.  
Also consider that for
$k = 1,2$, in the limit $m_k \to 0$, $a_{3-k}$ is a constant.
Physically, this result is sensible, as a negligible mass
should not perturb a larger mass off of its original orbit.  
In the case of multiple resonant arguments but no secular arguments,
the Hamiltonian can
be expressed as an appropriate linear combination of 
either $\Theta_1$ and $\Theta_2$, or $\Theta_3$ and $\Theta_4$.  Hence, the 
other four actions are constants and can be combined to reproduce
Eq. (\ref{constants3h}).  In this regime, Eqs.
(\ref{constants12}) and (\ref{constants12alt}) no longer represent constants
of the motion.  

\section{Single-term Systems}\label{single}

In special cases, a single resonant term may demonstrate
a system's important dynamical attributes.  This situation
occurs when one resonant object is much larger than the
other, and the larger object is on a circular orbit.  In 
this case, the more massive resonant object remains unperturbed
from its orbit, and the only nonzero resonant argument contains
only one nonzero value among the following coefficients:
$j_{1,\varpi}^{(1)}$,  
$j_{2,\varpi}^{(1)}$,  
$j_{1,\Omega}^{(1)}$,   
$j_{2,\Omega}^{(1)}$.

As a check on our model and a demonstration of its output, 
we reproduce some results from 
\citeauthor{winmur1997a} (\citeyear{winmur1997a})'s
study of Jovian asteroid motion.  Under the guises of the 
planar, circular restricted three-body problem, their
study integrates Hill's equations of motion for a Jovian-mass 
planet at $1$ AU and a massless asteroid at a range of 
values around the $2$:$1$ commensurability.  The mass 
ratio of the Jovian planet and the central object is taken 
to be $10^{-3}$, and the motion of an asteroid is computed 
for various $a_2$ and $e_2$ values chosen in order to 
demonstrate orbital behavior near the $2$:$1$ commensurability.  
The authors fixed the origin at the system's barycenter, 
and integrated the equations of motion in a uniformly 
rotating frame.  They also applied the common approximation 
of setting $f_d$, and hence, $\itC_{k}^{(1)}$, to constant 
values.  Our code does not rely on this approximation, which
has an observable effect, even in the restricted case, on 
the resulting orbital evolution
profiles.

Figure \ref{astecc} \marginpar{FIG. 1} illustrates an asteroid's 
motion with initial $a_2 = 0.602$ AU, $e_2 = 0.004$,
$\lambda_2 = 0^{\circ}$ and $\varpi_2 = 0^{\circ}$,
with $\phi_{\text{lib}} \equiv 
\lbrace 2, -1, 0, -1, 0, 0  \rbrace$. 
Direct comparison of our plots with
\citeauthor{winmur1997a} (\citeyear{winmur1997a})'s
Figure 14 reveal nearly equivalent profiles 
for the evolution of orbital 
elements.  Differences in $\varpi_2$ evolution 
and the presence of small $a_2$ modulation may be 
attributed to the different coordinate systems in 
which the asteroids evolved.  Varying the 
asteroid's semimajor axis by an amount comparable to 
the barycentric correction indeed induces drastically 
different behavior.  As to be expected in resonant systems, 
evolutionary behavior may be highly sensitive to the 
initial semimajor axes values.



As evidenced both by observations of the Solar System and by
the d'Alembert relations, one is less likely to come across
a dynamical system in a purely inclination-based resonance
where just one resonant angle is librating.  Such a resonance
must be of at least second-order, and eccentricity cannot
play a role in the resonant evolution.  
Figure \ref{astinc} \marginpar{FIG. 2} 
illustrates an asteroid's motion with initial 
$a_2 = 0.630$ AU, $I_2 = 10^{\circ}$,
$\lambda_2 = 0^{\circ}$ and $\Omega_2 = 60^{\circ}$,
with $\phi_{\text{lib}} \equiv \lbrace 4, -2, 0, 0, 0, -2  \rbrace$
evolved over the same time as the asteroid in Fig. \ref{astecc}.
Because this resonance is second-order, the so-called
``strength'' of the resonance scales as $\sim I^2$,
as opposed to $\sim e$ for a first-order eccentricity
resonance.


Figure \ref{N2con} \marginpar{FIG. 3} plots the relative error in 
the $N_2$ constants 
(Eq. \ref{constants12}) throughout the runs of the simulations 
in Figs. \ref{astecc} and \ref{astinc} for different user-inputted
integration accuracy parameters of our code.  For the eccentric
asteroid in Fig. \ref{astecc}, the solid line results from an
accuracy parameter three orders of magnitude smaller than that
from the dashed line.  Similarly, for the inclined asteroid of Fig.
\ref{astinc}, the dot-dashed and dot-dot-dot-dashed lines in 
Fig. \ref{N2con} result from accuracy parameters which 
differ by three orders of magnitude.  Figure \ref{N2con}
corroborates the validity of the expression for $N_2$,
and exhibits variation according only to computer precision.
The relative error 
in the $N_1$ constant, heliocentric 
Hamiltonian (Eq. \ref{hh}), Jacobi Hamiltonian 
(Eq. \ref{hj}), heliocentric $N_3$ constant (Eq. \ref{constants3h}),
Jacobi $N_3$ constant (Eq. \ref{constants3j}),
and angular momentum constants (Eqs. \ref{angmomgen}-\ref{angmomgenj})  
are all either indistinguishable from zero or are not
applicable to this specific problem.

The sole resonant argument in Figs. \ref{astecc} 
and \ref{astinc} librates about $0^{\circ}$ and $180^{\circ}$, respectively. 
One may attempt to model the analytical form of the libration of 
resonant angles by considering Eq. (\ref{phibarag2}).  Because
$D^{(i,u)}$ and $E^{(u)}$ vary with time, and because of the summation,
the equation is not easily integrated, and the evolution of 
the resonant arguments are sometimes by no means simple 
sinusoids (see, e.g. Fig. 17a of \citealt{winmur1997a}).  However, 
I find that, in some cases, the form of libration 
profiles can be Fourier 
decomposed into the dominant frequencies of the system
such that

\beq
\phi(t) = K_0 +\sum_{i=1}^{3} 
              \left[ K_{3i-2} \cos{ \po K_{3i} t \pc}
                   + K_{3i-1} \sin{ \po K_{3i} t \pc} 
              \right]
,
\label{foufit}
\eeq

\noindent{where} the summation can be taken to infinity.

For example, the dominant frequency of the asteroid in 
Figs. \ref{astecc} and \ref{astinc} correspond to periods of 
$7.1$ yr and $78$ yr, as can be evidenced by considering
the $K_3$ values from Table \ref{astfit}\marginpar{TABLE 1}.  
This table
provides a list of the coefficients, 
and their standard deviations, obtained
by fitting Eq. (\ref{foufit}) to the libration profiles
from Figs. \ref{astecc} and \ref{astinc}.  I 
performed the fits by using the
Levenberg-Marquardt algorithm \citep{preetal1992}.
We note also that the tiny standard deviations
associated with the $K_3$ values indicate excellent agreement
with the evolution from our code.
As more Fourier terms are included, the fit 
becomes better, although the number of terms needed
in some cases may be prohibitive.  The fit to the
libration profile for the inclined asteroid
is significantly better than that for the
eccentric asteroid when comparing the
ranges ($2.63^{\circ}$ vs. $26.02^{\circ}$)
and standard deviations 
($0.14^{\circ}$ vs. $8.16^{\circ}$)
of the residuals.  The reason for the
discrepancy has to do with the
different shapes of the libration
profiles.


The shape of some libration profiles suggest a
better equation to fit with fewer terms.  Such profiles
often resemble those seen in radar Doppler velocity curves
for exoplanet searches.  Further, the 
computation of the evolution of the true anomaly
from the eccentric anomaly 
(see Eq. \ref{fandE})
suggests that one can fit a libration profile
to the following equation:

\beq
\phi(t) = k_0 + k_1 \cos{\left[  k_2 + k_3 \arctan{\po k_4 
\tan{\po k_5 t \pc} \pc}  \right]}  
.
\label{dopfit}
\eeq

Table \ref{astfit2} \marginpar{TABLE 2} 
provides values 
for this equation's fit
to the libration profiles in 
Figs. \ref{astecc} and \ref{astinc}.
The fit to the libration profile for the
eccentric asteroid is a marked 
(one order of magnitude) improvement
over the Fourier fit.  Figure \ref{fitdevecc} \marginpar{FIG. 4}
displays the residuals of the new fit, and the 
residuals have a range of $3.87^{\circ}$ and a 
standard deviation of $0.77^{\circ}$.

\section{Multiple-term Systems}\label{multi}

A disturbing function composed of just one resonant argument is often too
simple a model for a realistic approximation to the evolution
of bodies in a known resonance.  Including the relevant secular terms
may improve the approximation significantly, even though with added terms
the constants of motion derived in 
Eqs. (\ref{constants12})-(\ref{constants3j}) strictly no longer exist.
Orbit-orbit resonances typically involve disturbing functions
with more than one resonant argument.  In this case, 
when no secular arguments are considered, the Hamiltonian can
be expressed as an appropriate linear combination of 
either $\Theta_1$ and $\Theta_2$, or $\Theta_3$ and $\Theta_4$.  Hence, the 
other four actions are constants and can be combined to reproduce
Eqs. (\ref{constants3h}) or (\ref{constants3j}).  In this regime, Eqs.
(\ref{constants12}) and (\ref{constants12alt}) no longer represent constants
of the motion. In the more general case, where multiple resonant and
secular arguments dominate the motion, the Hamiltonian and
total angular momentum of the system are the only constants
guaranteed to exist.

We expand our study of Jovian asteroid dynamics by including 
additional resonant and secular arguments, thus 
``perturbing'' our single-term models.  For the eccentric
asteroid of Fig. \ref{astecc} in the first-order
$2$:$1$ resonance, we now include, in Fig. \ref{difastecc}, \marginpar{FIG. 5}
all terms up to second-order associated with the  
$\lbrace 2,-1,0,-1,0,0 \rbrace$,
$\lbrace 4,-2,0,-2,0,0 \rbrace$ and 
$\aleph$ arguments (dashed line),
and all terms up to fourth-order associated
with the   
$\lbrace 2,-1,0,-1,0,0 \rbrace$,
$\lbrace 4,-2,0,-2,0,0 \rbrace$,
$\lbrace 6,-3,0,-3,0,0 \rbrace$,
$\lbrace 8,-4,0,-4,0,0 \rbrace$,
and $\aleph$ arguments
(dot-dashed line).  Similarly, for
the inclined
asteroid of Fig. \ref{astinc} in the second-order
$4$:$2$ resonance, we now include, in Fig. \ref{difastinc}, \marginpar{FIG. 6}
all terms up to second-order associated with the  
$\lbrace 4,-2,0,0,0,-2 \rbrace$
and $\aleph$ arguments (dashed line),
and all terms up to fourth-order associated
with the   
$\lbrace 4,-2,0,0,0,-2 \rbrace$,
$\lbrace 8,-4,0,0,0,-4 \rbrace$,
and $\lbrace 0,0,0,0,0,0 \rbrace$ arguments
(dot-dashed line).  The d'Alembert relations
preclude the existence of the 
$\lbrace 6,-3,0,0,0,-3 \rbrace$ argument.

These figures demonstrate that the
eccentricity and inclination profiles
change {\it significantly} depending
on the order of the terms included, 
{\it even for the circular restricted
three-body problem}.  \cite{veras2007b} 
demonstrates the {\it additional} consequences 
of eliminating the assumption that Jupiter 
remains on a circular orbit.
Doing so places doubt on the validity of approximating
the system by a single or a few terms.

Multiple-planet systems said to be in resonance often feature
one or more librating angles.  The smaller the amplitude of
these librating angles, the ``deeper'' into resonance the
system is purported to be.  Our model can make quantitative
statements about how the depth of resonance varies with 
orbital parameters, and how these variations differ
across different types of resonances.  Here we sample
phase space for regimes which could be applicable
to two-planet extrasolar planetary systems.  As a 
three-body problem, these systems are not likely
to be restricted in many ways.  However, one such restriction 
which we assume is that planets in these systems are
coplanar.

Given the extensive investigations of the $2$:$1$
resonance, and the investigation performed in Section \ref{gj876sec}
of this work, here we sample phase space for 
a couple other relevant resonances: the first-order $3$:$2$,
and third-order $4$:$1$ resonances.
For each, we have discovered configurations in the
relatively small region of phase space which allows
for libration of multiple arguments.  The inner planet
has a mass of $0.3 M_{Jup}$ ($\approx$ Mass of Saturn) 
and is set at $3$ AU away from a Solar-mass star.
The outer planet in the $3$:$2$ resonant system
has the same Saturn-like mass, while the outer
planet in the $4$:$1$ system is given a mass of 
$0.03 M_{Jup}$.
Table \ref{intparm} \marginpar{TABLE 3} displays the 
``nominal'' orbital configurations
from which variations result in 
Figs. \ref{libamp32} - \ref{libampcur41}.
\marginpar{FIG. 7}\marginpar{FIG. 8}\marginpar{FIG. 9}\marginpar{FIG. 10}The 
vastness of phase space allow us
to sample just a small slice, presented below, and 
force us to defer further
analysis to future studies.

These figures plot ``libration amplitude,'' defined
as half of the {\it range} of a resonant angle
taken over a few periods.
We must state such a definition because
the actual evolution of these angles is sometimes
heavily modulated and departs from a sinusoid, as in
the dot-dashed profile from Fig. \ref{difastecc}.
Angles are said to circulate if they
span $360^{\circ}$, and do so for the vast majority
of phase space.
Each large plot symbol represents the outcome
of a different simulation, and the curves
which are fit through these symbols hence 
represent just an approximation to the shapes
of these profiles.  Figure \ref{libamp32}
plots the outcomes for the $3$:$2$ resonance,
while Figs. \ref{libamp41}-\ref{libampcur41}
plot the outcomes for the $4$:$1$ resonance.
For the $3$:$2$ resonance, asterisks
and diamonds denote, respectively, the 
libration amplitude of
the $\lbrace 3,-2,-1,0,0,0 \rbrace$ and
$\lbrace 3,-2,0,-1,0,0 \rbrace$ arguments.
For the $4$:$1$ resonance, crosses,
squares, triangles, diamonds and asterisks,
denote, respectively, the libration 
amplitude of the 
$\lbrace 0,0,1,-1,0,0 \rbrace$,
$\lbrace 4,-1,0,-3,0,0 \rbrace$,
$\lbrace 4,-1,-1,-2,0,0 \rbrace$
$\lbrace 4,-1,-2,-1,0,0 \rbrace$, and
$\lbrace 4,-1,-3,0,0,0 \rbrace$ arguments.

Dashed lines in Figs. \ref{libamp32}-\ref{libamp41}
represent the ``nominal''
resonant semimajor axes.  For both 
resonances, this value proves to
be an excellent predictor for maximum depth
for nearly all resonance angles.
Note, however, the order-of-magnitude difference
in $\alpha$ values plotted in the two figures
due to the difference in order of the resonance.
The effective ``libration centers,'' which represent
the mean value of libration, hover around $180^{\circ}$
and $0^{\circ}$, respectively, for the asterisks
and diamonds in Fig. \ref{libamp32}.  For Figs. 
\ref{libamp41}-\ref{libampcur41}, the libration
centers are $180^{\circ}$ for the asterisks,
crosses and triangles, and $0^{\circ}$ for
the diamonds and squares.  We don't present
plots for the variation of $\lambda$ and
$\varpi$ for the $3$:$2$ resonance because
these variations produce differences in 
libration widths of less than a degree.

One can glean other results from the figures.
A comparison the vertical axes reveal that 
in no case does a $3$:$2$ resonant angle 
librate with an amplitude of less than
$100^{\circ}$, in stark contrast to the
$4$:$1$ resonance.  This phenomenon
most likely results from the difference
in mass of the outer planet in the two
systems; the more massive both resonant
objects are, the shallower the resonance.
Also, Figs. \ref{libamp41}-\ref{libampcur41}
suggest that multiple resonant arguments
librate simultaneously for most of the
phase space sampled, although occasionally
some of the arguments circulate.  Further,
the curves drawn through the asterisks
indicate that the libration width profiles
can very drastically for different librating
angles in the same system.  Figure \ref{libampcur41}
demonstrates that in order to produce similar
libration amplitude variations, $\varpi_2$
had to be varied by a factor of three greater
than $\varpi_1$ did.



\section{Dominant GJ 876 Terms}  \label{gj876sec}

We now consider the GJ 876 system with the outer 
(GJ 876 b) and inner (GJ 876 c)
giant planets, which are thought to reside in a $2$:$1$ resonance.
Ever improving orbital fits for initial system conditions abound; 
different sets may be found in  
\cite{fisetal2003}, \cite{lauetal2005}, \cite{rivetal2005} and
the on-line Extrasolar Planets Encyclopedia\footnote{
at http://vo.obspm.fr/exoplanetes/encyclo/catalog.php}.  
We chose
$m_0 = 0.32 m_{\odot}$, $m_1 = 1.929 m_{Jup}$, 
$m_2 = 0.617 m_{Jup}$, $a_1 = 0.20783$ AU, $a_2 = 0.13031$ AU, 
$e_1 = 0.0251$, $e_2 = 0.232$, $\lambda_1 = 351.1^{\circ}$,
$\lambda_2 = 147.1^{\circ}$, $\varpi_1 = 176.8^{\circ}$, and
$\varpi_2 = 198.3^{\circ}$ for the numerical integration.  
The small circularized non-resonant planet GJ 876 d,
at $a \approx 0.02$ AU, does not significantly affect the resonant evolution of
the other two planets \footnote{We have performed a full N-body 
simulation with
HNBody (Rauch and Hamilton 2007, in preparation) which demonstrates 
that the presence of 
planet GJ 876 d indeed has a negligible effect.}.  We evolve the resonant planets 
on coplanar orbits, and thereby neglect 
the inclinations and longitude of ascending nodes. 



Figure \ref{fullint} \marginpar{FIG. 11} illustrates 
how the resonant planets' orbital parameters evolve 
using the full N-body integrator, HNbody 
(Rauch and Hamilton 2007, in preparation).  We define $\phi_1$
as the resonant angle corresponding to the constants 
$\lbrace{ 2,-1,0,-1,0,0 \rbrace}$, and 
$\phi_2$ as the resonant angle corresponding to 
$\lbrace{ 2,-1,-1,0,0,0 \rbrace}$.  
Figure 1 demonstrates that both $\phi_1$ and $\phi_2$ 
librate about $0^{\circ}$, and $\varpi_1$ and $\varpi_2$ 
circulate, all with a period of $\approx 9$ years.  The 
eccentricities exhibit a periodicity commensurate with
that of the orbital angles' amplitudes of libration, whereas the 
semimajor axes exhibit little ($1$\% - $2$\%) variation.  The 
influence of the short-period terms are seen through the small 
modulations (or the noise) especially apparent in the
orbital elements of GJ 876 c because of its relatively large eccentricity.
The planets are said to be ``in resonance'' because of the simultaneous 
libration of $\phi_1$ and $\phi_2$, and GJ 876 c is said to be in 
``deep resonance'' because of the small ($<10^{\circ}$) amplitude 
of $\phi_1$.


A full treatment
of the GJ 876 system, even to first-order in eccentricities, requires 
multiple resonant and secular arguments.  Table \ref{termlist} 
\marginpar{TABLE 4} lists all relevant resonant and 
secular terms for this system up to fourth order, and 
Figs. \ref{order1}-\ref{order4} illustrate the planetary evolution caused
by all terms up to order one, two, three and four respectively.  Differences
from the true evolution exhibited by Fig. \ref{fullint} may be attributed to the
neglect of short-period terms and even higher-order resonant and secular
terms.  The simulations in Figs. \ref{order1}-\ref{order4} were 
initialized with {\it mean} 
orbital element values, obtained from averaging over the orbital evolution
of a few periods of the outer planet from the exact numerical integration.
The mean initial values used were
$a_1 = 0.20908$ AU, $a_2 = 0.13008$ AU, 
$e_1 = 0.03054$, $e_2 = 0.2304$, $\varpi_1 = 160.6^{\circ}$, 
$\varpi_2 = 178.6^{\circ}$, $\lambda_1 = 179.2^{\circ}$, and
$\lambda_2 = 180.3^{\circ}$.

Figure \ref{order1} \marginpar{FIG. 12} 
demonstrates that a full first-order treatment, which includes
the leading terms from the $\lbrace 2, -1, -1, 0, 0, 0  \rbrace$, 
$\lbrace 2, -1, 0, -1, 0, 0  \rbrace$, $\lbrace 0, 0, 1, -1, 0, 0  \rbrace$
and $\aleph$ arguments, forces $\phi_2$ to circulate rather than librate
and $\varpi_1$ to librate about $180^{\circ}$ rather than circulate.
These features are at odds with the full N-body integration.
Therefore, analytical manipulations of these terms alone have limited
utility when one attempts to analyze particular aspects of the system's
evolution.  

The combination of terms comprising the second-order solution 
showcased in Fig. \ref{order2} \marginpar{FIG. 13} is only 
marginally stable, and 
only for a high integration accuracy parameter ($\lesssim 10^{-7}$).
Nevertheless, amidst the chaos, both $\phi_1$ and $\phi_2$ do librate
about $0^{\circ}$, but on a longer timescale (note difference in $x$ axes).
Short-term evolution, however, showcases fictitious asymmetric libration 
of varying amplitude about $\sim 70^{\circ}$ or 
$\sim -70^{\circ}$.  The switch between
both libration centers occurs at extrema of the semimajor axes
and eccentricities, and the presence of asymmetric libration 
centers demonstrates
that the first-order solution is more accurate than the second-order
solution.  Similar results may be found in the restricted 3-body 
problem \citep{beauge1994} even though that problem is inherently different.
Further, \cite{lemhen1988} discuss the incorrect qualitative behavior
which arises from improper truncation of the disturbing function as applied
to asteroid studies.


Higher-order treatments, however, not only are stable but reproduce
expected features of the system.  The third-order solution in 
Fig. \ref{order3} \marginpar{FIG. 14} and the fourth-order solution in 
Fig. \ref{order4} \marginpar{FIG. 15} reproduce, to varying 
extents, the orbital
evolution shown from the N-body simulation of Fig. \ref{fullint}.  
However, in Fig. \ref{order3},
$e_1$ represents an unmodulated sinusoid, $\phi_1$ lacks any modulated
envelope, and $\phi_2$ acquires a slight sawtooth form.  These
minor features all differ from the N-body integration, but are all
partially remedied by the fourth-order treatment in Fig. \ref{order4}.  
However, this same fourth-order treatment demonstrates a decrease in precision
by altering the amplitudes and elongating the libration period by over
a factor of two.


The above considerations suggest that the so-called 
``Laplacian expansion'' (described in \citealt{ellmur2000}) expansion of the 
disturbing function best reproduces the evolution of GJ 876 for an
eccentricity order between $3$ and $4$.  The lack of precision
can be explained by the failure to satisfy the Sundman criterion 
\citep{sundman1916} for the coefficients of this expansion.  In the 
planar case, the criterion is typically
expressed as \citep{ferrazmello1994b,sidnes1994}: 

\beq
a_2 F(e_2) < a_1 f(e_1),
\eeq

\noindent{where}

\beq
\begin{split}
F(q) &= \sqrt{1 + q^2} \cosh{z} + q + \sinh{z}
,
\\
f(q) &= \sqrt{1 + q^2} \cosh{z} - q - \sinh{z}
,
\end{split}
\eeq

\noindent{such} that $w$ is implicitly defined as $z = q \cosh{z}$.  Applying
the criterion to representative GJ 876 orbital parameters yields Fig. 
\ref{sundcritreal}, \marginpar{FIG. 16} which demonstrates 
that the classic expansion
will show signs of divergence at particular points in the evolution.
As more resonant terms are included, the effect of this divergence 
might be amplified.  Hence,
the optimal order with which to model GJ 876 with this expansion is finite.
Taken as a sequence, Figs. \ref{order1}-\ref{order4} demonstrate that 
perhaps ``order'' is an inappropriate metric for classifying the 
accuracy of a model for systems with a phase space similar to that of GJ 876.
We highlight these perturbative issues in order to caution future
investigators regarding the validity of using particular resonant arguments
for analytical considerations.

Resonant systems may be affected, or even dominated, by perturbations external
to the detailed dynamical interplay described so far.  The next three sections
will present potentials for different types of perturbations and estimate
their effect on GJ 876.  Table \ref{effects} \marginpar{TABLE 5} 
summarizes these effects and what
orbital parameters they directly affect depending on the assumptions used.

\section{Central Body Oblateness} \label{oblsec}

\subsection{Overview}

The contribution of oblateness from a central body to a 
resonant system has shown to play a crucial role in ring 
and small satellite dynamics.  
\cite{gozmac1998} and \cite{shinkin2001} have
created analytical models for the dynamical evolution of satellites 
around oblate planets, and \cite{wiesel1982} has created a model of
rings in resonance with an oblate planet.
In this section, we incorporate the oblateness effect through {\it both}
an averaged (over short-period terms) and unaveraged oblateness potential.
Doing so helps determine 
specifically the error incurred when using the averaging approximation,
and helps clarify the seemingly contradictory expressions found in the
literature.  The oblateness potential, $\itR_{k}^{(obl)}$, has the form,

\beq
     \itR_{k}^{(obl)} = \f{\mathcal{G} M_{c}}{r_k} 
               \sum_{i=2}^{\infty} J_{i} 
               {\po  \f{R_{c}}{r_k} \pc}^i P_i 
                \po \sin{\delta_k} \pc
, 
\label{Vob}
\eeq

\noindent{where} 
$R_{c}$ is the equatorial radius of the central object,
$J_i$ are the oblateness coefficients, $P_i$ are Legendre polynomials,
and $\delta$ is the declination with respect to the central object's equatorial plane
\citep[][p. 563]{brocle1961}.  
In order to maintain consistency with the other equations in 
this work, we wish to express Eq. (\ref{Vob}) in terms of 
contact elements.  However, we may 
instead use expressions for osculating elements because every term in 
the expansion for the potential depends only upon the resonant body 
positions, and not velocities (expressions for positions are identical 
in both osculating and contact elements).

\subsection{Averaged expressions}

Various different expressions for the orbital average of the oblateness potential 
have appeared in the literature due to differences in meaning of the 
orbital elements.  \cite{greenberg1981} attempted to eliminate the confusion 
regarding these seemingly 
dissimilar formulas.  He wrote the following 
expressions for apsidal precession rates in osculating (subscript ``o''),
sidereal (subscript ``s''), and epicyclic (subscript ``e'') coordinates:

\beq
\dot{\varpi} \approx n_o 
\left[ \f{3}{2}  J_2     \po \f{R}{a_o} \pc^2 +
       \f{63}{8} J_{2}^2 \po \f{R}{a_o} \pc^4 -  
       \f{15}{4} J_4     \po \f{R}{a_o} \pc^4  \right]
,
\label{osculating}
\eeq

\beq
\dot{\varpi} \approx n_s 
\left[ \f{3}{2}  J_2     \po \f{R}{a_s} \pc^2 -
       \f{9}{8}  J_{2}^2 \po \f{R}{a_s} \pc^4 -  
       \f{15}{4} J_4     \po \f{R}{a_s} \pc^4  \right]
,
\label{sidereal}
\eeq

\beq
\dot{\varpi} \approx \sqrt{ \f{\itG M}{a_{e}^3}} 
\left[ \f{3}{2}  J_2     \po \f{R}{a_e} \pc^2 -
       \f{15}{4} J_4     \po \f{R}{a_e} \pc^4  \right]
.
\label{measured}
\eeq

Any one of Eqs. (\ref{osculating})-(\ref{measured}) may be correct depending on 
the {\it context} of the system studied.  Equation (\ref{osculating}) 
\citep{brouwer1959}
may be used when the orbital elements considered are all  
osculating Keplerian elements.  Equation (\ref{sidereal}) 
\citep[][p. 270]{brouwer1946,murder1999} may be 
used when dealing with observed values for the sidereal mean motion and the mean
distance from the central object (also called the ``observed semimajor axis,''
``apparent semimajor axis,'' or ``geometric semimajor axis'').  
Equation (\ref{measured}) \citep[][p. 269]{elletal1981, 
ellnic1984,borlon1994,murder1999} 
may be used when the {\it only} measured orbital element is the distance from the
central object, a typical characteristic of ring systems.  

Our model does not utilize truncated expressions for apsidal precession. Rather, we
wish to obtain an averaged disturbing function in terms of osculating elements.
Unfortunately, differing expressions for averaged potentials have also appeared in the
literature.  Therefore, understanding the manner in which 
Eqs. (\ref{osculating})-(\ref{measured}) are derived from disturbing functions is 
important in determining the correct form of the averaged potential.

\citet[][p. 269]{murder1999} and \cite{metris1991} 
write expressions for 
averaged potentials which contain a $J_{2}^2$ term.  As the potential is a 
linear combination of $J_i$ terms,
averaging only over the ``fast'' angle known as the true anomaly would not produce a 
product of the oblateness terms from which a $J_{2}^2$ term could be derived.
Typically, $J_{2}^2$ terms appear in the derivations of the variation of the 
orbital elements.  Such derivations include binomial expansions 
\citep{brouwer1946,greenberg1981}, Poincar\'{e}-von Zeipel
transformations \citep{brouwer1959,kozai1962b}, Taylor expansions of orbital elements
\citep{kozai1959}, or Poincar\'{e}-Lindstedt expansions 
\citep{borlon1994}
\footnote{The elimination of the $J_{2}^2$ term in Eq. (\ref{measured}) is a chance
cancellation; the Appendix of \cite{borlon1994} shows that
the $J_2 J_4$ and $J_{2}^3$ terms, for example, do not cancel.}.   
\cite{metris1991} uses the Hori-Deprit method in order to illustrate how a 
$J_{2}^2$ term may appear in the {\it potential}
if an {\it additional, different} averaging in the small parameters $J_i$ 
is performed over the often-neglected short-period terms.  
This procedure was applied to the osculating elements in order to yield the
mean osculating ones.  \citeauthor{metris1991}'s (\citeyear{metris1991}) resulting 
expression for $J_{2}^2$ differs from that of 
\citet[][p. 269]{murder1999}.

Further, the expression from \citet[][p. 269]{murder1999} does not 
contain $J_3$ terms
(which were argued to be typically negligible),
nor terms which are independent of eccentricity or inclination.  
These latter terms are not constant,
but rather depend on the variable semimajor axis.  The terms containing $J_3$ vanish
only if the disturbing function is averaged twice - once over the true anomaly, and once
over the argument of pericenter.  However, the latter orbital angle does not vary
``quickly'' in general.

Given the above considerations, we include 
\citeauthor{metris1991}' (\citeyear{metris1991}) $J_{2}^2$ 
short-period terms, and use 
\citeauthor{kozai1959}'s (\citeyear{kozai1959})
expression for the other terms
in order to obtain an averaged potential. 
The resulting expression, when truncated to ``fourth-order'' in 
eccentricities and inclinations (as is the disturbing function), reads:

\beq
\begin{split}
\langle \itR_{k}^{(obl)} \rangle &= 
\mu_k \sum_{i=1}^{2} 
\bigg[
a_{k}^{-2i-1}
\Big( \gamma_{i}^{(0)}
   +\gamma_{i}^{(1)} e_{k}^2
   +\gamma_{i}^{(2)} e_{k}^4
   +\gamma_{i}^{(3)} \sin^2{I_k}
   +\gamma_{i}^{(4)} \sin^4{I_k}
\\
   &+\gamma_{i}^{(5)} e_{k}^2 \sin^2{I_k}
   +\gamma_{i}^{(6)} e_{k}^2 \sin^2{I_k} \cos{\po 2\varpi_k - 2\Omega_k \pc}
   \Big) 
\bigg]
\\
&+
\mu_k a_{k}^{-4}
\left[
    \gamma_{i}^{(7)} e_{k} \sin{I_k}
   +\gamma_{i}^{(8)} e_{k}^3 \sin{I_k}
   +\gamma_{i}^{(9)} e_{k} \sin^3{I_k}
\right] 
\sin{\po \varpi_k - \Omega_k \pc}
,
\end{split}
\label{gamdef}
\eeq

\noindent{where},

\begin{equation}
\gamma_{1}^{(0)} = \f{1}{2} J_2 R_{c}^2 
\bbs \bbs \bbs 
\gamma_{2}^{(0)} = - \f{33}{56} J_{4} R_{c}^4 
                   - \f{9}{8} J_{2}^2 R_{c}^4
\bbs \bbs \bbs
 \gamma_{1}^{(1)} = \f{3}{4} J_2 R_{c}^2
\bbs \bbs \bbs
\gamma_{2}^{(1)} = - \f{165}{56} J_{4} R_{c}^4,
\label{gamexp1}
\end{equation}

\begin{equation}
\gamma_{1}^{(2)} = 0
\bbs \bbs \bbs
\gamma_{2}^{(2)} = - \f{99}{32} J_{4} R_{c}^4
\bbs \bbs \bbs
\gamma_{1}^{(3)} = - \f{3}{4} J_2  R_{c}^2
\bbs \bbs \bbs
\gamma_{2}^{(3)} = \f{165}{56} J_4  R_{c}^4 +
                   \f{45}{16} J_{2}^2  R_{c}^4,
\label{gamexp2}
\end{equation}

\begin{equation}
\gamma_{1}^{(4)} = 0
\bbs \bbs \bbs
\gamma_{2}^{(4)} = - \f{165}{64} J_{4} R_{c}^4
                   - \f{57}{32} J_{2}^2  R_{c}^4
\bbs \bbs \bbs
\gamma_{1}^{(5)} = - \f{9}{8} J_2  R_{c}^2
\bbs \bbs \bbs
\gamma_{2}^{(5)} = \f{825}{56} J_4  R_{c}^4 -
                   \f{3}{8} J_{2}^2  R_{c}^4,
\label{gamexp3}
\end{equation}

\begin{equation}
\gamma_{1}^{(6)} = 0
\bbs \bbs \bbs
\gamma_{2}^{(6)} = - \f{495}{224} J_{4} R_{c}^4
\bbs \bbs \bbs
\gamma_{1}^{(7)} = \f{3}{2} J_3  R_{c}^3
\bbs \bbs \bbs
\gamma_{1}^{(8)} = \f{15}{4} J_3  R_{c}^3 
\bbs \bbs \bbs
\gamma_{1}^{(9)} = -\f{15}{8} J_3  R_{c}^3 
.
\label{gamexp4}
\end{equation}

Incorporating disturbing function partial derivatives with Eqs. (\ref{lagagmod}) 
provides explicit expressions for the derivatives of the orbital elements 
for the oblateness
contributions.  In order to obtain the oblateness contribution to $\dot{\phi}$,
one can combine the derivatives of the orbital elements with Eq. (\ref{phidot}).

\subsection{Complete expressions}

In the unaveraged incorporation of oblateness into our model, we 
use Eq. (\ref{Vob}) explicitly.  The eccentric anomaly, $E_k$, is 
related to the mean anomaly through Kepler's equation:

\beq
E_k - e_k \sin{E_k} = M_k = \lambda_k - \varpi_k
\label{Kepeq}
\eeq

\noindent{such} that $E_k = E_k (\lambda_k, \varpi_k)$.

The true anomaly is related to the eccentric anomaly through
(Murray \& Dermott 1999, p. 33)

\beq
f_k = 
f_k (e_k, \lambda_k, \varpi_k)
=2 \arctan{ \left[
\sqrt{\f{1+e_k}{1-e_k}} \tan{\po \f{1}{2} E_k \pc} \right] }
.
\label{fandE}
\eeq

Let:

\beq
Y_k \equiv \sin{I_k} 
\sin{\po f_k + \varpi_k - \Omega_k \pc}
.
\label{auxX}
\eeq



Substituting Eq. (\ref{auxX}) into  Eq. (\ref{Vob}) and then
incorporating the appropriate partial derivatives into 
Eq. (\ref{lagagmod}) allows us to determine
the oblateness contribution for each orbital parameter.  In order
to compute these derivatives, the eccentric anomalies must be computed
through Kepler's Equation numerically at each timestep.

\subsection{Application to GJ 876}

Saturn's oblateness dominates the precession of the pericenters of the
system's close small satellites.  The Sun's oblateness, however, 
despite its significance in general relativistic computations 
\citep{iorio2005}, is thought to have a negligible effect on the 
evolution of the planets in the Solar System. Extrasolar systems, 
however, display a variety of orbital 
configurations which might admit the possibility of the central 
star's oblateness playing a role in the evolution.  
Massive ($ > M_{Saturn}$) planets are known to orbit 
their stars at distances an order of magnitude shorter than the Sun-Mercury
separation, and can evolve in orbit-orbit resonances absent from the inner
Solar System.  We further analyze the resonant planets in GJ 876 by attempting
to determine under what conditions can oblateness play a role, and how
likely their evolution is affected by the property of the star.  As the
effect of oblateness through inspection is typically negligible, we track
the contribution by explicitly plotting the contribution for the rate
of change of relevant orbital elements at each timestep.

GJ 876 is an $0.32 m_{\odot}$ M4 dwarf with a rotational velocity similar to that 
of the Sun ($\approx 2$ km/s) \citep{deletal1998} and a radius just a fraction
of the Sun's \citep{chabar1997}.  The star's oblateness 
is unknown, likely changes with time, and is a detailed function of the rotation 
profile, mass, and radius.
\cite{rozetal2001} compares quadrupole ($J_2$) and octopole ($J_4$)
Solar measurements from several references, and \cite{rozroe1997}
provide an upper bound to the Solar oblateness.  As the range of values estimated
for the Sun span two orders of magnitude ($\sim 10^{-8} - 10^{-6}$), and 
$J_2 \approx \Omega_{\odot}^2 R_{\odot}^3/m_{\odot}$
\citep{godroz2000}, where $\Omega_{\odot}$ is the rotational velocity of the Sun,
GJ 876's oblateness is likely to fall within this range.

Through simulations including the 11 arguments needed for a third-order treatment, 
we conclude that the GJ 876 planets are negligibly affected by stellar 
oblateness.  
Figure \ref{oblna} \marginpar{FIG. 17} plots the fractional contribution of 
oblateness at each timestep of $\dot{\varpi}_1$ (solid line)
and $\dot{\varpi}_2$ (dotted line) in the 
{\it unaveraged} planar problem given 
$J_2 = -10^{-6}$, $J_4 = 10^{-12}$ and $R_{\odot} = 6 \times 10^5$ km; 
Fig. \ref{obla} \marginpar{FIG. 18} is equivalent except in
the {\it averaged} problem.
Both figures indicate an upper bound for the fractional oblateness
contribution is $10^{-6} - 10^{-7}$.  The librational periodicity
is more evident in Fig. \ref{obla} than 
Fig. \ref{oblna} due to the averaged nature of
the system simulated there.
The magnitude of the 
fractional contributions from Figs  \ref{oblna} and \ref{obla} 
demonstrate that any planet nonnegligibly affected by 
stellar oblateness must 
be much closer to the
central star, at a radius which is 
likely to be inside the tidal circularization limit.  

If GJ 876 was a fast rotator ($30-50$ km/s) like
other observed M dwarfs (Delfosse {\it et al.} 1998), 
then the effect of oblateness would be at least two orders 
of magnitude greater, and likely affect the long-term, if not
short-term, evolution of the system.  As several tens of 
extrasolar planets have been discovered within $0.1$ AU of 
their parent stars, including several multiple systems, prospects
for finding an oblate star with a tight resonant system are
promising.  Such stars would be Jupiter-like due to their mass,
radius, and rotational period of $\approx 10$ hr.


\section{Central-body Precession} \label{presec}

\subsection{Introduction}

An object's precession about its axis exerts a gravitational 
influence on orbiting masses.  Although much precession-based research
focuses on Mars' chaotic obliquity (e.g. \citealt{hilton1991, 
groetal1996,chrmur1997,bills1999,bousou1999})
and although occasionally general models applicable to different 
systems are developed (e.g. \citealt{blitzer1984, 
penna1999}), investigators of resonant systems do not
always consider the effect of precession. \cite{kozai1960} 
conducted one of the first studies of how a satellite's
orbital elements are affected by the precession of its parent planet.
\cite{kinoshita1993} studied the motion of a satellite relative to
the equatorial plane of its oblate parent parent, and
\cite{rubincam2000} discusses the possibility that Pluto may be in
``precession-orbit'' resonance.  Expressions for 
the precession contribution to the disturbing function
are given implicitly by \cite{
goldreich1965a} and explicitly
by \cite{kopal1969a}, \cite{bruetal1970}  
and \cite{efroimsky2005b}; using their results, we can write:

\beq
\itR_{k}^{(pre)} = \sqrt{\mu_k a_k \po 1 - e_{k}^2 \pc}
\po \omega_1 \sin{I_k} \sin{\Omega_k} - 
    \omega_2 \sin{I_k} \cos{\Omega_k} + 
    \omega_3 \cos{I_k} 
\pc
,
\label{prec1}
\eeq

\noindent{}where $\omega_1$, $\omega_2$, and $\omega_3$ are the 
components of the precessional frequency corresponding to increasing
moments of inertia of the central body.  

\subsection{Application to GJ 876}

As displayed in Table 2, a central object's precession does
not alter the semimajor axes nor eccentricities of the planets directly,
but rather only indirectly through $\dot{\phi}$.  For GJ 876, we assume
representative values of 
$\omega_1 = \omega_2 = \omega_3 = 10$ pHz   
and $R_{\odot} = 6 \times 10^5$ km.  These $\omega$ values
correspond to a precessional period of $\approx 2 \times 10^4$ yr,
which corresponds to a 
frequency significantly less than the mean motion of the
resonant bodies, a realistic regime at least in the 
context of precessing stellar jets \citep{namouni2005}.
Figure \ref{prena} \marginpar{FIG. 19} plots the fractional contribution of 
precession at each timestep of $\dot{\epsilon}_1$ (solid line), 
$\dot{\epsilon}_2$ (dotted line), $\dot{\varpi}_1$ (dashed line),
and $\dot{\varpi}_2$ (dash-dot line).  The effect on the orbital angles
ranges from one part in $10^{-4}$ to one part in $10^{-2}$,
suggesting that for planar extrasolar planetary systems,
precession may play a greater role than central-body oblateness.  Direct 
comparison with
Figs. \ref{oblna}-\ref{obla} demonstrates the effect of precession on the evolution of
$\dot{\varpi}_1$ and $\dot{\varpi}_2$ is roughly two orders of magnitude
higher than that from oblateness.


\section{A Surrounding Thin Disk} \label{disksec}

\subsection{Introduction}

The birthplace of planets and satellites, a protoplanetary or protosatellite disk 
ultimately determines the dynamical interactions which occur in the system.  Although 
capture into resonance is likely to and often thought to occur after disk dissipation,
and requires a dissipative medium, 
capture may be achieved while the disk is still present, for both protoplanetary and 
protosatellite disks.  \cite{tholis2003} find that disks can play a crucial 
role in the migration and capture of planets in resonance, and 
\cite{kleetal2004} performed numerical simulations of two planets embedded in a 
thin disk, and finds resonant capture to be a common occurrence.  

In order to approximate the mass density profile of the nascent Solar System nebular 
disk, a model known as the Minimum Mass Solar Nebula (MMSN) has been devised 
\citep{weidenschilling1977,hayashi1981}.  The MMSN represents the pair of values 
$(\Sigma_0, w)$ which, when inserted into the density profile, 
$\Sigma = \Sigma_0 (r_0/r)^w$, provides the minimum nebular mass necessary to explain 
the current masses and positions of the planets.  A variety of assumptions go into 
this calculation, including the fraction and location of nebular material that is 
eventually scattered out of the system and the possible presence and location of 
the snow line \citep{saslec2000}.  The commonly used MMSN prescription sets 
$w = 1.5$ \citep{weidenschilling1977,hayashi1981}. 
\footnote{\cite{kuchner2004} extended the analysis to extrasolar systems, and derived 
a Minimum Mass Extrasolar Nebula (MMEN) based on data from 26 exoplanets in 
multiple-planet systems.  He derived an overall steeper profile, with 
$w = 2.0 \pm 0.5$, and individually fitted $\Sigma_0$ and $w$ values to individual 
multi-planet exosystems.}

We consider only the gravity of thin steady disks with a power-law density 
profile $\Sigma = \Sigma_0 (r_0/r)^w$, where $\Sigma_0$ and $r_0$ are 
constant, arbitrary parameters; we take the latter to be $1$ AU.  Such 
profiles could represent protoplanetary {\it or} protosatellite disks.  
We also assume that the disk harbors 
two massive non-central bodies, each having already carved out a gap, 
as illustrated in Fig. \ref{cartoon}. \marginpar{FIG. 20} The relative 
sizes of the mass-filled regions labeled I, II, and III largely determine 
the dynamical evolution of the system.  Depending on the orbital 
properties of planets in the system, and their interaction with the gas, regions
I and II may be small or unoccupied with gas.  We assume the disk 
contains the same total mass ($\equiv m_n$) as a 
particular fraction, $\chi$, of the central mass, where

\begin{equation}
m_{n} = \chi m_0 = \int \Sigma \po r \pc dA
,
\label{totmas1}
\end{equation}

\noindent{such that}

\begin{equation}
\Sigma_0 = 
\begin{cases}
\f{\chi m_0 \po 2 - w \pc}{2 \pi r_{0}^w}
\left[ r_{b}^{2-w} - r_{a}^{2-w} +
       r_{d}^{2-w} - r_{c}^{2-w} +
       r_{f}^{2-w} - r_{e}^{2-w} 
\right]^{-1},
&w \ne 2 \\
\f{\chi m_0}{2 \pi r_{0}^2}
\left[ 
\ln{
\po 
\f{ r_{b} r_{d} r_{f}  }
{   r_{a} r_{c} r_{e}  }
\pc
}
\right]^{-1},
&w = 2
.
\end{cases}
\label{totmas2}
\end{equation}

\subsection{Horizontal Contributions}

\cite{ballabh1973} categorized expressions for gravitational potentials of circularly 
symmetric distributions of matter, in the form of homogeneous and heterogeneous 
disks, as polynomials in semimajor axis.  Focusing on the Solar nebula, \cite{ward1981}
computed averaged gravitational potentials due to a thin disk, and utilized 
Laplace coefficients in the expansions.  Using the notation in the Appendix, we 
express the Laplace coefficients as $\beta$ constants.  The potential at a 
point $r_k$ distance from the central body in the disk plane due to the outer
(``out'') and inner (``in'') parts of a disk, read, respectively, 

\beq
\itR_{k}^{(out)} = 2 \pi \itG \Sigma_0 r_{0}^w 
\sum_{l=0}^{\infty} \f{\beta_{l+1}^{(0,\f{1}{2})}}{2l+w-1}
r_{k}^{2l}
\left[
r_{ogap}^{-2l-w+1} - r_{oedge}^{-2l-w+1}
\right]
, \bbs\bbs\bbs w \ne 1-2l
,
\label{new1}
\eeq

\beq
\itR_{k}^{(in)} = 2 \pi \itG \Sigma_0 r_{0}^w 
\sum_{l=0}^{\infty} \f{\beta_{l+1}^{(0,\f{1}{2})}}{2l-w+2}
r_{k}^{-2l-1}
\left[
r_{igap}^{2l-w+2} - r_{iedge}^{2l-w+2}
\right]
, \bbs\bbs\bbs w \ne 2l+2
,
\label{new2}
\eeq

\noindent{where} $r_{iedge}$ and $r_{oedge}$ are the edges of the entire disk, and 
$r_{igap}$ and $r_{ogap}$ are the boundaries of a gap that surrounds the point at 
which the potential is measured.  Equations (\ref{new1}) and (\ref{new2}) 
represent just the leading-order term of the potential, {\em not} the entire 
potential itself.  Given the schematic of 
Fig. \ref{cartoon}, the disturbing functions for the outer and inner 
planet then become, 
respectively (for $w \ne 1 - 2l$ and $w \ne 2l + 2$),

\begin{equation}
\begin{split}
\itR_1 = &2 \pi \itG \Sigma_0 r_{0}^w \sum_{l=0}^{\infty} 
\beta_{l+1}^{(0,\f{1}{2})}
\bigg[
\f{r_{1}^{2l}}{2l+w-1} 
\po r_{e}^{-2l-w+1} - r_{f}^{-2l-w+1} \pc \\
&+ \f{r_{1}^{-2l-1}}{2l-w+2}
\po  r_{d}^{2l-w+2} - r_{c}^{2l-w+2} + r_{b}^{2l-w+2} - r_{a}^{2l-w+2} \pc
\bigg]
,
\end{split}
\label{new3}
\end{equation}

\begin{equation}
\begin{split}
\itR_2 = &2 \pi \itG \Sigma_0 r_{0}^w \sum_{l=0}^{\infty} 
\beta_{l+1}^{(0,\f{1}{2})}
\bigg[
\f{r_{2}^{2l}}{2l+w-1} 
\po r_{c}^{-2l-w+1} - r_{d}^{-2l-w+1} + r_{e}^{-2l-w+1} - r_{f}^{-2l-w+1} \pc \\
&+ \f{r_{2}^{-2l-1}}{2l-w+2}
\po  r_{b}^{2l-w+2} - r_{a}^{2l-w+2} \pc
\bigg]
,
\end{split}
\label{new4}
\end{equation}

\noindent{for} fixed values of $r_a$, $r_b$, $r_c$, $r_d$, $r_e$, and $r_f$.  
At this stage we may choose to average over the angles on which $r_1$ and $r_2$ 
are dependent; we present both approaches for the sake of completeness. 
Given the derivatives of $r_k$ in terms of orbital elements, disturbing 
function derivatives can be computed directly from Eqs. (\ref{new3})-(\ref{new4}) 
and incorporated into Lagrange's planetary equations without any averaging.  
Conversely, expressing $r_1$ and $r_2$ in terms of orbital elements and 
averaging yields:

\beq
\itR_k = \sum_{l=0}^{\infty} 
\left[
{\overset{oh}{T}}_{k}^{(l)} 
{\overset{oh}{\Upsilon}}_{k}^{(l)}
a_{k}^{2l}
+
{\overset{ih}{T}}_{k}^{(l)} 
{\overset{ih}{\Upsilon}}_{k}^{(l)}
a_{k}^{-2l-1}
\right]
,
\label{newpot}
\eeq

\noindent{where} ${\overset{oh}{\Upsilon}}_{k}^{(l)}$ and ${\overset{ih}{\Upsilon}}_{k}^{(l)}$ are functions of $e_k$ and $I_k$, and where ${\overset{oh}{T}}_{k}^{(l)}$ and ${\overset{ih}{T}}_{k}^{(l)}$ are independent of $a_k, e_k,$ and $I_k$.  The designations ``oh'' and ``ih'' represent ``outer horizontal'' and ``inner horizontal.''  Generalizing Eqs. (\ref{new1})-(\ref{new4}) to admit any value of $w$ yields:

\begin{equation}
{\overset{oh}{T}}_{1}^{(l)} = 
\begin{cases}
2 \pi \itG \Sigma_0 r_{0}^w 
\f{\beta_{l+1}^{(0,\f{1}{2})}}{2l + w - 1}
\po r_{e}^{-2l-w+1} - r_{f}^{-2l-w+1} \pc
,
&w \ne 1 - 2l \\
2 \pi \itG \Sigma_0 r_{0}^{1-2l} 
\ln{\po \f{r_f}{r_e} \pc}
,
&w = 1 - 2l
\end{cases}
,
\label{labToh}
\end{equation}

\begin{equation}
{\overset{oh}{T}}_{2}^{(l)} = 
\begin{cases}
2 \pi \itG \Sigma_0 r_{0}^w 
\f{\beta_{l+1}^{(0,\f{1}{2})}}{2l + w - 1}
\po r_{c}^{-2l-w+1} - r_{d}^{-2l-w+1} + r_{e}^{-2l-w+1} - r_{f}^{-2l-w+1} \pc
,
&w \ne 1 - 2l \\
2 \pi \itG \Sigma_0 r_{0}^{1-2l} 
\ln{\po \f{r_d r_f}{r_c r_e} \pc}
,
&w = 1 - 2l
\end{cases}
,
\label{labToh2}
\end{equation}

\begin{equation}
{\overset{ih}{T}}_{1}^{(l)} = 
\begin{cases}
2 \pi \itG \Sigma_0 r_{0}^w 
\f{\beta_{l+1}^{(0,\f{1}{2})}}{2l - w + 2}
\po  r_{d}^{2l-w+2} - r_{c}^{2l-w+2} + r_{b}^{2l-w+2} - r_{a}^{2l-w+2} \pc
,
&w \ne 2 + 2l \\
2 \pi \itG \Sigma_0 r_{0}^{2+2l} 
\ln{\po \f{r_b r_d}{r_a r_c} \pc}
,
&w = 2 + 2l
\end{cases}
,
\label{labTih}
\end{equation}

\begin{equation}
{\overset{ih}{T}}_{2}^{(l)} = 
\begin{cases}
2 \pi \itG \Sigma_0 r_{0}^w 
\f{\beta_{l+1}^{(0,\f{1}{2})}}{2l - w + 2}
\po  r_{b}^{2l-w+2} - r_{a}^{2l-w+2} \pc
,
&w \ne 2 + 2l \\
2 \pi \itG \Sigma_0 r_{0}^{2+2l} 
\ln{\po \f{r_b}{r_a} \pc}
,
&w = 2 + 2l
\end{cases}
.
\label{labTih}
\end{equation}

Using the binomial theorem under assumption of small $e$ yields:

\beq
 {\overset{oh}{\Upsilon}}_{k}^{(l)} \approx 1 - \po 2l^2 + l \pc e_{k}^4 +
\mathcal{O} \po e_{k}^6 \pc
,
\label{Palmdef1}
\eeq

\beq
 {\overset{ih}{\Upsilon}}_{k}^{(l)} \approx 
\f{\po 2 l + 1 \pc e_{k}^2 - 1}{\po e_{k}^2 - 1 \pc^{2 l + 1}}
+
\mathcal{O} \po \f{e_{k}^4}{ \po e_{k}^2 - 1  \pc^{2l + 1}} \pc
,
\label{Palmdef2}
\eeq

With the above equations, one may now determine all of the partial derivatives of the 
averaged disk disturbing functions and incorporate them into Lagrange's planetary
equations.

\subsection{Vertical Contributions}

\subsubsection{Overview}

\cite{campin1973} derive expressions of potential due to 
a cylindrical shell, but evaluate the expressions at the midplane.
\cite{ward1981} evaluates the potential above or below the midplane by performing an
inclination expansion and assuming that the maximum vertical
(above the midplane) displacement of a body is much less than
the distance between the disk edge and the body's semimajor axis
($2 r r' s_{k}^2 \ll \left[ a_k - r' \right]^2$). We adopt the more general 
assumption that $2 r r' s_{k}^2 \ll (r_k - r')^2$, which is
nearly equivalent except at high eccentricities.  With this
assumption, we can directly add the vertical disk contributions
to the horizontal ones.  In the expansion of the vertical disk contribution, 
the first term is independent of inclination and equals the corresponding
horizontal term exactly. The next term in the expansion is
of order $\mathcal{O} (I_{k}^2)$ and is 
incorporated into our model.

The vertical disk contribution is computed along similar lines as for 
the horizontal contribution, except the outer and inner potentials now read,

\beq
\itR_{k}^{(out)} = -2 \pi \itG \Sigma_0 r_{0}^w s_{k}^2
\po 1 - \cos^2 f_k  \pc
\sum_{l=1}^{\infty} \f{\beta_{l}^{(1,\f{1}{2})}}{2l+w-1}
r_{k}^{2l}
\left[
r_{ogap}^{-2l-w+1} - r_{oedge}^{-2l-w+1}
\right]
,
\label{vertdisk1}
\eeq

\beq
\itR_{k}^{(in)} = -2 \pi \itG \Sigma_0 r_{0}^w s_{k}^2
\po 1 - \cos^2 f_k  \pc
\sum_{l=1}^{\infty} \f{\beta_{l}^{(1,\f{1}{2})}}{2l-w+2}
r_{k}^{-2l-1}
\left[
r_{igap}^{2l-w+2} - r_{iedge}^{2l-w+2}
\right]
,
\label{vertdisk2}
\eeq

\noindent{where} as previously stated,  $s_k = (\sin{I_k/2})$, and $f_k$ 
is the true anomaly. These expressions represent the leading order of the 
portion of the potential which contains 
inclination.  Note importantly that the summations begin with $l=1$, as 
opposed to those from Eqs. (\ref{new1})-(\ref{new2}).  Given the 
expressions for the disturbing functions similar to 
those in Eqs. (\ref{new3}) and (\ref{new4}), and the derivatives of 
$r_k$, $s_k$, {\it and} $f_k$ in 
terms of orbital elements, disturbing function derivatives can be 
incorporated into Lagrange's planetary equations without any 
averaging.  Conversely, the averaged potential reads, with ``ov'' 
and ``iv'' referring to ``outer vertical'' and ``inner vertical,''  

\beq
\itR_k = \sum_{l=1}^{\infty} 
\left[
{\overset{ov}{T}}_{k}^{(l)} 
{\overset{ov}{\Upsilon}}_{k}^{(l)}
a_{k}^{2l}
+
{\overset{iv}{T}}_{k}^{(l)} 
{\overset{iv}{\Upsilon}}_{k}^{(l)}
a_{k}^{-2l-1}
\right]
,
\label{newpotv}
\eeq

\noindent{with,} 

\begin{equation}
{\overset{ov}{T}}_{1}^{(l)} = 
\begin{cases}
-2 \pi \itG \Sigma_0 r_{0}^w 
\f{\beta_{l}^{(1,\f{1}{2})}}{2l + w - 1}
\po r_{e}^{-2l-w+1} - r_{f}^{-2l-w+1} \pc
,
&w \ne 1 - 2l \\
-2 \pi \itG \Sigma_0 r_{0}^{1-2l} 
\ln{\po \f{r_f}{r_e} \pc}
,
&w = 1 - 2l
,
\end{cases}
\label{labTov}
\end{equation}

\begin{equation}
{\overset{ov}{T}}_{2}^{(l)} = 
\begin{cases}
-2 \pi \itG \Sigma_0 r_{0}^w 
\f{\beta_{l}^{(1,\f{1}{2})}}{2l + w - 1}
\po r_{c}^{-2l-w+1} - r_{d}^{-2l-w+1} + r_{e}^{-2l-w+1} - r_{f}^{-2l-w+1} \pc
,
&w \ne 1 - 2l \\
-2 \pi \itG \Sigma_0 r_{0}^{1-2l} 
\ln{\po \f{r_d r_f}{r_c r_e} \pc}
,
&w = 1 - 2l
,
\end{cases}
\label{labTov2}
\end{equation}

\begin{equation}
{\overset{iv}{T}}_{1}^{(l)} = 
\begin{cases}
-2 \pi \itG \Sigma_0 r_{0}^w 
\f{\beta_{l}^{(1,\f{1}{2})}}{2l - w + 2}
\po  r_{d}^{2l-w+2} - r_{c}^{2l-w+2} + r_{b}^{2l-w+2} - r_{a}^{2l-w+2} \pc
,
&w \ne 2 + 2l \\
-2 \pi \itG \Sigma_0 r_{0}^{2+2l} 
\ln{\po \f{r_b r_d}{r_a r_c} \pc}
,
&w = 2 + 2l
,
\end{cases}
\label{labTiv}
\end{equation}

\begin{equation}
{\overset{iv}{T}}_{2}^{(l)} = 
\begin{cases}
-2 \pi \itG \Sigma_0 r_{0}^w 
\f{\beta_{l}^{(1,\f{1}{2})}}{2l - w + 2}
\po  r_{b}^{2l-w+2} - r_{a}^{2l-w+2} \pc
,
&w \ne 2 + 2l \\
-2 \pi \itG \Sigma_0 r_{0}^{2+2l} 
\ln{\po \f{r_b}{r_a} \pc}
,
&w = 2 + 2l
,
\end{cases}
\label{labTiv2}
\end{equation}

\noindent{and,}

\beq
{\overset{ov}{\Upsilon}}_{k}^{(l)}
\approx s_{k}^2 \po \f{1}{2} - \f{4l+3}{8} e_{k}^2 - \f{1}{16} e_{k}^4 \pc
+ \mathcal{O} \po s_{k}^2 e_{k}^6 \pc
,
\label{defupov}
\eeq

\beq
{\overset{iv}{\Upsilon}}_{k}^{(l)}
\approx s_{k}^2 \f{\f{1}{2} - \f{4l+1}{8} \po e_{k}^2 + \f{1}{2} e_{k}^4 \pc}
{\po e_{k}^2 - 1 \pc^{2l}}  + \mathcal{O} \po \f{s_{k}^2 e_{k}^6}{\po e_{k}^2 - 1 \pc^{2l}}\pc
.
\label{defupiv}
\eeq

\subsection{Application to GJ 876}

Due to the close sweeping orbits of the two resonant planets in GJ 876
and their proximity to the star, we consider only a disk exterior to the
outer planet ($r_a = r_b$, $r_c = r_d$, so that only the region labeled ``III''
exists).  The presence of the ``common gap'' in between the planets
has been shown to be the result of some numerical simulations of two planets 
and a disk \citep{bryetal2000b,kley2000}.  An exterior disk could 
represent the equilibrium state remnant from the last instance of giant 
planet migration.  
Because an accurate treatment of the short-period variations of the planetary
semimajor axes and eccentricities would entail the 
inclusion of Lindblad and corotation resonances for disk feedback, we focus
just on the disk-induced variations of the resonant angles.    
We take $r_e = 0.3$ AU and $r_f = 20$ AU, but through additional 
simulations find that the results are robust against the value 
of $r_f$, as little mass resides toward the outer disk edge 
for $w \ge 1$.  The value of $r_e$ was chosen to lie several 
Hill Radii away from the apocenter of the outer planet.
As the disk theory presented here is viable to third order,
all 11 resonant and secular arguments up to third-order were 
included in the disk simulations.

We find that disks as massive as the MMSN alters
libration widths of the dominant resonant angles on the order
of degrees and the circulation rates of the planets' longitude
of pericenters on the order of degrees per year.  The steeper
the surface density profile, the greater the effect.
Most disks perturb the planetary system enough to markedly
affect the orbital angle values, but not enough
to change their character.  However, a massive or steep
enough disk will warp the $\varpi$ profiles and transform
libration into circulation.  Figure \ref{libcir} \marginpar{FIG. 21}
demonstrates how this phenomenon occurs with $\chi = 1/3$,
$w = 3/2$, $r_e = 0.3$ AU and $r_f = 20$ AU.  The background
line represents the evolution of $\phi_2$ without a disk present,
and the foreground crosses were computed for the presence
of a disk.  Of further note is that the averaged
disk problem fails to reproduce this circulation.  The
circulation is achieved by the increase in amplitude of the
short-term oscillations in the modulated envelope at almost
$4$ yrs into the orbit, and repeats with subsequent librational
periods.

As $m_n \propto \Sigma_0 \propto \itR_k$, and hence ultimately
$\dot{\varpi}_k \propto m_n$, a direct relationship exists
between disk mass and the prospects for transforming libration
into circulation.  Equations (\ref{defA}), (\ref{ntotdervn6}) and
(\ref{Bval5}) indicate that $\dot{\varpi}_k \propto m_{0}^{-1/2}$
and Eqs. (\ref{defC2}) suggest $\dot{\varpi}_k \propto m_{3-k}$.
Hence, over time, as the disk primarily loses mass to the
central star, the effect on apsidal libration is muted, and
significant qualitative changes in apsidal behavior become
more unlikely.  However, transfer of disk material to either
or both resonant bodies might maintain a significant perturbation
on the apsidal angle.  Detailed explorations of the phase space
of regimes of mass loss with respect to apsidal libration
represents a possible future avenue of study.

\section{Conclusion} \label{conc}

We present a model which evolves any two objects in resonance, whose orbits 
don't cross, using only the resonant and secular argument defined by the user, and, 
optionally, including
central-body oblateness, central-body precession, and the presence of a surrounding
nascent disk.  Many careful studies of particular resonances rely on assumptions about 
what aspects of a system are relevant and necessary for inclusion.  Our model provides 
a useful tool for determining quantitative estimates of the contributions from each
argument and effect, and helps determine what must or should be included
when analytically studying an orbit-orbit resonance.  We also provide 
libration width analyses of relevant regions of phase space
(Figs. \ref{libamp32}-\ref{libampcur41}),
general explicit expressions for the variation of each resonant or secular argument 
considered (Eqs. \ref{phibarag2}-\ref{Edef}), constants of resonant motion entirely 
in terms of orbital elements (Eqs. \ref{hh}, 
\ref{constants12}-\ref{constants3h}),
averaged oblateness potential terms that were previously a source of confusion
in the literature (Eqs. \ref{gamdef}-\ref{gamexp4}), 
and gap-ridden disk potentials entirely in terms of orbital elements 
(Eqs. \ref{newpot}-\ref{Palmdef2} and \ref{newpotv}-\ref{defupiv}).

We apply several aspects of this model to the GJ 876 extrasolar planetary system, 
and conclude that at least a third-order treatment is necessary in order 
to mimic the qualitative evolution of the system.  A fourth-order treatment
improves the approximation in some but not all respects due to Sundman's
convergence criteria for Laplacian expansions, bringing into question
the viability of using ``order'' as the metric for quantifying the accuracy
of a resonant system.  The GJ 876 planets are negligibly
influenced by the oblateness and precession of the central star, but suggest
that given the right conditions in other exosystems, these effects may
play a role in the dynamical evolution.  A protoplanetary disk varies,
sometimes significantly, librational amplitudes and circulation rates
of extrasolar systems, and may even convert one type of motion into another.
Such dynamical flags may constrain unknown orbital parameters in
newly discovered exosystems.


\newpage

\appendix
\section{Appendix}

The quantity $f_{d}^{(i,p)}$ (Eq. \ref{fd}) is a function of both semimajor 
axes through the Laplace coefficients, $b_{s}^{(j)}$.  When expressed as
an infinite hypergeometric series, the coefficients may be
expressed as, for $j \ne 0$, 
(\citealt{brocle1961}, p. 495; \citealt{murder1999}, p. 237):

\begin{equation}
b_{s}^{(j)} = 
\beta_{1}^{(|j|,s)} \alpha^{|j|} + \beta_{2}^{(|j|,s)} \alpha^{|j|+2} + 
\beta_{3}^{(|j|,s)} \alpha^{|j|+4}   \dots =
\sum_{v=1}^{\infty} \beta_{v}^{(|j|,s)} 
                    \alpha^{|j|+2v-2}
\label{lapdef}
\end{equation}

\bigskip

\noindent{such} that

\beq
\beta_{i}^{(j,s)} = \f{ 2^{2-i} \po 2s + 2i - 4 \pc!!}
                       {\po i-1 \pc! \po |j| + i - 1 \pc! \po 2s-2 \pc!!}
                       \left[ \prod_{v = s}^{v = s + i + |j| - 2} v \right]
.
\label{betadef}
\eeq

For the case of $j = 0$ we can better compute the coefficients
by using the following formula:

\beq
\beta_{i}^{(0,s)} = 2 \left[    
\prod_{v = 2}^{v = i}
\f{s + v - 2}{v - 1}
\right]^2 
,
\bs
 i > 2
\label{betadef2}
\eeq

\noindent{with} $\beta_{1}^{(0,s)} = 2$.

By expressing $f_d$ as a polynomial in
$\alpha \equiv a_2/a_1$, we can 
then take the necessary analytical partial
derivatives of $\itC_{k}^{(i,p)}$ and
avoid integration when computing Laplace
coefficients.  The differential operator 
$\itD$ acts on $\alpha$ such that for $Y$th-order  
eccentricity-type resonances,

\beq
\sum_{y=0}^{Y} \Phi_{A,j}^{(y)}
\alpha^y \itD^y b_{s}^{(j)}
=
\sum_{y=0}^{Y} \Phi_{A,j}^{(y)}
\sum_{v=1}^{\infty} 
\left[
\prod_{i=2}^{y+1} \po |j| + 2v - i \pc
\right]
\beta_{v}^{(|j|,\f{1}{2})} \alpha^{|j|+2v-2}
,
\eeq

\noindent{where} $\Phi_{A,j}^{(y)}$ represents the function of $j$ corresponding
to the particular term in the resonance which may be read off from Appendix
B of \cite{murder1999}.  
The term in square brackets 
equals unity when $y = 0$.  For each value of $v$, the corresponding 
coefficient of $\alpha$ may be computed.  For inclination resonances, we
also need to compute:

\beq
\sum_{t=1}^T
\sum_{y=1}^{Y-1} \Phi_{B,j}^{(y-1),t}
\alpha^y \itD^{y-1} b_{s}^{(j)}
=
\sum_{t=1}^T
\sum_{y=0}^{Y-1} \Phi_{B,j}^{(y-1),t}
\sum_{v=1}^{\infty} 
\left[
\prod_{i=2}^{y} \po |j| + 2v - i \pc
\right]
\beta_{v}^{(|j|,\f{3}{2})} \alpha^{|j|+2v-1}
,
\eeq

\noindent{}where $t$ is a counter for the
linear combination of up to $T$ terms which
may appear in the expression for a given
inclination term (for example, $f_9$).  Similarly,
we have:

\beq
\sum_{t=1}^T
\sum_{y=2}^{Y-2} \Phi_{C,j}^{(y-2),t}
\alpha^y \itD^{y-2} b_{s}^{(j)}
=
\sum_{t=1}^T
\sum_{y=2}^{Y-2} \Phi_{C,j}^{(y-2),t}
\sum_{v=1}^{\infty} 
\left[
\prod_{i=2}^{y-1} \po |j| + 2v - i \pc
\right]
\beta_{v}^{(|j|,\f{5}{2})} \alpha^{|j|+2v}
.
\eeq

We can finally express $\kappa_{l}^{(i,p)}$, for a particular term, as:

\begin{equation}
\kappa_{l}^{(i,p)} =
     \begin{cases}
           0,
           &\text{for $l < |j|$} \\
                                 \\
           \sum_{y=0}^{Y} \Phi_{A,j}^{(y)} 
           \beta_{1}^{(|j|,\f{1}{2})}
           \prod_{i=2}^{y+1} \po l + 2 - i \pc
           ,
           &\text{for $l = |j|$}     \\
                                    \\
           \sum_{t=1}^T
           \sum_{y=1}^{Y-1} \Phi_{B,j}^{(y-1),t} 
           \beta_{\f{l-|j|+1}{2}}^{(|j|,\f{3}{2})}
           \prod_{i=2}^{y} \po l + 1 - i \pc
           ,
           &\text{for $l - |j| > 0$ and odd} \\
                                   \\
           \sum_{y=0}^{Y} \Phi_{A,j}^{(y)} 
           \beta_{\f{l-|j|+2}{2}}^{(|j|,\f{1}{2})}
           \prod_{i=2}^{y+1} \po l + 2 - i   \pc + 
            \\
           \sum_{t=1}^T
           \sum_{y=2}^{Y-2} \Phi_{C,j}^{(y-2),t} 
           \beta_{\f{l-|j|}{2}}^{(|j|,\f{5}{2})}
           \prod_{i=2}^{y-1} \po l - i \pc
           ,
           &\text{for $l - |j| > 0$ and even}
.
\end{cases}
\label{fpandpk2}
\end{equation}

\newpage

\section*{Acknowledgments}

I wish to thank Michael Efroimsky, Cristi\'{a}n Beaug\'{e}, Alessandro Morbidelli and an anonymous referee for valuable guidance and advice, my advisor Phil Armitage for affording me the time to pursue this endeavor and for reviewing the manuscript, Larry Esposito, Glen Stewart, and the rest of the Colorado Rings Group for their constant support, David Nesvorn\'{y} for introducing me to the Sundman criterion, James Meiss and the Dynamical Systems Group for entertaining my idea, Juri Toomre for reading the manuscript, and Re'em Sari for a beneficial discussion.  I gratefully acknowledge support from the National Science Foundation under grant AST~0407040, and from NASA under grant NAG5-13207 issued through the Office of Space Science.

\newpage

\bibliography{refs}

\newpage

\section*{Figure and Table Captions}

\begin{figure}

\centerline{\psfig{figure=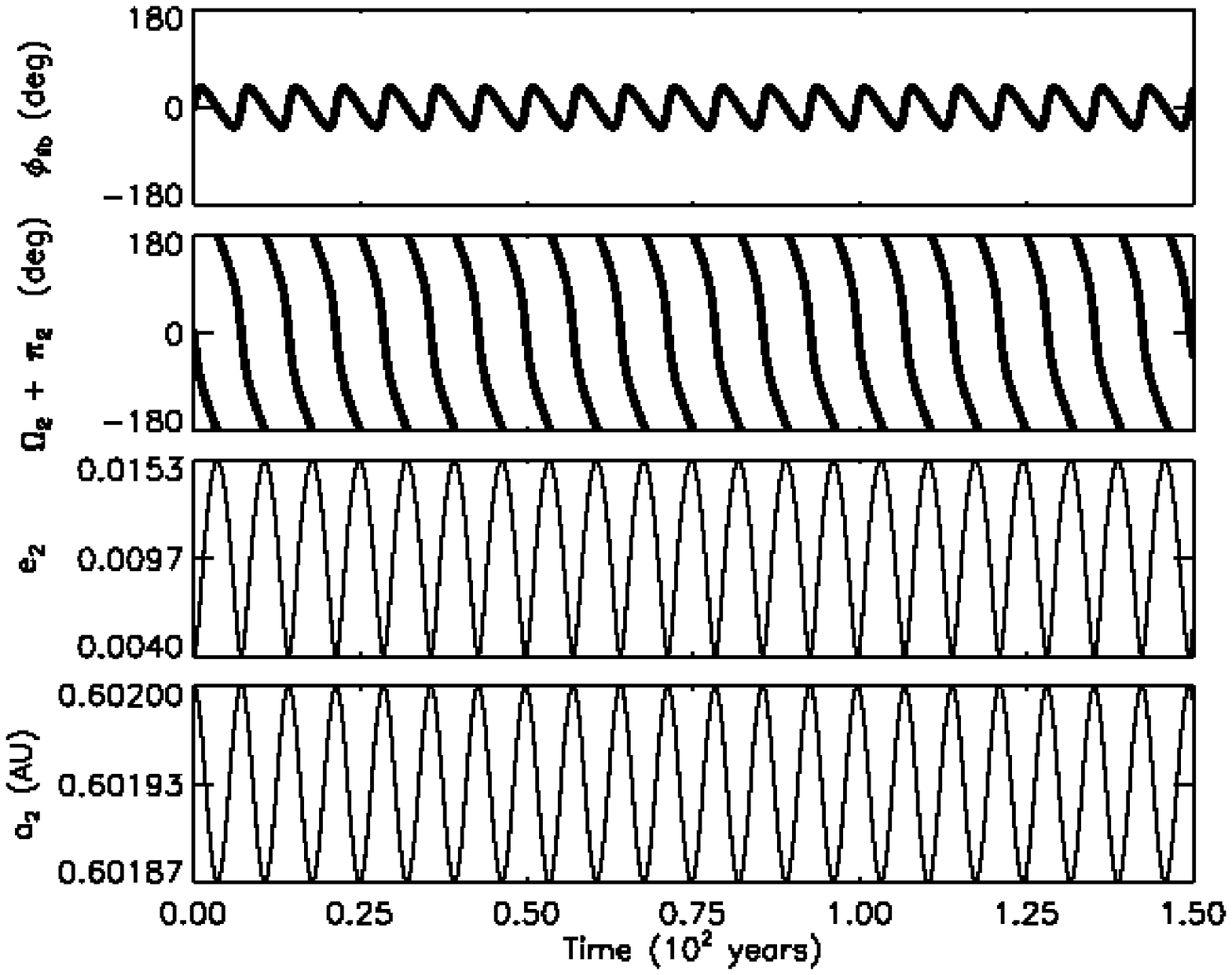,width=8.0truein,height=8.0truein}}

\caption{}\label{astecc}
\end{figure}

Figure \ref{astecc}: Evolution of a Jovian asteroid on an eccentric orbit.
The resonant angle $\phi_{\text{lib}} \equiv 
{\lbrace 2, -1, 0, -1, 0, 0  \rbrace}$.

\bigskip

\begin{figure}

\centerline{\psfig{figure=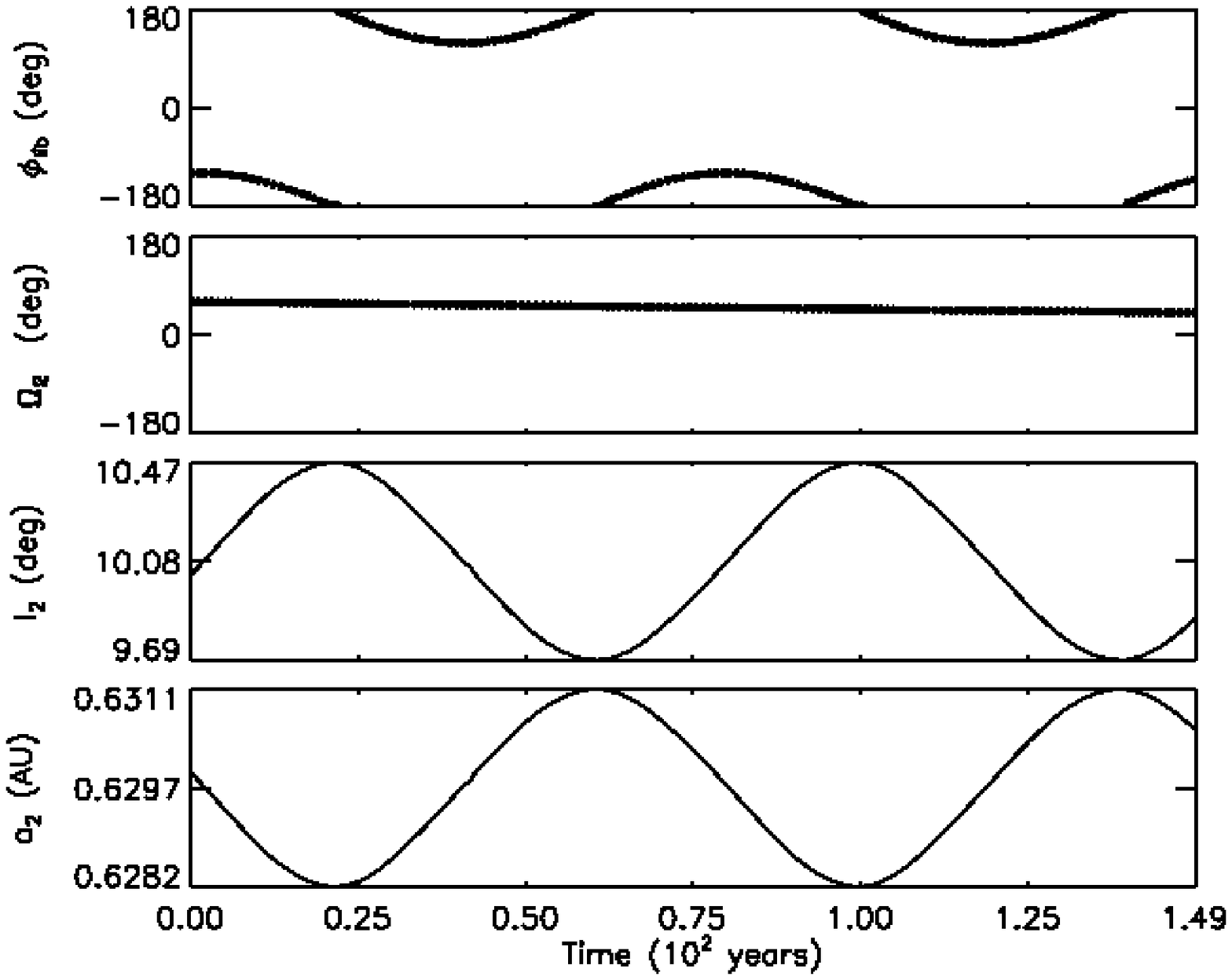,width=8.0truein,height=8.0truein}}

\caption{}\label{astinc}
\end{figure}

Figure \ref{astinc}: Evolution of a Jovian asteroid on an inclined orbit.
The resonant angle $\phi_{\text{lib}} \equiv 
{\lbrace 4, -2, 0, 0, 0, -2  \rbrace}$.

\bigskip

\begin{figure}

\centerline{\psfig{figure=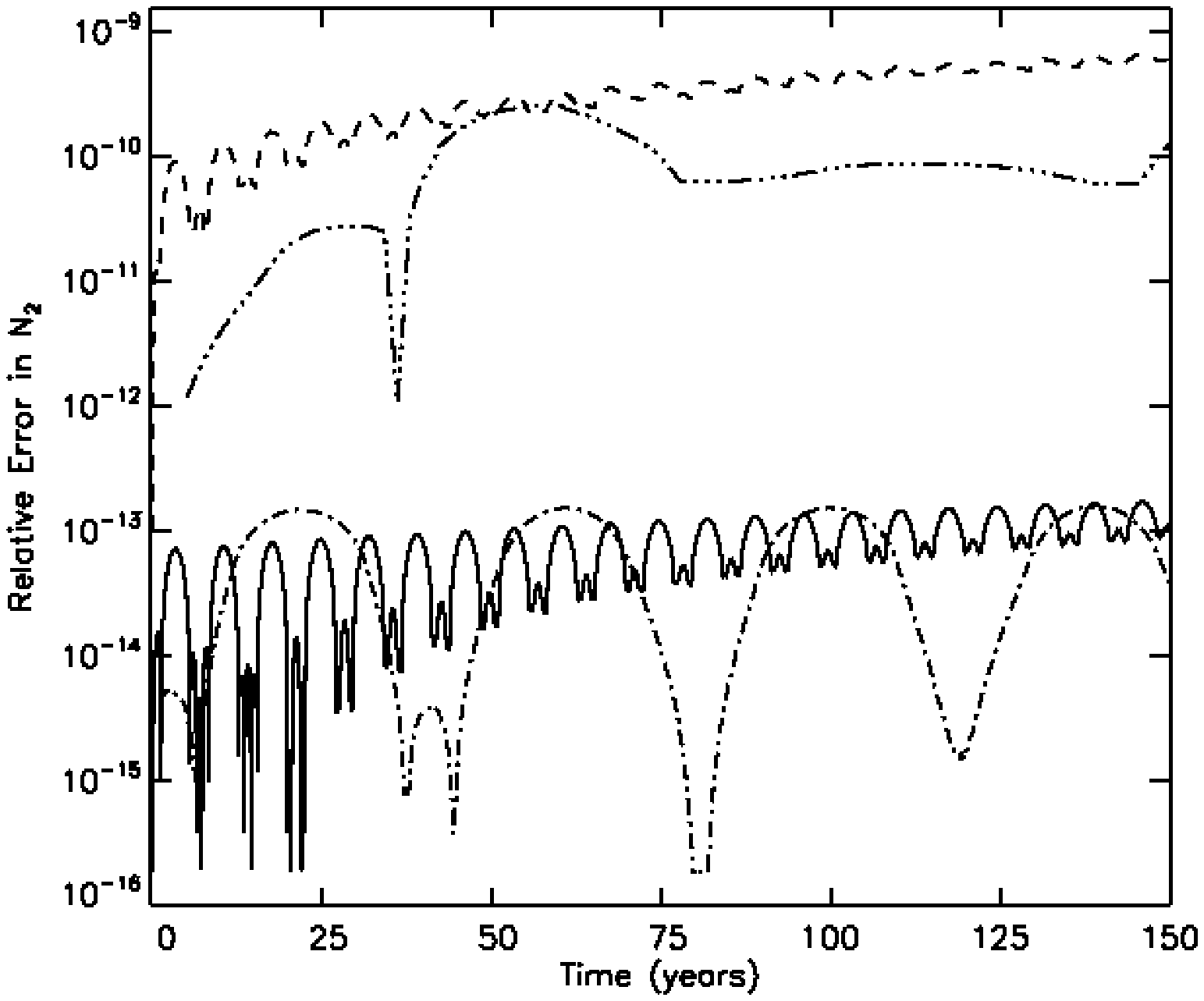,width=8.0truein,height=8.0truein}}

\caption{}\label{N2con}
\end{figure}

Figure \ref{N2con}: Relative error in $N_2$ for the asteroid evolution in 
Fig. \ref{astecc} for user-inputted integration accuracy parameters of $10^{-8}$ 
(solid) and $10^{-5}$ (dashed), and for the evolution in
Fig. \ref{astinc} for accuracy parameters of $10^{-8}$ 
(dashed-dot) and $10^{-5}$ (dashed-dot-dot-dot).

\bigskip

\begin{figure}

\centerline{\psfig{figure=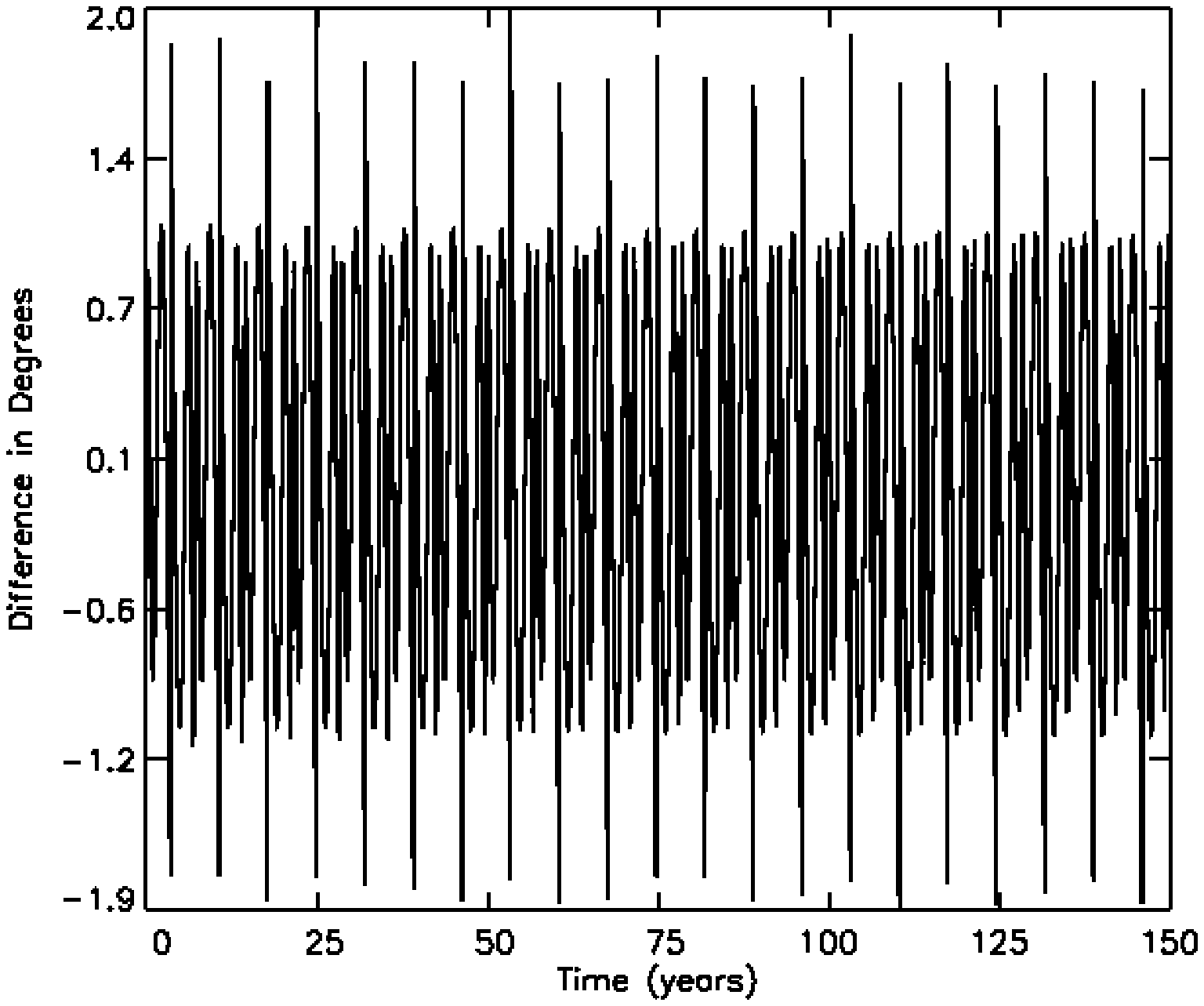,width=8.0truein,height=8.0truein}}

\caption{}\label{fitdevecc}
\end{figure}

Figure \ref{fitdevecc}: Deviation of the libration profile in 
Fig. \ref{astecc} from the model of Eq. (\ref{dopfit}) using
coefficients in Table \ref{astfit2}.

\bigskip

\begin{figure}

\centerline{\psfig{figure=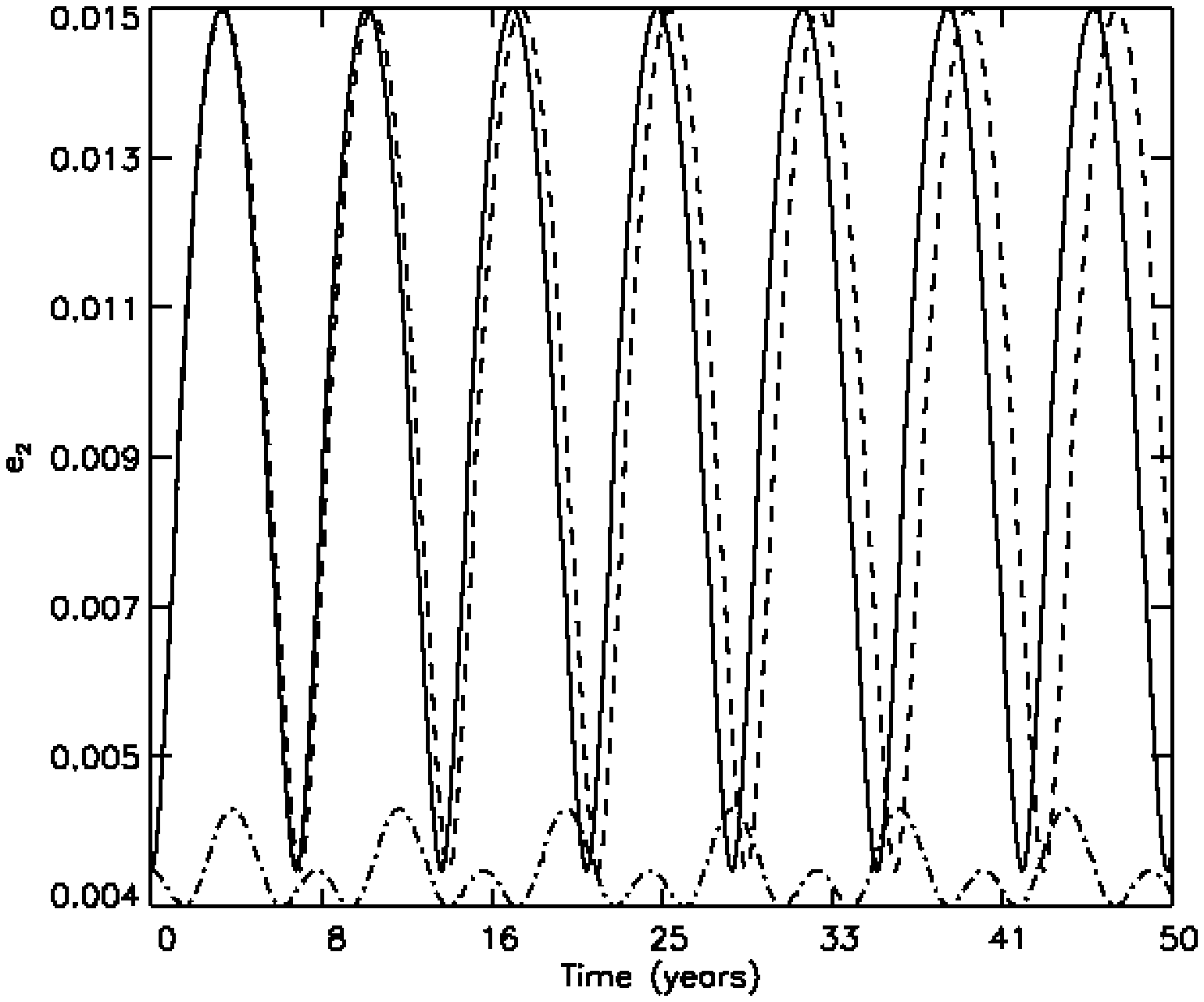,width=8.0truein,height=8.0truein}}

\caption{}\label{difastecc}
\end{figure}

Figure \ref{difastecc}: The eccentricity profile
in Fig. \ref{astecc} (solid line) and those derived with all terms
up to second-order from the arguments $\lbrace 2,-1,0,-1,0,0 \rbrace$,
$\lbrace 4,-2,0,-2,0,0 \rbrace$,
and $\aleph$ (dashed line) and with all 
terms up to fourth-order from the arguments
$\lbrace 2,-1,0,-1,0,0 \rbrace$,
$\lbrace 4,-2,0,-2,0,0 \rbrace$,
$\lbrace 6,-3,0,-3,0,0 \rbrace$,
$\lbrace 8,-4,0,-4,0,0 \rbrace$,
and $\aleph$ (dot-dashed line).

\bigskip

\begin{figure}

\centerline{\psfig{figure=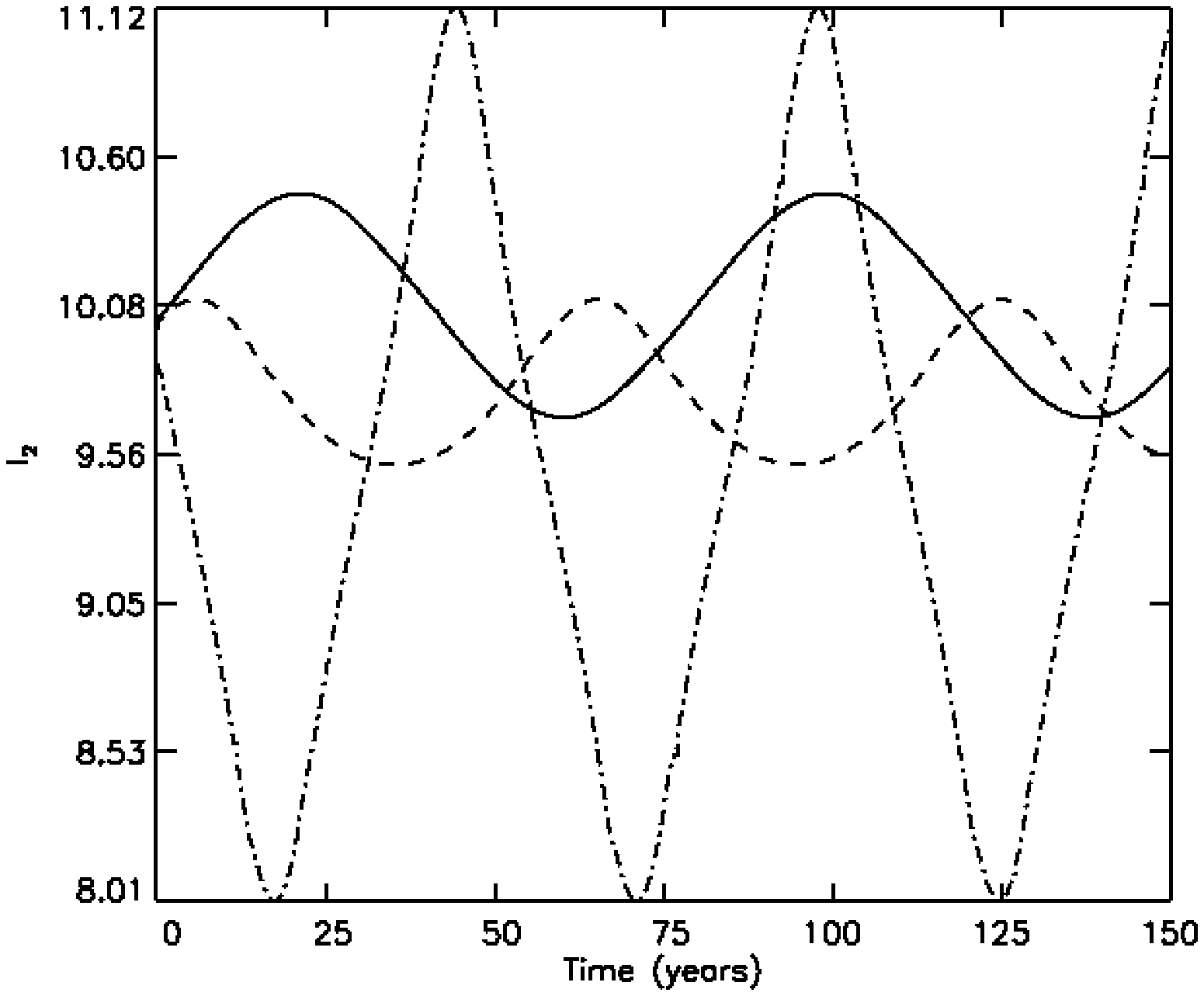,width=8.0truein,height=8.0truein}}

\caption{}\label{difastinc}
\end{figure}

Figure \ref{difastinc}: The inclination profile
in Fig. \ref{astinc} (solid line) and those derived with all 
terms up to second-order from the arguments 
$\lbrace 4,-2,0,0,0,-2 \rbrace$ and   
$\aleph$ (dashed line) and with 
all terms up to fourth-order from the arguments
$\lbrace 4,-2,0,-2,0,0 \rbrace$,
$\lbrace 8,-4,0,-4,0,0 \rbrace$,
and $\aleph$ (dot-dashed line).

\bigskip

\begin{figure}

\centerline{\psfig{figure=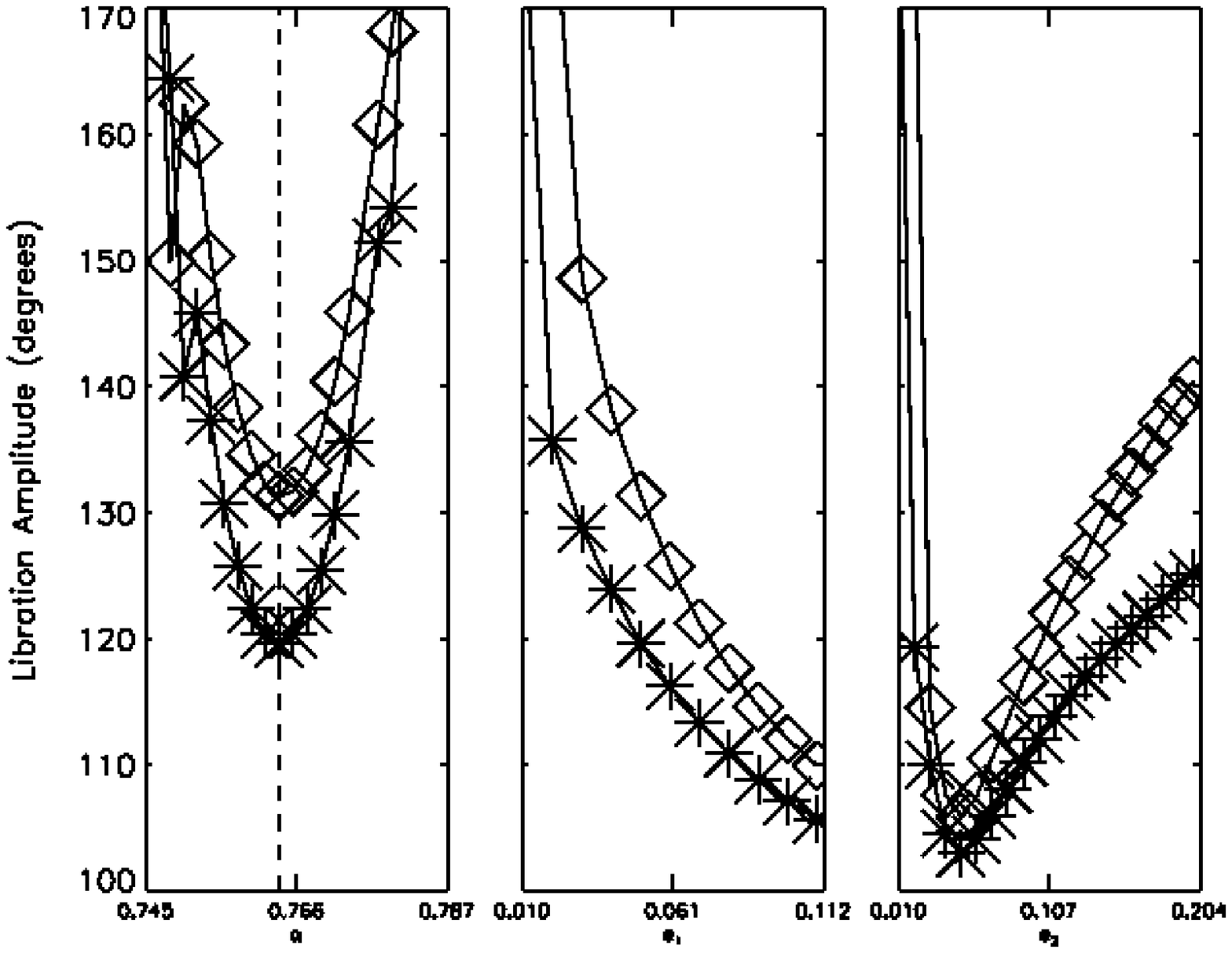,width=8.0truein,height=8.0truein}}

\caption{}\label{libamp32}
\end{figure}

Figure \ref{libamp32}:  Libration amplitudes for 
two planets in a $3$:$2$ resonance with parameters that vary
from the state given in Table \ref{intparm}.  
Diamonds and asterisks represent half the variation
exhibited by the $\lbrace 3,-2,0,-1,0,0 \rbrace$
and $\lbrace 3,-2,-1,0,0,0 \rbrace$ arguments
respectively for individual simulations.  The
dashed line indicates a ``nominal'' 
$\alpha$ value of $(3/2)^{(-2/3)}$.

\bigskip

\begin{figure}[htbp]

\centerline{\psfig{figure=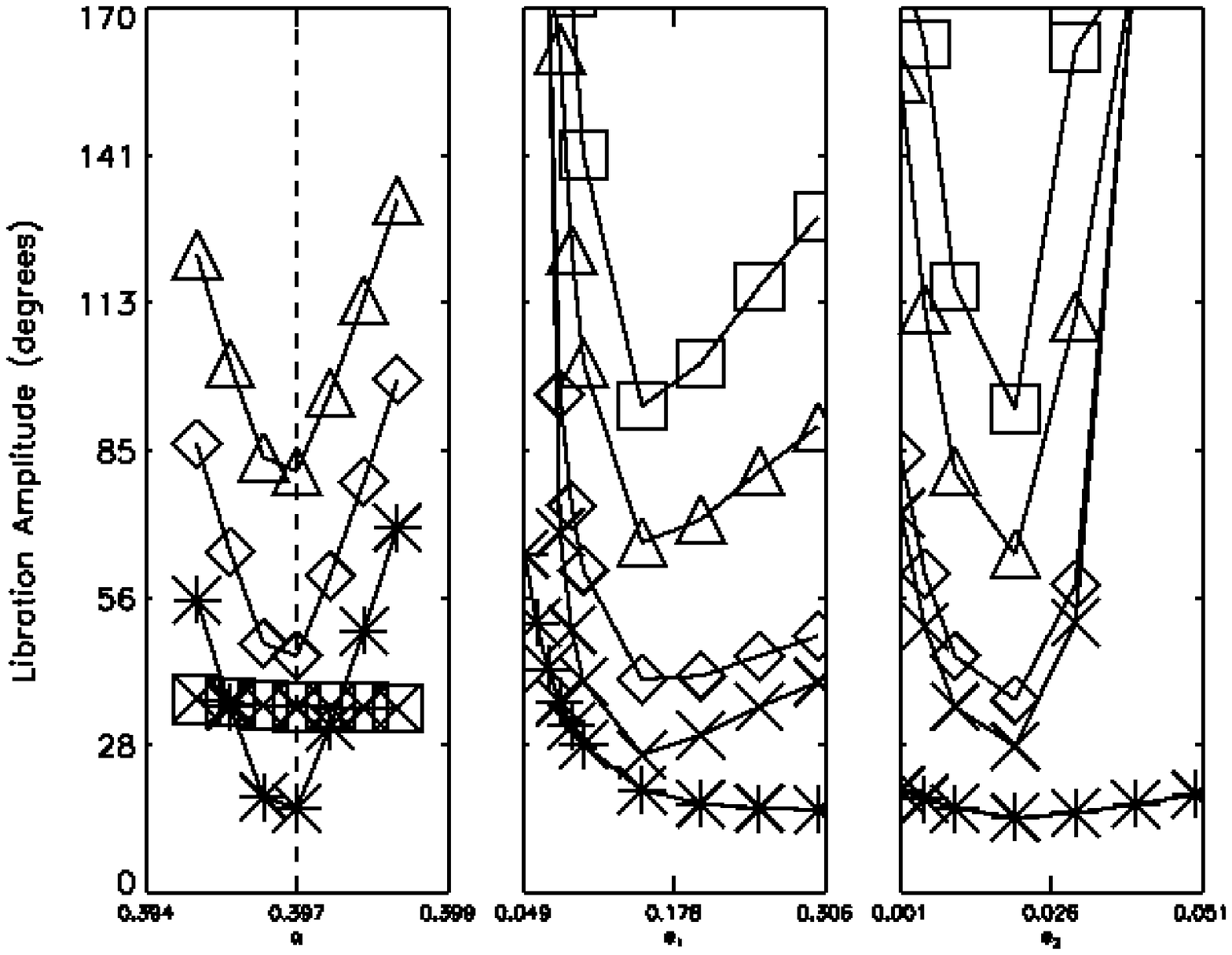,width=8.0truein,height=8.0truein}}

\caption{}\label{libamp41}
\end{figure}

Figure \ref{libamp41}:  Libration amplitudes for 
two planets in a $4$:$1$ resonance with parameters that vary
from the state given in Table \ref{intparm}.  
Crosses, squares, triangles, diamonds and
asterisks represent half the variation
exhibited by the 
$\lbrace 0,0,1,-1,0,0 \rbrace$,
$\lbrace 4,-1,0,-3,0,0 \rbrace$,
$\lbrace 4,-1,-1,-2,0,0 \rbrace$,
$\lbrace 4,-1,-2,-1,0,0 \rbrace$,
$\lbrace 4,-1,-3,0,0,0 \rbrace$
 arguments
respectively for individual simulations.  The
dashed line indicates a ``nominal'' 
$\alpha$ value of $(4/1)^{(-2/3)}$.

\bigskip

\begin{figure}

\centerline{\psfig{figure=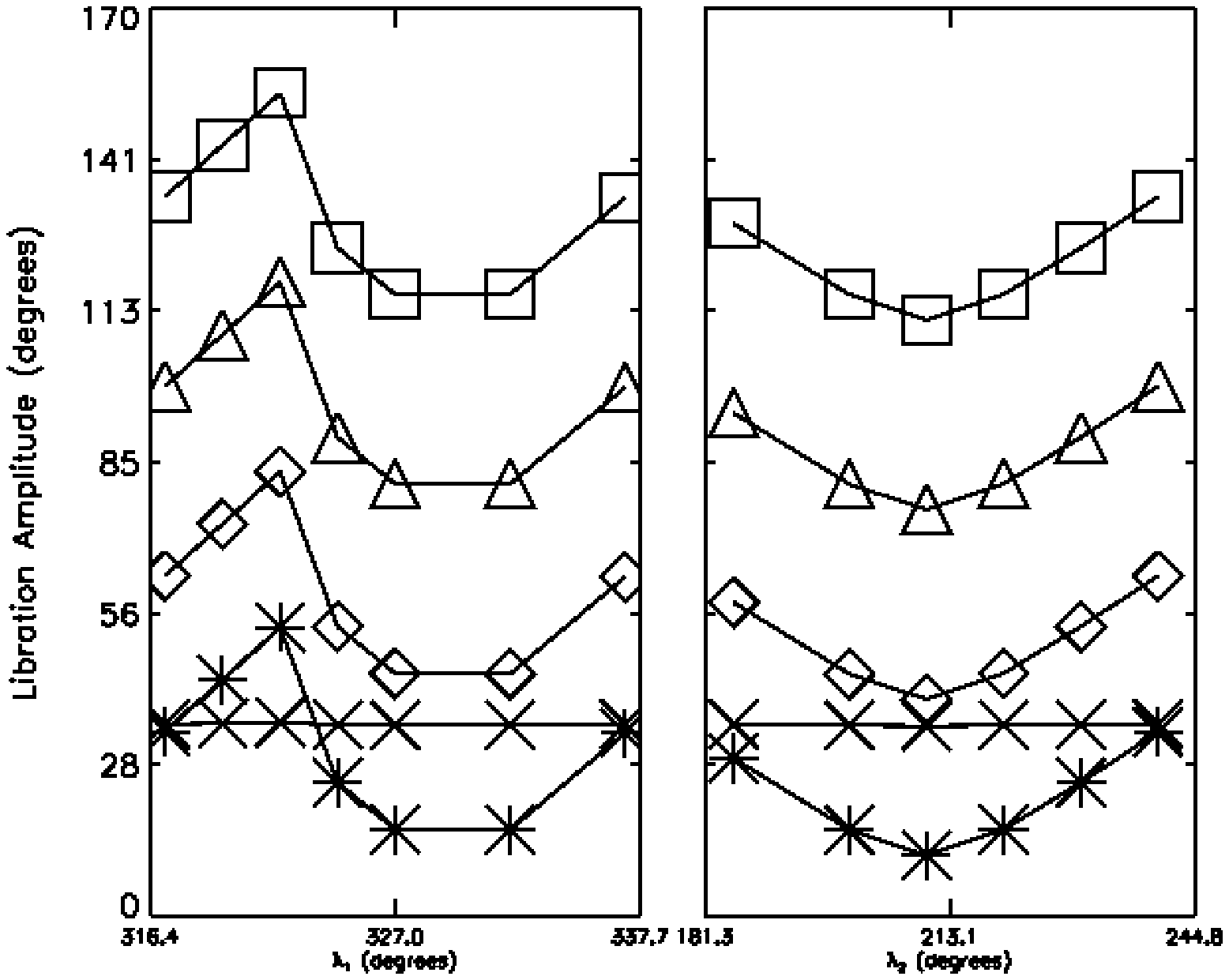,width=8.0truein,height=8.0truein}}

\caption{}\label{libamplam41}
\end{figure}

Figure \ref{libamplam41}:  Same as Fig. \ref{libamp41} but
where $\lambda_1$ and $\lambda_2$ are varied.

\bigskip

\begin{figure}

\centerline{\psfig{figure=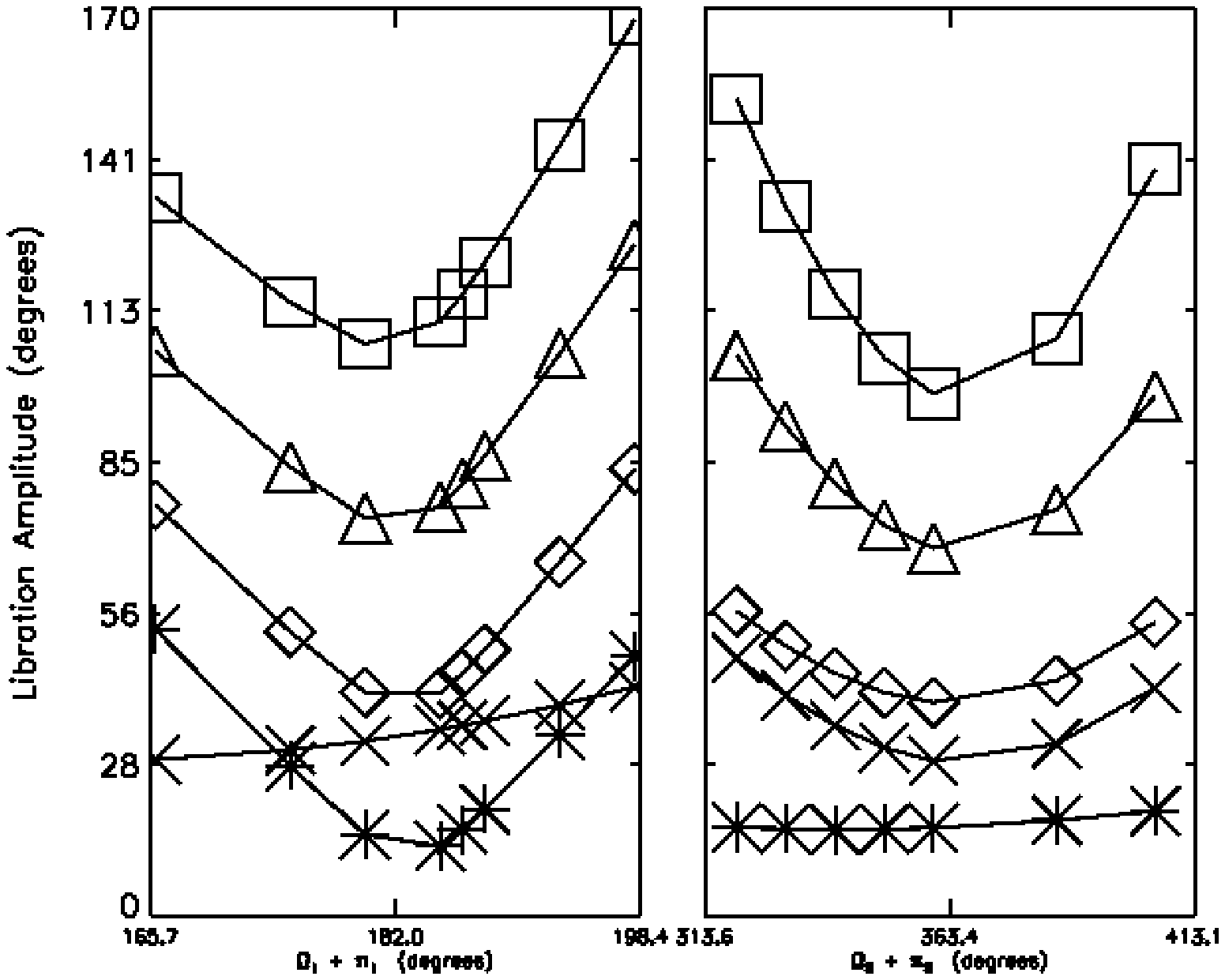,width=8.0truein,height=8.0truein}}

\caption{}\label{libampcur41}
\end{figure}

Figure \ref{libampcur41}:  Same as Fig. \ref{libamp41} but
where $\varpi_1$ and $\varpi_2$ are varied.

\bigskip

\begin{figure}

\centerline{\psfig{figure=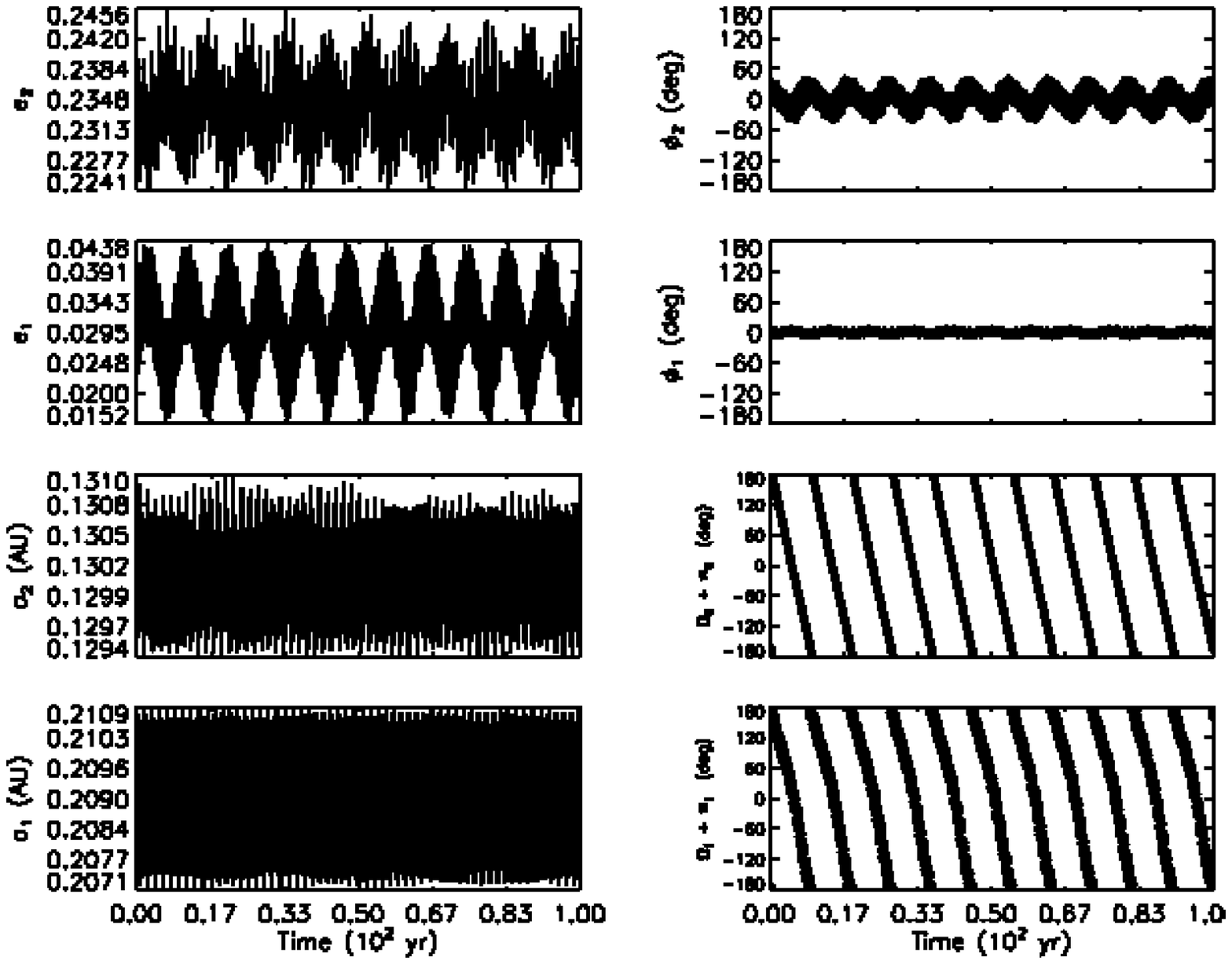,width=8.0truein,height=8.0truein}}

\caption{}\label{fullint}
\end{figure}

Figure \ref{fullint}: GJ 876 b's and c's evolution from a full N-body integration.  
The angles $\phi_1$ and $\phi_2$ undergo libration about $0^{\circ}$,
and $\varpi_1$ and $\varpi_2$ undergo circulation.

\bigskip

\begin{figure}

\centerline{\psfig{figure=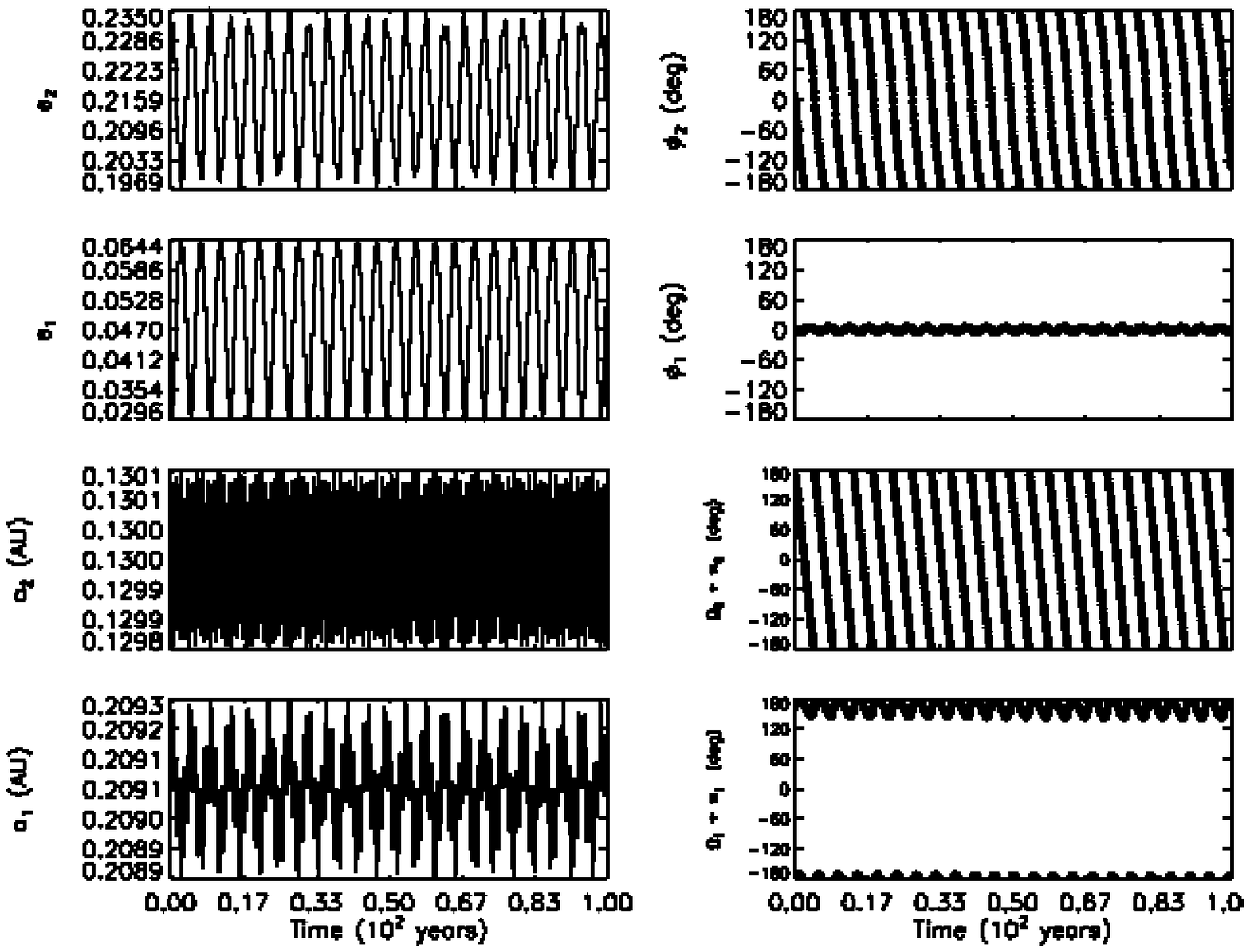,width=8.0truein,height=8.0truein}}

\caption{}\label{order1}
\end{figure}

Figure \ref{order1}: GJ 876 b's and c's evolution due to all resonant and secular
terms up to first order in eccentricities.

\bigskip

\begin{figure}

\centerline{\psfig{figure=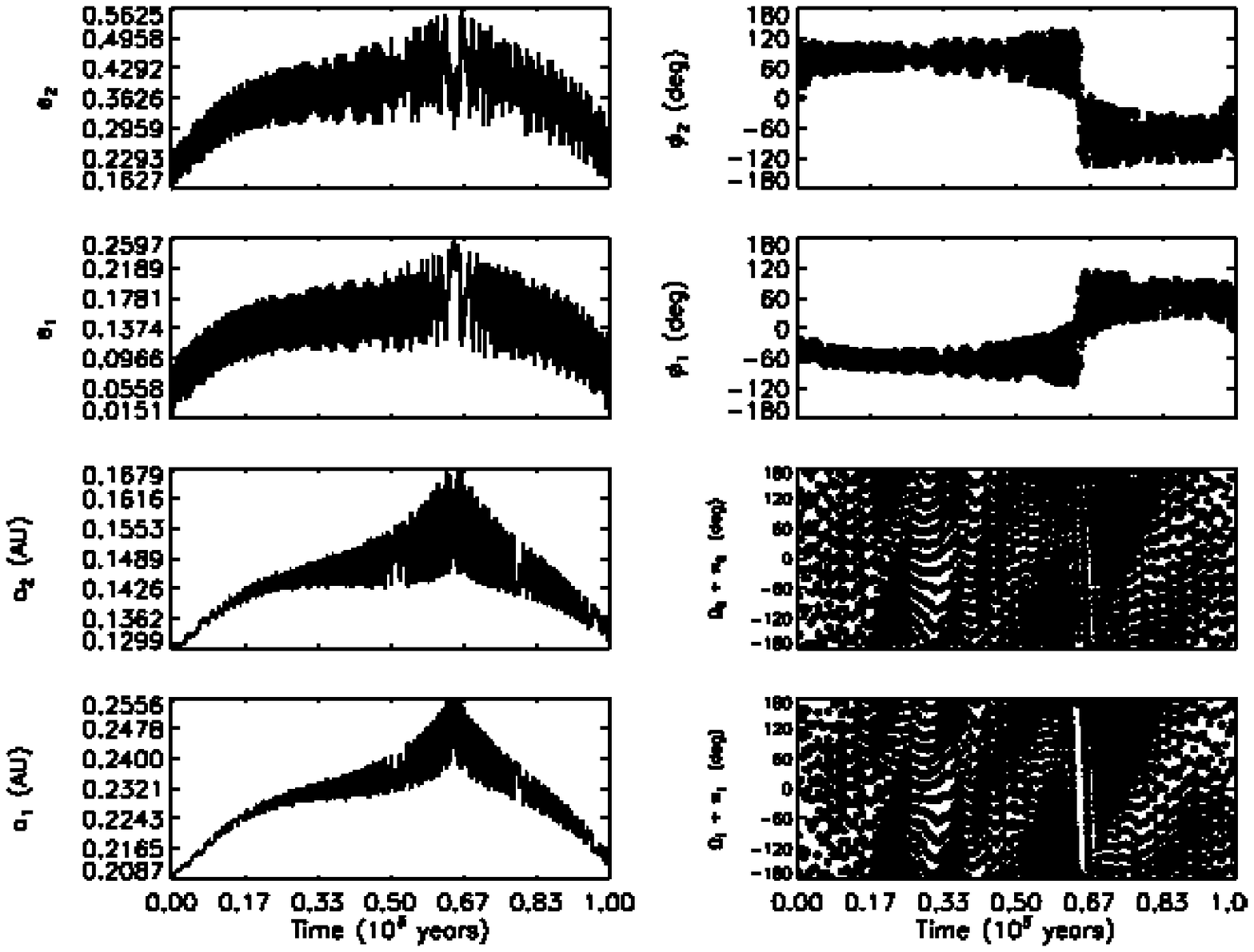,width=8.0truein,height=8.0truein}}

\caption{}\label{order2}
\end{figure}

Figure \ref{order2}: GJ 876 b's and c's evolution due to all resonant and secular
terms up to second order in eccentricities.

\bigskip

\begin{figure}

\centerline{\psfig{figure=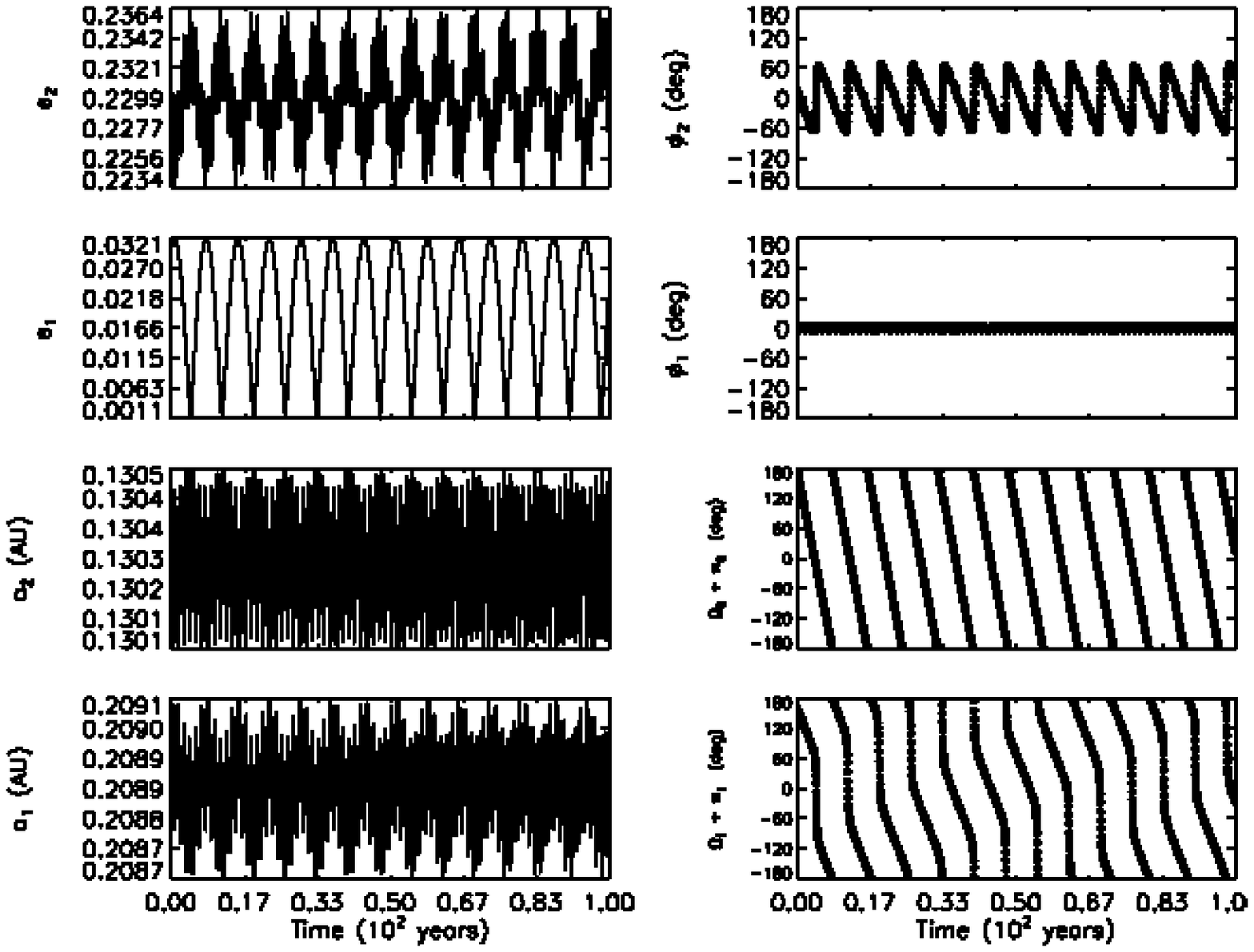,width=8.0truein,height=8.0truein}}

\caption{}\label{order3}
\end{figure}

Figure \ref{order3}: GJ 876 b's and c's evolution due to all resonant and secular
terms up to third order in eccentricities.

\bigskip

\begin{figure}

\centerline{\psfig{figure=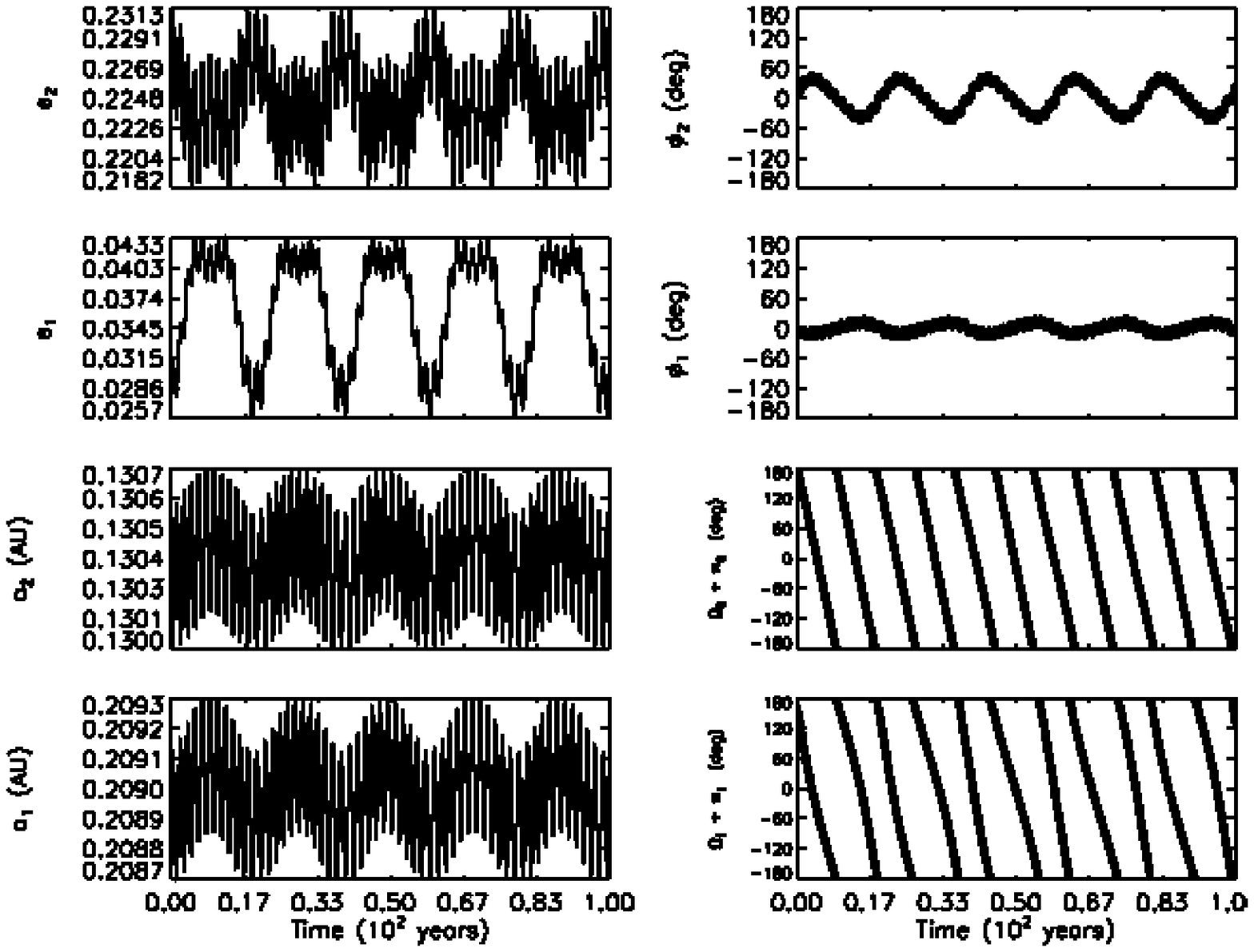,width=8.0truein,height=8.0truein}}

\caption{}\label{order4}
\end{figure}

Figure \ref{order4}: GJ 876 b's and c's evolution due to all resonant and secular
terms up to fourth order in eccentricities.

\bigskip

\begin{figure}

\centerline{\psfig{figure=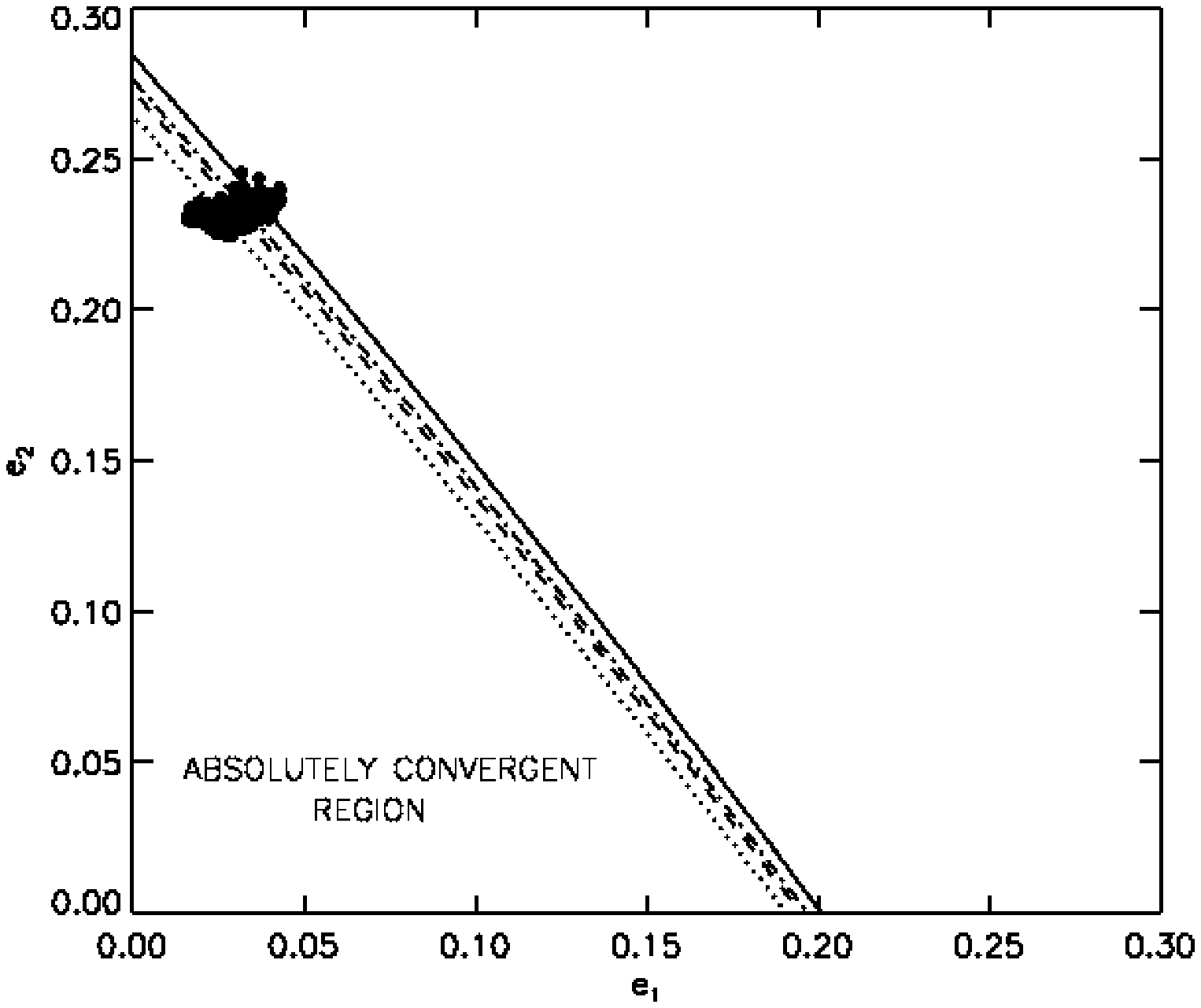,width=8.0truein,height=8.0truein}}

\caption{}\label{sundcritreal}
\end{figure}

Figure \ref{sundcritreal}: Laplacian convergence region for 
minimum values of $a_1$ and $a_2$ (dashed lines),
maximum values of $a_1$ and $a_2$ (dash-dot lines),
minimum value of $a_1$ and maximum value of $a_2$ (dotted lines),
and
maximum value of $a_1$ and minimum value of $a_2$ (solid lines),
for eccentricity values (dots) achieved during the simulation.

\bigskip

\begin{figure}

\centerline{\psfig{figure=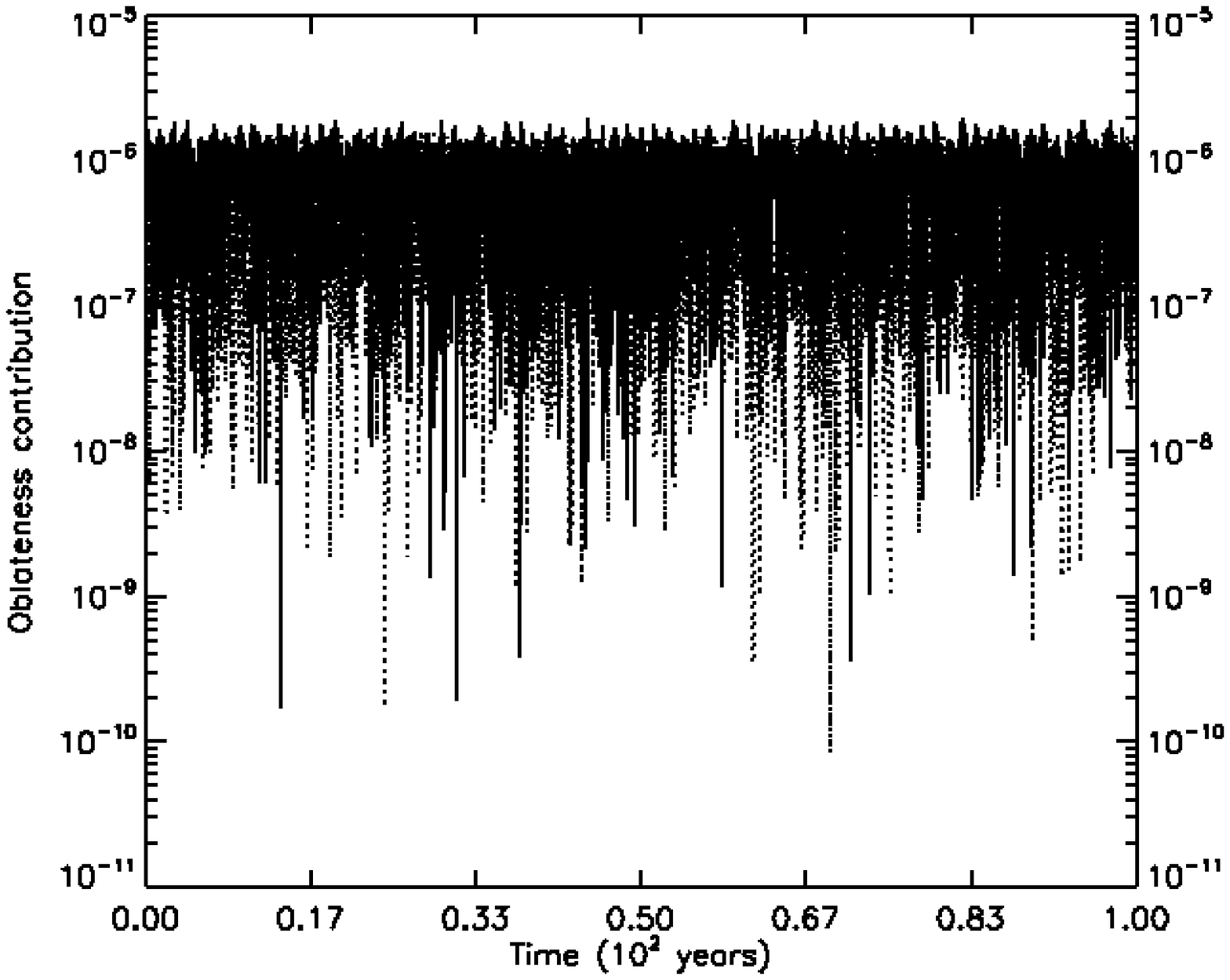,width=8.0truein,height=8.0truein}}

\caption{}\label{oblna}
\end{figure}

Figure \ref{oblna}: Central-body oblateness contribution to system evolution
per timestep expressed as a fraction of the unperturbed values of 
$\dot{\varpi}_1$ (solid line)
and $\dot{\varpi}_2$ (dotted line) assuming values of
$J_2 = -10^{-6}$, $J_4 = 10^{-12}$ and $R_{\odot} = 6 \times 10^5$ km
in the unaveraged problem.

\bigskip

\begin{figure}

\centerline{\psfig{figure=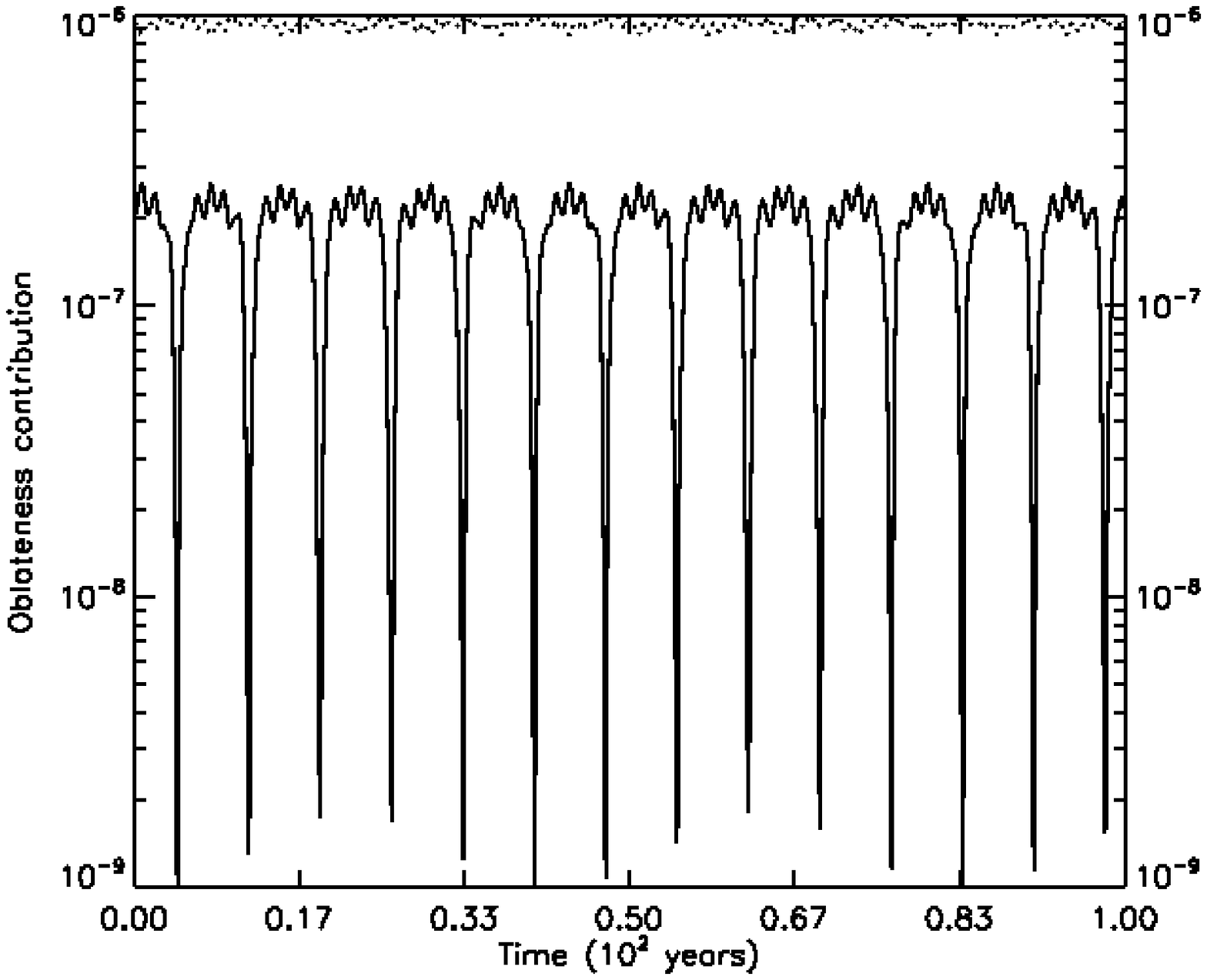,width=8.0truein,height=8.0truein}}

\caption{}\label{obla}
\end{figure}

Figure \ref{obla}: Central-body oblateness contribution to system evolution
per timestep expressed as a fraction of the unperturbed values of 
$\dot{\varpi}_1$ (solid line)
and $\dot{\varpi}_2$ (dotted line) assuming values of
$J_2 = -10^{-6}$, $J_4 = 10^{-12}$ and $R_{\odot} = 6 \times 10^5$ km
in the averaged problem.

\bigskip

\begin{figure}

\centerline{\psfig{figure=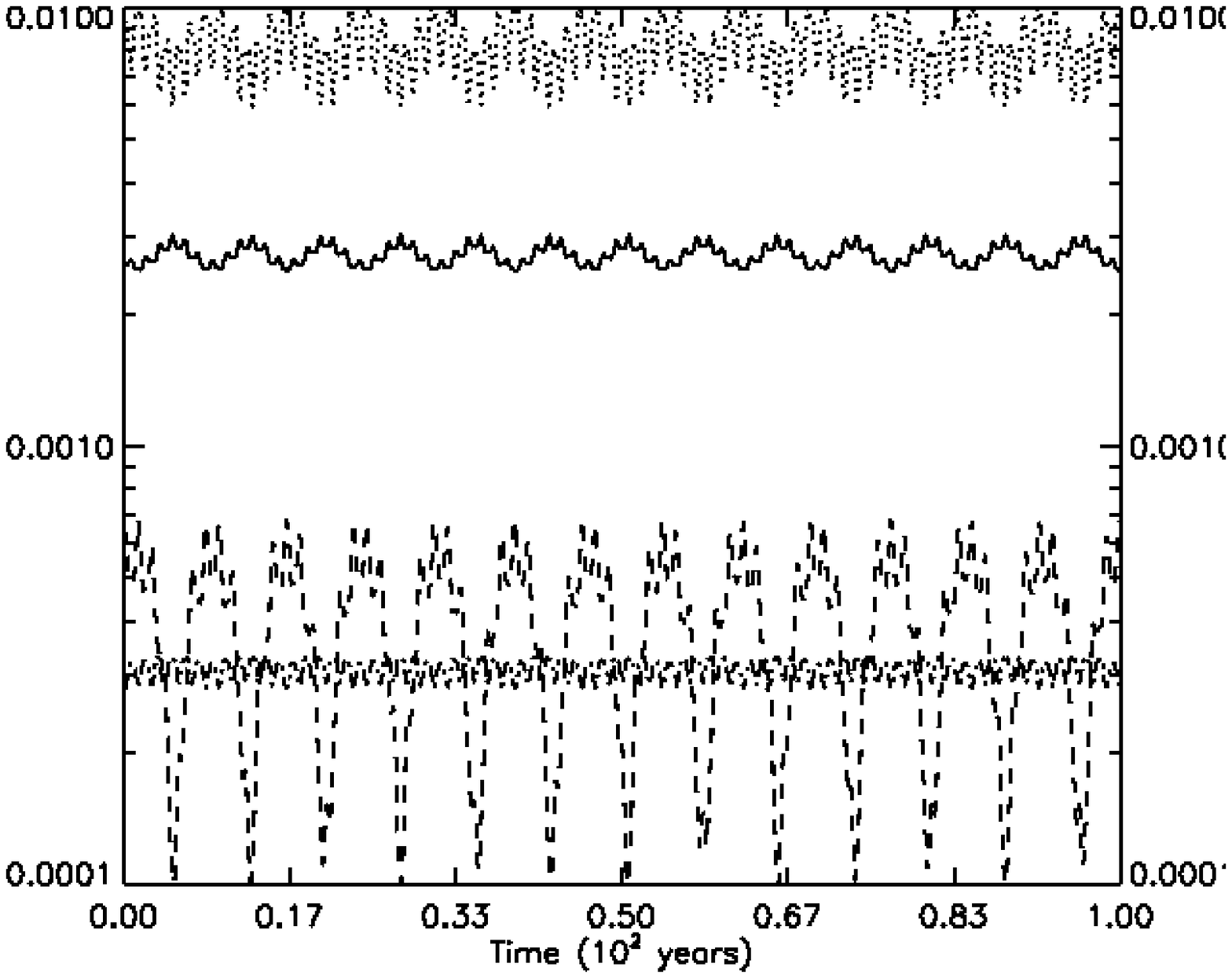,width=8.0truein,height=8.0truein}}

\caption{}\label{prena}
\end{figure}

Figure \ref{prena}: Precessional contribution to system evolution per timestep
expressed as a 
fraction of the unperturbed values of $\dot{\epsilon}_1$ (solid line), 
$\dot{\epsilon}_2$ (dotted line), $\dot{\varpi}_1$ (dashed line),
and $\dot{\varpi}_2$ (dash-dot line).  Values of 
$\omega_1 = \omega_2 = \omega_3 = 10$ pHz and $R_{\odot} = 6 \times 10^5$ km
are assumed.

\bigskip

\begin{figure}

\centerline{\psfig{figure=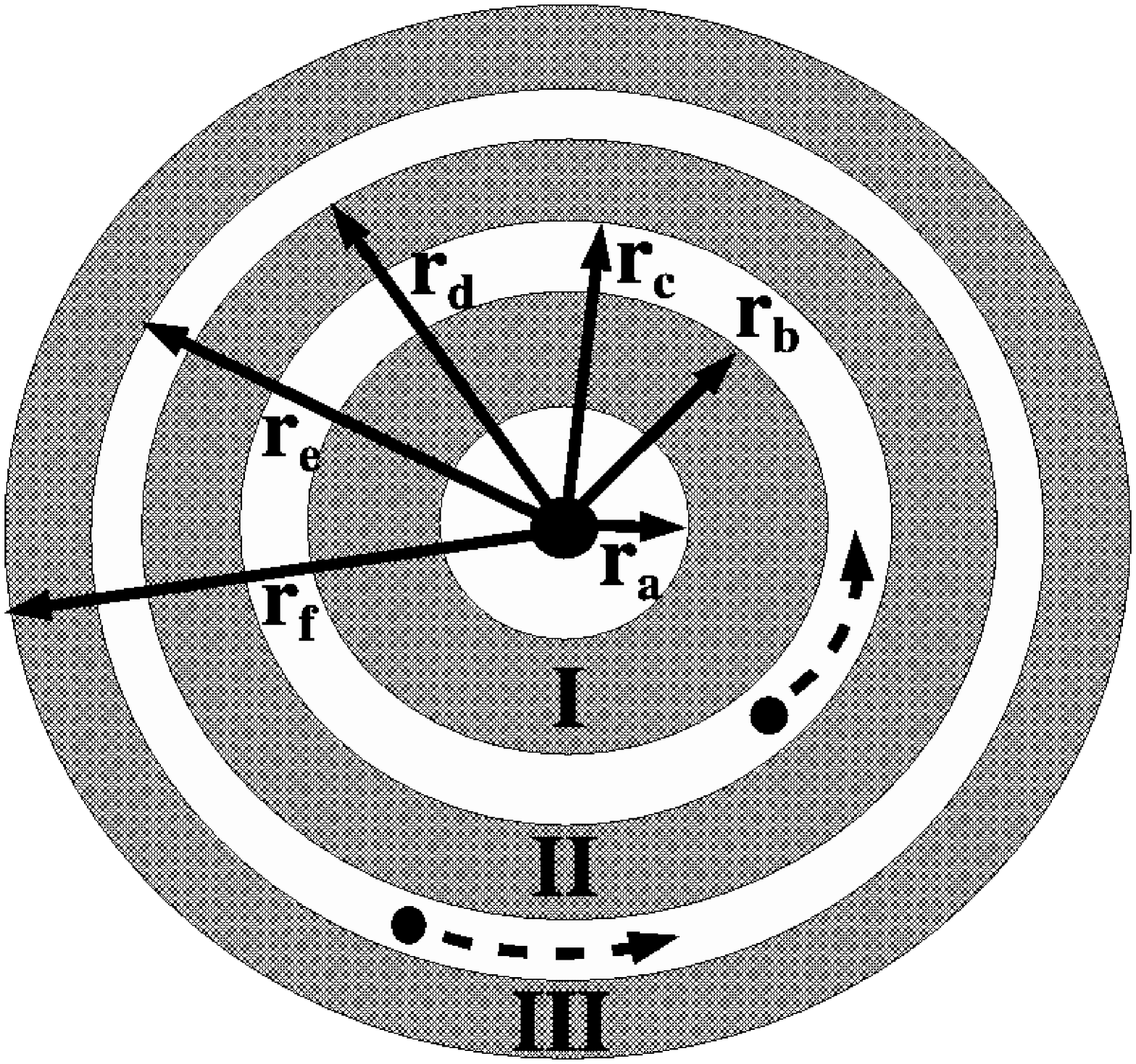,width=8.0truein,height=8.0truein}}

\caption{}\label{cartoon}
\end{figure}

Figure \ref{cartoon}: Cartoon of two planets embedded in a surrounding thin disk with gaps.

\bigskip

\begin{figure}

\centerline{\psfig{figure=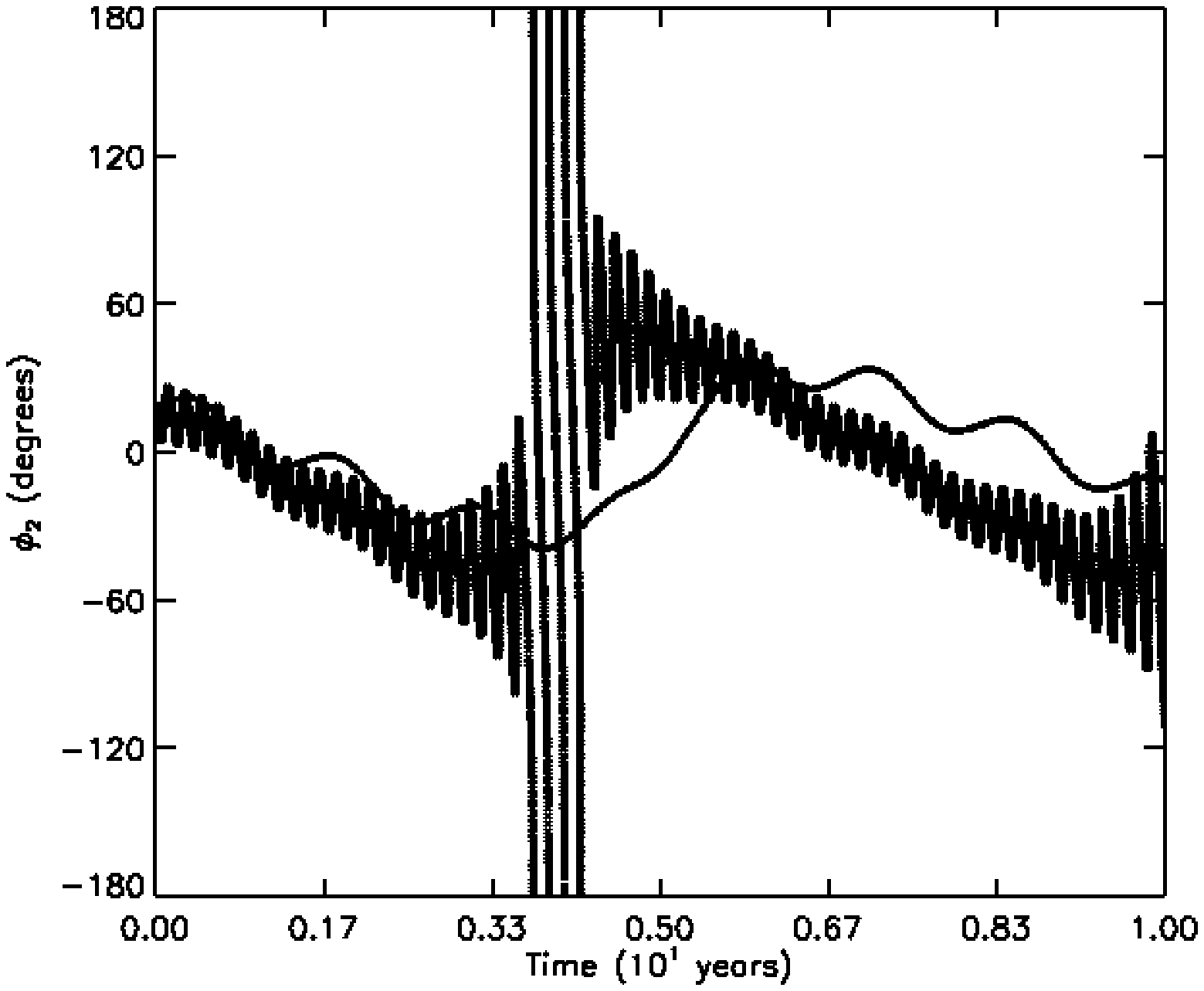,width=8.0truein,height=8.0truein}}

\caption{}\label{libcir}
\end{figure}

Figure \ref{libcir}: Evolution of $\phi_2$ without a disk (background curve) and
with a disk (foreground crosses) that is $1/3$ the mass of the central star
with a power law exponent of $w = 1.5$ and with $r_e = 0.3$ AU, $r_f = 20$ AU.
Note how the disk converts resonant libration into circulation.

\bigskip

Table 1: Constants and their standard deviation
obtained in Eq. (\ref{foufit})'s analytical fit
to the libration profile for the eccentric asteroid
from Fig. \ref{astecc} and the inclined asteroid from 
Fig. \ref{astinc}.  The range and standard deviation of
the difference between the fit and the model for the 
eccentric asteroid is  
$26.02^{\circ}$ and 
$8.16^{\circ}$, and for the inclined asteroid is
$2.63^{\circ}$ and
$0.14^{\circ}$.

\bigskip

Table 2: Constants and their standard deviations
obtained in Eq. (\ref{dopfit})'s analytical fit
to the libration profile for the eccentric asteroid
from Fig. \ref{astecc} and the inclined asteroid from 
Fig. \ref{astinc}.  The range and standard deviation of
the difference between the fit and the model for the 
eccentric asteroid is  
$3.87^{\circ}$ and 
$0.77^{\circ}$, and for the inclined asteroid is
$2.95^{\circ}$ and
$0.18^{\circ}$.

\bigskip

Table 3:  Initial parameters which are varied to produce
Figs. \ref{libamp32}-\ref{libampcur41}.

\bigskip

Table 4:  Summary of resonant and secular disturbing function cosine arguments, each of which may admit more than one term, relevant to the GJ 876 system up to fourth-order in eccentricities.

\bigskip

Table 5:  Summary of the orbital elements directly affected ($\star$) by central-body oblateness and precession, and a nascent disk, under the assumptions of orbit averaging and planarity.

\newpage

\centerline{Table 1}

\begin{table}[h]
\begin{center}
\renewcommand{\arraystretch}{1.5}
\begin{tabular}{| c | c | c | c | c |} \cline{1-5}
Constants & Eccentric asteroid & Eccentric asteroid $\sigma$ 
          & Inclined asteroid & Inclined asteroid $\sigma$ 
\\ \hline
$K_0$ & -0.204 & 0.10 & 180 & 0.0053 
\\ \hline
$K_1$ & 1.15 & 0.24 & 60.6 & 0.0087     
\\ \hline
$K_2$ & 38.0 & 0.14 & 11.5 & 0.017   
\\ \hline
$K_3$ & 0.883 & 0.000061 & 0.0805 & 0.0000028
\\ \hline
$K_4$ & 0.870 & 0.127 & -0.290 & 0.015
\\ \hline
$K_5$ & 0.041 & 0.253 & -0.556 & 0.010
\\ \hline
$K_6$ & 1.60 & 0.0030 & 0.238 & 0.00027
\\ \hline
$K_7$ & 0.0494 & 0.127 & -0.409 & 0.0077
\\ \hline
$K_8$ & -0.00379 & 0.25 & 0.0924 & 0.016
\\ \hline
$K_9$ & 3.23 & 0.051 & 0.406 & 0.00041 \\ \hline
\end{tabular}
\label{astfit}
\end{center}
\end{table}

\newpage

\centerline{Table 2}

\begin{table}[h]
\begin{center}
\renewcommand{\arraystretch}{1.5}
\begin{tabular}{| c | c | c | c | c |} \cline{1-5}
Constants & Eccentric asteroid & Eccentric asteroid $\sigma$ 
          & Inclined asteroid & Inclined asteroid $\sigma$ 
\\ \hline
$k_0$ & -0.0694 & 0.0098 & 180 & 0.014    
\\ \hline
$k_1$ & 36.8 & 0.013 & -61.3 & 0.015       
\\ \hline
$k_2$ & -1.56 & 0.00053 & 2.95 & 0.00039     
\\ \hline
$k_3$ & 1.97 & 0.00047 & 2.03 & 0.00086  
\\ \hline
$k_4$ & 2.05 & 0.0020 & 0.964 & 0.00061  
\\ \hline
$k_5$ & 0.442 & 0.0000020 & 0.0402 & 0.0000018 \\ \hline
\end{tabular}
\label{astfit2}
\end{center}
\end{table}

\newpage

\centerline{Table 3}

\begin{table}[h]
\begin{center}
\renewcommand{\arraystretch}{1.5}
\begin{tabular}{| c | c | c | c | c | c | c | c | c | c | } \cline{1-10}
Resonance  & $m_1$ & $m_2$ & $\alpha$ & $e_1$ &  $e_2$ 
                   & $\lambda_1$ & $\lambda_2$ 
                   & $\varpi_1$ & $\varpi_2$
\\ \hline
$3$:$2$ & $0.3 m_{Jup}$ & $0.3 m_{J}$ & 0.763 & 0.05 
        & 0.15 & 281 & 210 & 140 & 340 
\\ \hline
$4$:$1$ & $0.03 m_{Jup}$ & $0.3 m_{J}$ & 0.397 & 0.25 
        & 0.01 & 327 & 220 & 186.5 & 340 \\ \hline
\end{tabular}
\label{intparm}
\end{center}
\end{table}

\newpage

\centerline{Table 4}

\begin{table}[h]
\begin{center}
\renewcommand{\arraystretch}{1.5}
\begin{tabular}{| c | c | c | c | c | c |} \cline{1-6}
``resonant'' (R) or ``secular'' (S) & argument  & 1st-order & 2nd-order &  3rd-order & 4th-order \\ \hline\hline
S & $0$ &     &  $e_{1}^2$, \bbs\bbs  $e_{2}^2$  &   & 
 $e_{1}^4$, \bbs\bbs  $e_{2}^4$, \bbs\bbs $e_{1}^2 e_{2}^2$  \\ \hline
S & $\varpi_1 - \varpi_2$  &   &  $e_1 e_2$  &  & $e_{1}^3 e_2$, \bbs\bbs $e_1 e_{2}^3$   \\ \hline
S & $2 \varpi_1 - 2 \varpi_2$ &  &  &  &  $e_{1}^2 e_{2}^2$  \\ \hline
R & $2  \lambda_1 - \lambda_2 - \varpi_1 \equiv \phi_2$  & $e_1$ &   & $e_{1}^3$, \bbs\bbs $e_{1} e_{2}^2$ & \\ \hline
R & $2  \lambda_1 - \lambda_2 - \varpi_2 \equiv \phi_1$  & $e_2$ &   & $e_{2}^3$, \bbs\bbs $e_{1}^2 e_{2}$   & \\ \hline
R & $4  \lambda_1 - 2 \lambda_2 - 2 \varpi_1$  &    & $e_{1}^2$  &    
 & $e_{1}^4$,  \bbs\bbs  $e_{1}^2 e_{2}^2$ \\ \hline
R & $4  \lambda_1 - 2 \lambda_2 -  \varpi_1 - \varpi_2$  &     & $e_1 e_2$ &    &  $e_{1} e_{2}^3$,  \bbs\bbs  $e_{1}^3 e_{2}$ 
\\ \hline
R & $4  \lambda_1 - 2 \lambda_2 - 2 \varpi_2$ &    & $e_{2}^2$ &   &  $e_{2}^4$,  \bbs\bbs  $e_{1}^2 e_{2}^2$  
\\ \hline
R & $6  \lambda_1 - 3 \lambda_2 - 3 \varpi_1$  &    &   & $e_{1}^3$  &  
\\ \hline
R & $6  \lambda_1 - 3 \lambda_2 - 2 \varpi_1 - \varpi_2$ &    &    & $e_{1}^2 e_2$ &   \\ \hline
R & $6  \lambda_1 - 3 \lambda_2 - \varpi_1 - 2 \varpi_2$  &    &    & $e_{1} e_{2}^2$    &   \\ \hline
R & $6  \lambda_1 - 3 \lambda_2 - 3 \varpi_2$  &    &   &  $e_{2}^3$  &    \\ \hline
R & $8  \lambda_1 - 4 \lambda_2 - 4 \varpi_1$  &   &    &     &  $e_{1}^4$    \\ \hline
R & $8  \lambda_1 - 4 \lambda_2 - 3 \varpi_1 - \varpi_2$  &  &  &  &  $e_{1}^3 e_2$  \\ \hline
R & $8  \lambda_1 - 4 \lambda_2 - 2 \varpi_1 - 2 \varpi_2$  &   &   &   &  $e_{1}^2 e_{2}^2$   \\ \hline
R & $8  \lambda_1 - 4 \lambda_2 - \varpi_1 - 3 \varpi_2$  &   &   &   & $e_1 e_{2}^3$   \\ \hline
R & $8  \lambda_1 - 4 \lambda_2 - 4 \varpi_2$  &  &   &   & $e_{2}^4$ \\ \hline
\end{tabular}
\label{termlist}
\end{center}
\end{table}

\newpage

\centerline{Table 5}

\begin{table}[h]
\begin{center}
\renewcommand{\arraystretch}{1.5}
\begin{tabular}{c | c | c | c | c | c | c | c | c | c | c | } \cline{1-11}
effect: &  & \multicolumn{4}{c |}{oblateness}  & \multicolumn{4}{c |}{disk}  
 & precession  
\\ \hline
averaged? & & yes & yes & no & no & yes &
yes & no & no & no 
\\ \hline
planar? & & yes & no & yes & no & yes &
no & yes & no & no
\\ \hline\hline
& $\dot{a}$  &  &  &  $\star$ & $\star$  &   &
 & $\star$ & $\star$ &  
\\ \hline
& $\dot{e}$ &  & $\star$  & $\star$ & $\star$ &  & 
 & $\star$ & $\star$ &   
\\ \hline
& $\dot{I}$ &  & $\star$ &  & $\star$ &  &
 &  & $\star$ &  $\star$
\\ \hline
& $\dot{\epsilon}$ &  &  & $\star$ & $\star$ &  &
 & $\star$  & $\star$ &  $\star$
\\ \hline
& $\dot{\varpi}$ & $\star$ & $\star$ & $\star$ & $\star$ & $\star$ &
$\star$ & $\star$  & $\star$ & $\star$
\\ \hline
& $\dot{\Omega}$ &  & $\star$ &  & $\star$ &  &
$\star$ &  & $\star$   & $\star$  
\\ \hline
& $\dot{\phi}$ & $\star$  & $\star$ & $\star$  & $\star$  & $\star$ &
$\star$ & $\star$  & $\star$ & $\star$ 
\\ \hline
\end{tabular}
\label{effects}
\end{center}
\end{table}

\newpage

\newpage

\newpage

\newpage

\newpage

\newpage

\newpage

\clearpage
\newpage

\newpage

\newpage

\newpage

\newpage

\newpage

\newpage

\newpage

\newpage

\newpage

\newpage

\newpage

\newpage

\newpage

\end{document}